\documentclass[fontsize=12pt,a4paper,headings=normal,
twoside=false,leqno,parskip=half-,abstract=true]{scrartcl}
\usepackage[english]{babel}
\usepackage[utf8]{inputenc}
\usepackage{hyperref}
\hypersetup{
 pdftitle={Rebels under permutation symmetry},
 pdfauthor={Bernold Fiedler},
 colorlinks=true,
 linkcolor=blue,
 citecolor=blue,
 filecolor=blue,
 urlcolor=blue}
 \usepackage{placeins}
 
\usepackage{afterpage}
\usepackage{subcaption}

\usepackage{graphicx}
\usepackage[format=plain,labelfont=bf,font=small]{caption}
\usepackage{xcolor}
\usepackage[arrow, matrix, curve]{xy}
\usepackage{float}
\usepackage{caption}
\usepackage{enumitem}  
\usepackage{tabulary}
\usepackage{array}
\newcolumntype{N}[1]{>{\centering\arraybackslash}m{#1}}

\usepackage{amsmath,amsthm}
\swapnumbers % Switch number/label style
\usepackage{amssymb} 
\makeatletter
\newcommand{\tpitchfork}{%
  \vbox{
    \baselineskip\z@skip
    \lineskip-.52ex
    \lineskiplimit\maxdimen
    \m@th
    \ialign{##\crcr\hidewidth\smash{$-$}\hidewidth\crcr$\pitchfork$\crcr}
  }%
}
\makeatother
\usepackage{latexsym}

\usepackage[notref,notcite,color,   final %make final instead of draft, by % %
]{showkeys}
\definecolor{refkey}{rgb}{1,0,0}
\definecolor{labelkey}{rgb}{1,0,0}
\usepackage{tikz}

\usepackage{ upgreek }

  \mathchardef\ordinarycolon\mathcode`\:
  \mathcode`\:=\string"8000
  \begingroup \catcode`\:=\active
    \gdef:{\mathrel{\mathop\ordinarycolon}}
  \endgroup

\theoremstyle{plain}

\makeatletter

\@addtoreset{equation}{section}
\makeatother

\begin{document}

\title{\LARGE{
Global heteroclinic rebel dynamics\\
among large 2-clusters\\
in permutation equivariant systems\\
}}

\author{
 \\
Bernold Fiedler*, Sindre W.~Haugland**, \\
Felix P.~Kemeth***, Katharina Krischer** \\
{~}\\
%\emph{Dedicated to Alexander Mielke} \\
%\emph{on the occasion of his sixtieth birthday}\\
\vspace{2cm}}

\date{version of \today}
\maketitle
\thispagestyle{empty}

\vfill

*\\
Institut für Mathematik\\
Freie Universität Berlin\\
Arnimallee 3\\ 
14195 Berlin, Germany\\
\\
***\\
Institut für Physik\\
Technische Universität München\\
James-Franck-Straße 1\\
85748 Garching, Germany\\
\\
****\\
Department of Chemical and Biomolecular Engineering\\
Whiting School of Engineering\\
Johns Hopkins University\\
3400 North Charles Street\\
Baltimore, MD 21218, USA\\
% weitere Adressen hier eintragen

%%%%%%%%%%%%%%%%%%%%%%%%%%%%%%%%%%%%%%%%%%%%%%%%%%%%%%%%%%%

\newpage
\pagestyle{plain}
\pagenumbering{roman}
\setcounter{page}{1}

\begin{abstract}
\parindent 0cm
\parskip 4pt
We explore equivariant dynamics under the symmetric group $S_N$ of all permutations of $N$ elements.
Specifically we study one-parameter vector fields, up to cubic order, which commute with the standard real $(N-1)$-dimensional irreducible representation of $S_N$. The parameter is the linearization at the trivial 1-cluster equilibrium of total synchrony.

All equilibria are cluster solutions involving up to three clusters.
The resulting global dynamics is of gradient type: all bounded solutions are cluster equilibria and heteroclinic orbits between them.
In the limit of large $N$, we present a detailed analysis of the web of heteroclinic orbits among the plethora of 2-cluster equilibria.
Our focus is on the global dynamics of 3-cluster solutions with one rebel cluster of small size.
These solutions describe slow relative growth and decay of 2-cluster states. 
For $N\rightarrow\infty$, the limiting heteroclinic web defines an integrable \emph{rebel flow} in the space of 2-cluster equilibrium configurations.
We identify and study the seven qualitatively distinct global rebel flows which arise in this setting.

Applications include oscillators with all-to-all coupling, and electrochemistry. 
For illustration we consider synchronization clusters among $N$ complex Stuart-Landau oscillators with complex linear global coupling.

\end{abstract}

%\newpage
%\vspace{2cm}
\vfill
{\small \tableofcontents}

%%%%%%%%%%%%%%%%%%%%%%%%%%%%%%%%%%%%%%%%%%%%%%%%%%%%%%%%%%%

\newpage
\pagenumbering{arabic}
\setcounter{page}{1}

\section{Introduction}\label{intro}

\numberwithin{equation}{section}
\numberwithin{figure}{section}
\numberwithin{table}{section}
Networks of identical oscillators, with identical global all-to-all (e.g.~mean-field) coupling are a ubiquitous source of dynamical systems which are equivariant under the \emph{symmetric group}$S_N$ of all permutations of $N$ elements $\{1,\ldots, N\}$.
It is not our ambition here to survey those very extensive parts of the literature which present numerous, if scattered and often anecdotal, evidence based on simulations and, less frequently, experiments.
See our companion paper \cite{kf20} aimed at that community, for such a more applied focus.

Here, we rather develop a novel mathematical description, and analysis, of the gradient-like dynamics of large 2-clusters driven by heteroclinic orbits of small rebel clusters, which switch their cluster affiliation.
In the limit of large $N$, but restricted to polynomial vector fields of at most cubic order, we describe and study the resulting global heteroclinic dynamics as an integrable \emph{rebel flow} in the two-dimensional space of all 2-cluster equilibrium configurations.
Seven distinct rebel flows arise, in the one-parameter bifurcation setting \eqref{ODExn} below.
We describe these results in sections \ref{resul1} -- \ref{resul7}.

In section \ref{sl}, we address the specific example of identical Stuart-Land oscillators with identical all-to-all coupling, for general complex parameters. 
We include some references to earlier work on that specific problem, there.
See also our companion paper \cite{kf20}, for a specific parameter setting.
We will eliminate a global averaged phase oscillation.
Near the trivial periodic solution of total synchrony, we consider loss of synchrony, and of stability, through bifurcation at a zero transverse eigenvalue.
We reduce the complex ODE dynamics from $\mathbb{C}^N= \mathbb{R}^{2N}$ to a local center manifold of real dimension $N-1$.
In particular we study the resulting reduced dynamics of 2-cluster periodic solutions and their heteroclinic transitions, up to and including third order.

One main tool is equivariance under the full permutation group $S_N \, $.
For a general background on dynamics and equivariance see for example \cite{gost86, gost02, guho, van}.
For a more specific background on $S_N\,$-equivariance, equilibria, and their stability see \cite{elm, gost02, stelm, DiSt}.
An application to the evolutionary biology of sympatric speciation is outlined there.
For complementary mathematical perspectives on coupled phase oscillators and a focus on local Hopf bifurcation, in the spirit of equivariance, we refer to  \cite{Ash92, DMR, Ash07, Ash16} and the references there.
Global equivariant Hopf bifurcation of periodic solutions, not limited to $S_N\,$-equivariance, has been addressed in \cite{FieHabil}.

We recall the abstract setting of  \cite{elm, gost02, stelm, DiSt} next.
Permutations $\pi\in S_N$ act linearly on vectors $\mathbf{x}\in X:=\mathbb{R}^N$ by permutations of their components $x_n$. This linear representation of $S_N$ is given by
\begin{equation}
	\label{SN}
	(\pi \mathbf{x})_n := x_{\pi^{-1}(n)}\,.
\end{equation}

Group \emph{invariants} $I:\mathbb{R}^N\rightarrow \mathbb{R}$ satisfy 
\begin{equation}
	\label{inv}
	I(\pi \mathbf{x})=I(\mathbf{x})
\end{equation}
for all $\pi\in S_N$ and all $\mathbf{x}\in \mathbb{R}^N$, by definition.
The ring of polynomial $S_N$ invariants $I$ is freely generated by the power sums
\begin{equation}
	\label{pm}
	p_m := \sum_{n=1}^N x_n^m\,,
\end{equation}
for $m=1,\ldots,N$. We may subsume the case of constant $I$ as $m=0$.

\emph{Equivariant} vector fields $\mathbf{f}: \mathbb{R}^N \rightarrow \mathbb{R}^N$, here under the group $S_N\,$, commute with the linear group action:
\begin{equation}
	\label{eqv}
	\mathbf{f}(\pi \mathbf{x}) = \pi \mathbf{f}(\mathbf{x})\,,
\end{equation}
for all $\pi\in S_N$ and all $\mathbf{x}\in \mathbb{R}^N$.
For Lipschitz continuous equivariant $\mathbf{f}$, the solutions $\mathbf{x}=\mathbf{x}(t)$ of the associated ordinary differential equation (ODE) 
\begin{equation}
	\label{ODE}
	\dot{\mathbf{x}} = \mathbf{f}(\mathbf{x})
\end{equation}
are unique. 
Therefore the equivariance condition \eqref{eqv} means, equivalently, that $\pi \mathbf{x}(t)$ is a solution of \eqref{ODE}, whenever $\mathbf{x}(t)$ itself is a solution.

One example of group equivariant vector fields $\mathbf{f}(\mathbf{x})$ are the (negative) gradients
\begin{equation}
	\label{grainv}
	f_n(\mathbf{x}) := -\partial_n \,I(\mathbf{x})
\end{equation}
of group invariants $I$. Here $\partial_n$ denotes the partial derivative with respect to $x_n,$ for $ n=1,\ldots,N.$

The consequences of a \emph{gradient structure} \eqref{grainv} are striking, even without any group invariance.
Stationary solutions, alias equilibria $\mathbf{f}(\mathbf{x})=0$ of the ODE \eqref{ODE}, become \emph{critical points} $\nabla I(\mathbf{x})=-\mathbf{f}(\mathbf{x})=0$.
The \emph{energy}, or \emph{Lyapunov, function} $I(\mathbf{x}(t))$ decreases strictly with time $t$, along any nonstationary solution $\mathbf{x}(t)$.
In particular, any nonstationary solution $\mathbf{x}(t)$ which remains bounded for all real times $-\infty<t<+\infty$ is \emph{heteroclinic} between equilibria, i.e.~$\mathbf{x}(t)$ becomes stationary for $t\rightarrow\pm\infty$.
The energy $I$ at the target equilibrium (or equilibria), for $t\rightarrow +\infty$, is always strictly lower than at the source, i.e.~for $t\rightarrow -\infty$.
Any nonstationary recurrences and, in particular, any periodic, homoclinic, or chaotic  solutions, as well as any heteroclinic cycles, are therefore excluded, a priori.

The resulting ``webs'' of heteroclinic orbits (in the terminology of \cite{Ash07}) however, the main object of study in our present paper, can be quite intriguing.
For some illustrations in a different setting, involving parabolic partial differential equations and Jacobi systems of monotone nearest neighbor coupling, we refer to the (heteroclinic) directed connection graphs of \cite{FieTatra}, the geometric extensions in \cite{firo18}, and the references there.
Acyclicity, i.e.~the absence of directed cycles in the connection graphs, is a consequence of the gradient structure.
In the presence of continua of equilibria substantial further complications may arise, even under gradient-like dynamics.
See for example elliptic Hopf and Takens-Bogdanov bifurcation without equilibria, as discussed in \cite{FiLiAl, AfFiLi, LiebHabil}.
In a setting of five coupled Stuart-Landau oscillators, but going beyond gradient structure, Ashwin et al have also investigated heteroclinic cycles; see for example \cite{Ash07}.

Note that the linear zero sum space
\begin{equation}
	\label{X0}
	X_0:=\{\mathbf{x}\in X\,|\,p_1:= x_1+\ldots+x_N=0\}
\end{equation}
is an $(N-1)$-dimensional linear subspace of $X=\mathbb{R}^N$ which is invariant under the action \eqref{SN} of $S_N \,$.
The \emph{standard representation} of $S_N$ on $X_0$ is given by the restriction of the linear representation \eqref{SN} to $X_0$.
That representation is \emph{irreducible}: there does not exist any nontrivial proper subspace of $X_0$ which would also be invariant under all $S_N$.
In coupled oscillator settings with all-to-all coupling, for example like section \ref{sl} below, the space $X_0=(1,\ldots,1)^\perp$ often describes the simplest onset of asynchrony.

The following cubic $S_N\,$-equivariant one-parameter vector fields, with arbitrary real parameter $\lambda$ and fixed cubic coefficient $c$, are the main object of our present study:
\begin{equation}
	\label{ODExn}
	\dot{x}_n = f_n(\mathbf{x}) := (\lambda+c\cdot \langle x^2\rangle) x_n+ \widetilde{x^2_n}+ \widetilde{x^3_n}\,.
\end{equation}
The parameter $\lambda$ is the linearization at the trivial equilibrium $\mathbf{x}=0$ of total synchrony. We use the abbreviations 
\begin{equation}
	\label{avg}
	\langle x^m \rangle:= \tfrac{1}{N}p_m(\mathbf{x}), \qquad \widetilde{x^m_n}:= x^m_n- \langle x^m\rangle
\end{equation}
for the averages and the deviations of $m$-th powers.
It is a simple, but useful, exercise to check that the zero sum space $X_0$ is indeed invariant, not only under the linear action \eqref{SN} of the group $S_N$ but also under the nonlinear dynamics of \eqref{ODExn}.
Indeed $\langle x \rangle= \langle \widetilde{x_n^m} \rangle=0$.

It turns out that, up to scaling and possible time-reversal, the ODE \eqref{ODExn} on $X_0$ represents the most general cubic vector field which is equivariant under the standard representation \eqref{SN}, \eqref{X0}
Actually \cite{gost02}, 2.4--2.7, in the notation \eqref{avg}, provide the seemingly more general form
\begin{equation}
	\label{ODEgen}
	\dot{x}_n=f_n(\mathbf{x})= \lambda x_n + A \widetilde{x^2_n}+ B\widetilde{x^3_n}+C \langle x^2 \rangle x_n
	\end{equation}
of \eqref{ODExn}.
Here we have inserted $\pi_1:=p_1=\sum x_n=0$, in their notation, using definition \eqref{X0} of $X_0$.
See also \cite{stelm, DiSt}.
A much more detailed resource, which is difficult to obtain, is the thesis \cite{elm}.

Let us reduce \eqref{ODEgen} to the form \eqref{ODExn}, for general nonzero $A,B$.
Linear rescalings $t \rightarrow \tau t$, $x_n \rightarrow \sigma x_n$ amount to the replacements
\begin{equation}
	\label{scale}
	\lambda \rightarrow \tau \lambda, \quad A\rightarrow \tau \sigma A, \quad B \rightarrow \tau \sigma^2 B, \quad C \rightarrow c:= \tau \sigma^2C \, .
	\end{equation}
Renaming $\tau \lambda$ as $\lambda$, the choices $\sigma= A/B, \ \tau= B/A^2$ then lead to \eqref{ODExn} with 
\begin{equation}
	\label{cdef}
	c:=C/B \, .
	\end{equation}
Note that negative $B$, in particular, are associated with \emph{time reversal} in \eqref{ODExn}.

Actually, the cubic case \eqref{ODExn} possesses a \emph{gradient structure} \eqref{grainv}. 
Explicitly, $f_n(\mathbf{x}) := - \partial_n \,I(\mathbf{x})$ holds on $X_0 \, $, as required in \eqref{grainv}, for the quartic $S_N\,$-invariant polynomial
\begin{equation}
\label{gracub}
-I(\mathbf{x}):= (\tfrac{1}{2} \lambda\cdot p_2  + \tfrac{1}{4N}c\cdot p_2^2) +(\tfrac{1}{3} p_3 -\tfrac{1}{N} p_1p_2)
+ (\tfrac{1}{4} p_4-\tfrac{1}{N} p_1p_3)\,.
\end{equation}
Here we have used $p_1=0$ on $X_0$\,.
We caution the fast reader, however, that naive \emph{a priori} insertion of $p_1=0$ in \eqref{gracub} does not provide the correct \emph{gradient flow} \eqref{grainv} of $I$ on $X_0$.

For a broader perspective, let us comment on the gradient structure \eqref{ODE}, \eqref{grainv} from a slightly more abstract point of view.
It turns out that  \emph{all} $S_N\,$-equivariant polynomial vector fields $\mathbf{f}$ of the standard representation on $X_0 \,$, up to and including order three, are in fact gradients of scalar invariants.
The gradient property of polynomial equivariant vector fields fails, in contrast, from order four upwards.
This observation follows, for example, by direct inspection of the dimensions of the pertinent spaces of invariants and equivariants.
Alternatively, and more abstractly, these are classical results on dimension counts based on the Molien function; see for example \cite{DMR}.

The importance of the dynamics \eqref{ODExn} reaches far beyond any direct interpretation as a network of $N$ identical scalar ``cells'' with all-to-all coupling via power sums. 
Indeed, the bifurcation analysis of any fully permutation-symmetric network, at eigenvalue 0, typically leads to irreducible eigenspaces.
Beyond total synchrony $x_1=\ldots=x_N$ the standard representation on $X_0$ provides the simplest interesting case.
Any center manifold reduction, and subsequent truncation to cubic terms will then lead to our reference bifurcation problem \eqref{ODExn} with one or the other fixed value of the remaining cubic coefficient $c$.
See  \cite{elm, stelm} for an example concerned with the biological evolution of sympatric speciation.
In section \ref{sl} we choose the oscillatory cluster dynamics of identical Stuart-Landau oscillators with identical all-to-all coupling, for illustration.

The remaining sections are organized as follows.

In section \ref{clust} we study \emph{3-cluster solutions}, i.e.~solutions  $\mathbf{x}(t)$ of our reference ODE \eqref{ODExn} which feature at most three different values of the components $x_n\,$.
More generally, an $M$\emph{-cluster} features at most $M < N$ values
\begin{equation}
	\label{1Mcl}
	\{x_1,\ldots,x_N\} = \{\xi_1,\ldots,\xi_M\}\,.
\end{equation}
Note how $M$-clusters degenerate to $M'$-clusters, for some $M'<M$, when some of the $\xi$-components still coincide.
Dynamically this may happen asymptotically, in the limit $t\rightarrow \pm\infty$.

As Kuramoto noticed long ago \cite{Naku1994}, all equilibria of \eqref{ODExn} are (at most) 3-clusters.
The reason is simple: any equilibrium component $\xi=\xi_k=x_n$ must satisfy the same cubic equation
\begin{equation}
\label{eqc}
0=f_n(\mathbf{x}) = (\lambda+c\cdot \tfrac{1}{N}p_2) \xi+(\xi^2-\tfrac{1}{N}p_2)+(\xi^3-\tfrac{1}{N}p_3)\,,
\end{equation}
with the same coefficients $c,\lambda, p_2,p_3\,.$
This admits at most three distinct cluster values $\xi=\xi_1, \xi_2, \xi_3 \,$.

Cognizant of the gradient structure \eqref{grainv}, we aim at the dynamics of certain 3-cluster solutions which become heteroclinic between 2-cluster equilibria.
In section \ref{clust} we simplify this task as follows.
For $k=1,2,3$, let $N_k$ count the number of components $n$ which satisfy $x_n=\xi_k \, $; note $N_1+N_2+N_3=N$.
We then pass to the limit $N\rightarrow \infty$ of large clusters $N_1+ N_3$ with a remaining \emph{rebel cluster} $N_2$ of uniformly bounded size; for example we may focus on single rebels $N_2=1$.
Heteroclinic orbits between 2-cluster equilibria are then characterized by 
\begin{equation}
	\label{cluhet }
	\xi_2(t)-\xi_1(t)\rightarrow 0 \quad \mathrm{or} \quad \xi_3(t)-\xi_2(t)\rightarrow 0\,,
\end{equation}
for $t\rightarrow \pm \infty$.

In section \ref{large} our heteroclinic objective gets simplified, in the limit $N=\infty$, by the somewhat surprising appearance of a skew product structure \eqref{ODEsq}, \eqref{ODEy1s} over the scalar quantity $s(t) := (\xi_3(t)-\xi_1(t)) /(\alpha+1)$.
Here $\alpha:=N_1/N_3$ denotes the relative population fraction of components $x_n$ in the two large clusters,
and $s(t)$ describes the dynamics of these clusters relative to each other.
The gradient structure \eqref{gracub} leads to asymptotically stationary $s_* \,,$
\begin{equation}
	\label{sum}
	s(t) := (\xi_3(t)-\xi_1(t)) /(\alpha+1) \longrightarrow s_*=\mathrm{const},
\end{equation}
for $t\rightarrow \pm\infty$.
See section \ref{sumdy} for a detailed analysis of this dynamics, which drives the skew product.

In section \ref{rebdy} we pass to the asymptotic states of stationary $s=s_*=\mathrm{const}$.
In suitable coordinates $y= \xi_2- \xi_1 \, $ for the dynamics of the rebels $\xi_2\,$, this reduces our task to the discussion of a single scalar, cubic ODE 
\begin{equation}
	\label{block}
	\dot{y}\,=\, y \, (y-(\alpha+1)s_*)\,(y-\bar{y}(s_*))
\end{equation}
on the real line; see \eqref{ODEy1s}, \eqref{ODEys}.

Rebel heteroclinic solutions between 2-clusters will easily be identified, in terms of $y(t)$.
Indeed, the 2-cluster equilibria $\xi_2=\xi_1$ and $\xi_2=\xi_3$ correspond to the equilibria $y=0$ and $y=(\alpha+1)s_*\,$, respectively.
At the crucial 3-cluster equilibrium $\bar{y}(s_*)$, the small rebel cluster $\xi_2(t)$ might get stuck in its transition between the two major clusters $\xi_1, \xi_3 \, $. We call this  phenomenon  \emph{blocking} of 2-cluster heteroclinicity.
We thus arrive at the alternative of 2-cluster heteroclinicity, versus blocking of heteroclinicity by a 3-cluster.
In the non-blocking regions, the 2-cluster heteroclinicity will be encoded, globally, in the novel concept of a rebel flow, which is central to our subsequent results.

Based on the general procedures explained in section \ref{rebdy}, our main results on the heteroclinic rebel dynamics of \eqref{ODExn} between 2-cluster equilibria are presented and discussed in section \ref{resul}.
For the resulting rebel flows, each with the bifurcation parameter $\lambda$ as a first integral,
we distinguish seven intervals of qualitatively different global behavior.
The six critical cubic coefficients
\begin{equation}
	\label{cri}
	c\quad =\quad -2,\ -\tfrac{3}{2},\ -\tfrac{4}{3},\ -\tfrac{5}{4},\ -1,\ -\tfrac{1}{2}
\end{equation}
mark transitions between these qualitatively different rebel flows.
In sections \ref{resul1} -- \ref{resul7} we illustrate the resulting seven inequivalent rebel flows in the plane $(N_1/N,s)$ of 2-cluster equilibrium configurations; see diagrams \ref{fig_6_1}--\ref{fig_6_8}.
Each interval of $c$ is illustrated for a particular value of $c$ which qualitatively represents the rebel dynamics, for any fixed $c$ in that interval.

Each diagram is foliated by the parameters $\lambda$, as level curves, where such 2-cluster equilibria appear.
In the non-blocking regions, the rebel heteroclinic migration induce a slow drift of the population fraction $\alpha=N_1/N_3\,$, along constant parameter levels $\lambda$.
Heteroclinic transitions between the major clusters $N_1,N_3$ are in fact achieved by single rebels, or by rebel populations of relatively small size $N_2$.

In the limit $N\rightarrow\infty$, we represent this rebel dynamics by a formal \emph{rebel flow}, along each level curve of constant $\lambda$, in the $(\alpha,s)$- or $(N_1/N,s)$-plane.
In other words, the rebel flow encodes the heteroclinic dynamics of single rebels, towards the preferred 2-cluster, in the  $(\alpha,s)$-plane of all 2-cluster equilibrium configurations.
Simply because the parameter $\lambda$ is constant, along each heteroclinic orbit of \eqref{ODExn}, the constant value of $\lambda$ is a conserved quantity of the rebel flow.
In the terminology of \cite{ArnODE}: the rebel flow is integrable with \emph{first integral} $\lambda$.
Conversely, any discretization of the rebel flow by a grid of size ratios $\alpha=N_1/N_3\,$, for cluster sizes $N_1+N_3=N$ compatible with $N$, provides an approximation of the web of heteroclinic rebel dynamics, for large $N$.

Even for gradient flows \eqref{ODE}, \eqref{grainv} the connection graph of heteroclinic orbits between isolated equilibria cannot, in general, be described by the discretization of a rebel \emph{flow}, i.e.~by a rebel map.
Indeed the rebel flow description requires that the connection graph, with isolated equilibria as vertices and rebel heteroclinic orbits as directed edges, features a unique outgoing edge, for each equilibrium.
Similarly, by time reversal of the flow, there should be a unique incoming edge.
Kasner maps, in Bianchi cosmologies of general relativity, are a prominent example where this property is violated; see \cite{HeLaUg} for a recent survey.

On the discrete level of fixed large $N$, in our setting, the very possibility of a discretized rebel flow therefore hinges on the fact that each 2-cluster equilibrium configuration of sizes $(N_1,N_3)$ possesses a ``unique'' equilibrium target cluster, under outgoing heteroclinic orbits driven by a single rebel $N_2=1$, for fixed parameter $\lambda$.
Here ``uniqueness'' is understood after identification of symmetry related equilibria, by factoring out the full permutation group symmetry $S_N$.
Indeed, our focus on cluster sizes and cluster dynamics achieves just that symmetry reduction.
Then, the standalone rebellion either leads to ``the'' neighboring 2-cluster $(N_1+1,N_3-1)$, for increasing $\alpha=N_1/N_3\,$, or else to $(N_1-1,N_3+1)$, for decreasing $\alpha$.
Exceptions arise at the boundaries $N_1=1$ or $N_1=N-1$, alias $\alpha=1/(N-1) \rightarrow 0$ or $\alpha=N-1 \rightarrow \infty$, of course, and at the blocking boundary $y=\bar{y}(s_*)$ of \eqref{block}.
Rebellions there can lead to synchrony and to stationary 3-clusters, respectively.

As a corollary we observe that any nonvanishing component $\alpha=N_1/N_3$ of the rebel flow indicates instability of the particular 2-cluster configuration.
In particular, all 2-clusters outside the blocking region are unstable.
Their instability may lead towards 2-clusters with smaller, or larger, size ratios $\alpha$, depending on the direction of the rebel drift in $\alpha$.
Eventually, this leads to the blocking region, to 3-cluster equilibria, or to total 1-cluster synchrony as the only options for (multi-)stability.

For a more thorough mathematical discussion of unstable dimensions of 2-cluster equilibrium solutions, but not of the heteroclinic dynamics between them, we refer to Elmhirst's thesis \cite{elm}.
A concise, and more easily accessible, summary and an extension to 3-clusters is available by \cite{DiSt}.
A partial extension to, and unfolding of the 3-cluster $N/3$ degeneracy by, quintic vector fields has been achieved in \cite{DiRo06}.
For numerical (multi-)stability results on the Stuart-Landau system of section \ref{sl} with $N=16$ oscillators, we refer to \cite{KeDiss, KeHaKr2019} and our companion paper \cite{kf20}.

Contrary to standard intuition, the rebel transitions do not always favor the larger cluster.
The seven cases which we discuss in sections \ref{resul1} -- \ref{resul7} below in fact indicate how rebellion is an exceedingly subtle phenomenon, even in our simplistic cubic setting.

In section \ref{sl} we discuss the promised application to clustering in Stuart-Landau oscillators with global complex linear coupling. Section \ref{conc} provides a brief summary.

So, where are the theorems?
The present paper is a detailed case study of $S_N\,$-equivariant 3-cluster dynamics in the standard representation on $X_0 \,, $ as is.
Our main focus is the rebel dynamics among the 2-dimensional plethora of coexisting 2-cluster solutions of size ratios $ \alpha = N_1 / N_3 \,, $ for $N\rightarrow\infty$ and with a single bifurcation parameter $\lambda$.
The main novelty is our systematic, if unusual, presentation of the heteroclinic rebel dynamics as a formal rebel flow on the level contour diagrams $\lambda = \lambda (\alpha, s)$, in section \ref{rebdy}, where $s=(\xi_3-\xi_1)/(\alpha + 1)$ measures asynchrony.
All of section \ref{resul} can then be read as a long theorem, which establishes the pairwise inequivalence of these rebel flows in the seven complementary intervals
\begin{equation}
	\label{ccrif}
	c\not\in \{-2, -\tfrac{3}{2}, -\tfrac{4}{3}, -\tfrac{5}{4}, -1, -\tfrac{1}{2}\} \, .
\end{equation}
The six critical values of $c$ where rebel flows change are identified in sections \ref{sumc=-1} and \ref{rebdy}.
We conjecture, conversely, equivalence of the rebel flows in each of the seven intervals.
Alas, we did not embark on the, more cumbersome than enlightening, proof of this somewhat academic question.

\textbf{Acknowledgment.} 
The first author gratefully acknowledges the deep inspiration by, and hospitality of, his coauthors at München who initiated this work.
Ian Stewart personally provided us with a copy of the extensive thesis \cite{elm}, which saved us quite some duplication of effort. 
Extensive corrections of ever so many revisions were most diligently typeset by Patricia Habasescu.
This work has also been supported by the Deutsche Forschungsgemeinschaft, SFB910, project A4 ``Spatio-Temporal Patterns: Control, Delays, and Design", and by KR1189/18 ``Chimera States and Beyond".

\section{Cluster dynamics}\label{clust}

Let $\dot{\mathbf{x}}=\mathbf{f}(\mathbf{x})$ denote any vector field which is equivariant under the standard irreducible action of the symmetric group $S_N$ on the zero sum space $\mathbf{x}\in X_0\,$. 
See \eqref{SN}--\eqref{ODE} and \eqref{X0}.
The $M$-clusters are defined as those vectors $\mathbf{x}\in X_0$ which possess at most $M$ distinct components $x_n=\xi_k$; see \eqref{1Mcl}. After applying a suitable permutation $\pi\in S_N$ to $\mathbf{x}$ if necessary, we may assume without loss of generality that the indices are sorted as
\begin{equation}
\label{2Mcl}
x_1=\ldots=x_{N_1}\,,\quad \ldots\,, \quad x_{N_1+\ldots+N_{M-1}+1}=\ldots=x_N\,.
\end{equation}
We call $N_k$ the \emph{cluster size}, and $\xi_k$ the \emph{cluster value}, of cluster $k$, for $k=1, \ldots, M$. In other words, $\mathbf{x}$ is fixed under the direct product $S_\mathbf{N}:=S_{N_1}\times\ldots\times S_{N_M}$ of permutation subgroups, where the first factor $S_{N_1}$ acts on the first $N_1$ components of $\mathbf{x}$, and so on.
Any other $M$-cluster is fixed under a group suitably conjugate to $S_\mathbf{N}$.

By \eqref{eqv}, the linear space of $S_\mathbf{N}$-fixed vectors $\mathbf{x}$ is invariant under the ODE flow of $\mathbf{f}$.
In particular, nondegenerate $M$-clusters $\mathbf{x}(t)$ remain nondegenerate $M$-clusters, for all finite times $t$.
Only in the limit $t\rightarrow\ \pm \infty$, an $M$-cluster $\mathbf{x}(t)$ may possibly degenerate to an $M'$-cluster with fewer clusters, i.e.~$M'<M$.
Since any equilibria are at most 3-clusters, by \eqref{eqc}, this is precisely the situation which we plan to study, for $M=3$ and $M'=2$.

Specifically, consider the dynamics of any nondegenerate 3-cluster
\begin{equation}
	\label{3cl}
	\{x_1,\ldots,x_n\} = \{\xi_1,\xi_2,\xi_3\}
\end{equation}
in our all-to-all coupled system \eqref{ODExn}.
Then the power sums $p_m$ of \eqref{pm} become
\begin{equation}
	\label{pxi}
	p_m=N_1 \xi_1^m+N_2 \xi_2^m+N_3 \xi_3^m \ .
\end{equation}
The cluster sizes $N_k\geq1$, respectively, count the number of times the distinct cluster values $\xi_k$ occur among the $x_n \,$.

With these weighted power sums $p_m\,$, the resulting dynamics of the cluster values $\xi_k\,$, for $k=1,2,3$, is of course given by the 3-cluster system
\begin{equation}
	\label{ODExi}
	\dot{\xi_k} = (\lambda+c\cdot \tfrac{1}{N}p_2) \xi_k+(\xi_k^2-\tfrac{1}{N}p_2)+(\xi_k^3-\tfrac{1}{N}p_3)\,.
\end{equation}
Here we have simply replaced  $x_n$ by $\xi_k$\,, in \eqref{ODExn}.

Taking differences $\xi_j-\xi_k$ of any two equations in \eqref{ODExi} we obtain
\begin{equation}
	\label{ODEjk}
	\tfrac{d}{dt}(\xi_j-\xi_k) = (\xi_j-\xi_k)\left(\lambda+c\cdot \tfrac{1}{N}p_2 +(\xi_j+\xi_k)+(\xi_j^2+\xi_j \xi_k + \xi_k^2)\right).
\end{equation}
Consider the redundant scaled difference variables
\begin{equation}
\label{xi2y}
	y_1:=\frac{N_3}{N}(\xi_2-\xi_1), \quad y_2:=\frac{N_3}{N}(\xi_3-\xi_2), \quad y_3:=\frac{N_3}{N}(\xi_1-\xi_3)=-(y_1+y_2)\,.
\end{equation}
The flow invariant zero sum space $X_0$ of \eqref{X0} becomes planar, in the variables $\xi_k\,$:
\begin{equation}
	\label{xi0}
	0 = p_1 = N_1 \xi_1+N_2 \xi_2+N_3 \xi_3 \,.
\end{equation}
Therefore it is not surprising that we can invert the transformation from the redundant coordinates $(\xi_1,\xi_2,\xi_3)$ on $X_0$ to $(y_1,y_2)\in\mathbb{R}^2$ by
\begin{equation}
	\label{y2xi}
	\begin{aligned}
		\xi_1 &=& -(1+\tfrac{N_2}{N_3})&y_1 &-\ y_2\,;\\
		\xi_2 &=&  \tfrac{N_1}{N_3}&y_1 &-\ y_2\,;\\
		\xi_3 &=& \tfrac{N_1}{N_3}&y_1 &+\ (\tfrac{N_1}{N_3}+\tfrac{N_2}{N_3})y_2\,.
	\end{aligned}
\end{equation}
In principle, this allows us to rewrite the 3-system \eqref{ODExi}, i.e.~a planar system on $X_0$, in terms of the two new variables $y_1,y_2$.
The explicit calculation is a little messy.
In the limit $N\rightarrow\infty$ of large symmetric groups $S_N\,$, however, the calculation will simplify.

\section{The limit of large symmetric groups $\mathbf{S_N}$}\label{large}

As announced in the introduction, we now consider the $S_N\,$-equivariant 3-cluster dynamics \eqref{ODExi} of \eqref{ODExn}, in the limit of large $N$.
Fix a finite asymptotic size ratio
\begin{equation}
\label{alpha}
N_1/N_3 \rightarrow \alpha \in (0,\infty)
\end{equation}
of the two large clusters sizes $N_1$ and $N_3\,$, for $N\rightarrow\infty$.
We assume that the size $N_2$ of the cluster $\xi_2$ remains small compared to $N_1+N_3=N-N_2 \,,$ i.e.
\begin{equation}
\label{rebel}
N_2/N \rightarrow 0\,,\qquad \mathrm{for} \ \ N\rightarrow\infty\,.
\end{equation}
Note how \eqref{rebel} is equivalent to $N_2/N_3\rightarrow 0$, and likewise to $N_2/N_1 \rightarrow 0,$ for $N\rightarrow\infty$. 
We therefore call the comparatively tiny cluster $(N_2,\xi_2)$ the \emph{rebel cluster}.
In fact, we may well consider the minimal size $N_2=1$ of single, standalone rebels.

Inserting the limits \eqref{alpha}, \eqref{rebel} into the transformation \eqref{y2xi} above provides the simplified expressions
\begin{equation}
	\label{y2xi-infty}
	\begin{aligned}
		\xi_1 &=& -&y_1 &- y_2\,;\\
		\xi_2 &=&  \alpha&y_1 &- y_2\,;\\
		\xi_3 &=& \alpha&y_1 &+ \alpha y_2\,.
	\end{aligned}
\end{equation}
In the above limit $N\rightarrow\infty$, this allows us to rewrite the 3-cluster ODE \eqref{ODExi} in the still slightly unwieldy planar form
\begin{align}
	\label{ODEy1y2}
&\resizebox{.91\hsize}{!}{$\displaystyle{
	\dot{y}_1 = y_1\Big(\lambda+(\alpha-1)y_1-2y_2+(\alpha^2-\alpha+1)y_1^2+3(1-\alpha)y_1y_2+3y_2^2+\alpha c(y_1+y_2)^2 \Big)
}$}\\
	\label{ODEy2y1}
&\resizebox{.91\hsize}{!}{$\displaystyle{
	\dot{y}_2 = y_2\Big(\lambda+2\alpha y_1+(\alpha-1)y_2+3\alpha^2y_1^2+3\alpha(\alpha-1)y_1y_2+(\alpha^2-\alpha+1)y_2^2+\alpha c(y_1+y_2)^2 \Big)
}$}	
\end{align}
Just for academic completeness -- or so it seems at first -- let us also write the resulting ODE for the sum
\begin{equation}
\label{3sum}
s:=y_1+y_2=-y_3=(\xi_3-\xi_1)/(\alpha+1)
\end{equation}
which appeared in \eqref{xi2y}, redundantly:
\begin{equation}
	\label{ODEsq}
	\dot{s} = s\left(\lambda+(\alpha-1)s+q s^2 \right)\,; \qquad q:=\alpha^2+(c-1)\alpha+1\,.
\end{equation}
This is an at most cubic scalar ODE for the sum $s$ alone.
In particular, bounded solutions $ s(t)$ converge to some equilibria $s \equiv s_*$ of \eqref{ODEsq} for $ t \rightarrow \pm \infty \,$, respectively.
Substitution of $y_2=s-y_1$ in \eqref{ODEy1y2} provides the complementing ODE
\begin{equation}
\label{ODEy1s}
\dot{y}_1 = (\alpha+1)^2 \,y_1\,(y_1-s)\,(y_1-\bar{y}_1(s)) + y_1 \dot{s}/s \,.
\end{equation}
Here we have abbreviated 
\begin{equation}
	\label{y1bar}
	\bar{y}_1(s) := (\alpha+1)^{-1}((2-\alpha)s-1) \, .
\end{equation}
The polynomial $\dot{s}/s$ abbreviates the quadratic expression in the parenthesis of \eqref{ODEsq}.

In conclusion, we observe a \emph{skew product structure}, in the limit $N\rightarrow$ of two large clusters $N_1,N_3\,$, and one comparatively small cluster $N_2$.
Indeed the two ODEs \eqref{ODEsq} and \eqref{ODEy1s} identify the $N$-asymptotic 3-cluster dynamics \eqref{ODExi} in the zero sum subspace $X_0$ of \eqref{xi0} as a system where the autonomous dynamics \eqref{ODEsq} of $s$ drives the scalar dynamics \eqref{ODEy1s} of $y_1\,$. 

Perhaps the skew product structure of our restricted 3-cluster problem should not surprise us, after all.
In fact  \eqref{ODEsq} describes the relative dynamics $s=(\xi_3-\xi_1)/(\alpha+1)$ of the two large clusters $\xi_1$ and $\xi_3\,$, which is not affected by the comparatively small number $N_2$ of rebels $\xi_2 \,$.
Ignoring $N_2 \,,$ indeed, the zero sum condition \eqref{xi0} implies conservation of $N_1\xi_1+N_3\xi_3=0$, and hence a one-dimensional autonomous dynamics for the difference variable $s$ of \eqref{3sum}.
Because $s=0$ indicates synchrony of the two large clusters, i.e.~effectively a one-cluster dynamics, we also call $s$ the \emph{asynchrony variable}.
The rebel dynamics \eqref{ODEy1s} describes the remaining deviation $y_1=(\xi_2-\xi_1)/(\alpha+1)$ of the rebels $\xi_2$ in the small cluster $(N_2,\xi_2)$ from the state $\xi_1$ of the large cluster $(N_1,\xi_1)$, once the two large clusters $(N_1, \xi_1), (N_3, \xi_3)$ have reached a status quo equilibrium  $s=s_*$ according to their size ratio $\alpha = N_1/N_3 \,$.

\section{Two-cluster dynamics}\label{sumdy}

In this section we discuss the autonomous dynamics of the two large clusters $(N_1,\xi_1)$ and $(N_3,\xi_3)$.
By \eqref{ODEsq}, we only have to study  the asynchrony sum $s=(\xi_3-\xi_1)/(\alpha+1)=y_1+y_2$ defined in \eqref{3sum}, i.e.
\begin{equation}
	\label{ODEs}
	\dot{s} = s\left(\lambda+(\alpha-1)s+q s^2 \right)\,.
\end{equation}
Here the asymptotic ratio $0<\alpha= \lim N_1/N_3 < \infty$ of the sizes $N_1$ and $N_3$ of the two large clusters, for $N\rightarrow\infty$, is a fixed parameter, in addition to the cubic coefficient $c$ and the bifurcation parameter $\lambda$.
Also from \eqref{ODEsq}, we recall the abbreviation
\begin{equation}
	\label{ODEq}
	q=q(\alpha):=\alpha^2+(c-1)\alpha+1
\end{equation}
for the quadratic coefficient $q$ in the parenthesis of \eqref{ODEs}.

The scalar ODE \eqref{ODEs} is cubic in $s$ with trivial equilibrium $s=0$ of total synchrony.
The remaining equilibria $s=s_*$ are characterized by the vanishing quadratic parenthesis in \eqref{ODEs} at bifurcation parameters $\lambda$, i.e.~at parameters
\begin{equation}
	\label{lamas}
	\lambda=\lambda(\alpha,s):= s(1-\alpha -q s)\,.
\end{equation}

Explicit and elementary calculations reveal the standard bifurcation diagrams with respect to $\lambda$, for fixed size ratio $\alpha$ and cubic coefficient $c$.
For example we obtain
\begin{align}
	\label{a=0}
	\dot{s}&=s\left(\lambda-s(1-s)\right)&\mathrm{at}\  \alpha=0\, , &\ q=1 ;\\
	\label{a=1}
	\dot{s}&=s\left(\lambda+(c+1)s^2\right)&\mathrm{at}\  \alpha=1, &\ q=c+1\,.
\end{align}
We discuss three cases depending on the sign of $c+1$, below.
See section \ref{resul} for many additional relevant examples.

\subsection{The degenerate transition case $\mathbf{c=-1}$}\label{sumc=-1}

%%%%%%%%%%%%%%%%%%%%%%%%%%%%%%
\begin{figure}[p]
\centering
	\includegraphics[width=0.8\textwidth]{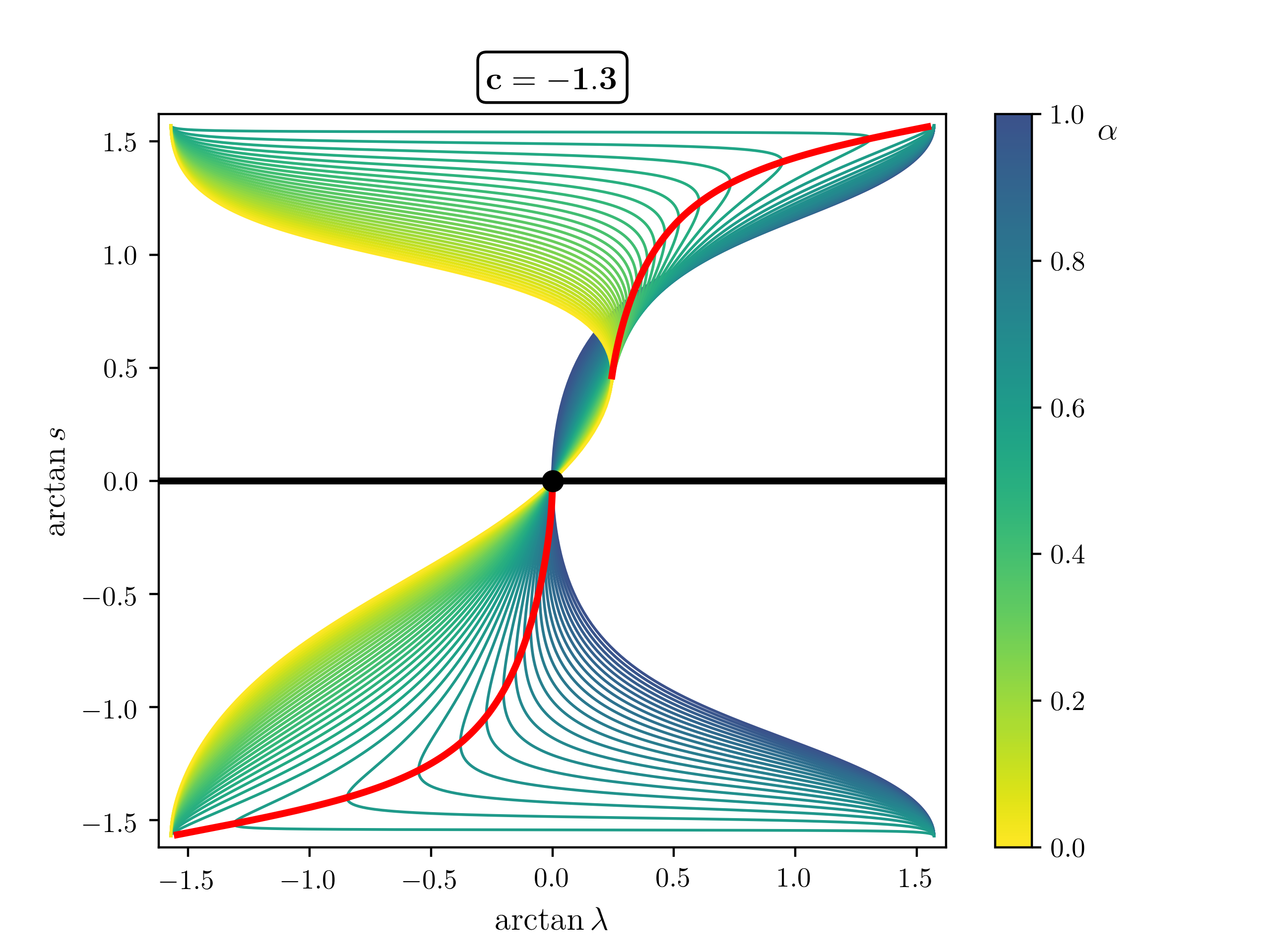}
	\caption{
	Global bifurcation diagrams for the compactified asynchrony $s$ of 2-cluster equilibria in the ODE flow \eqref{ODEs}, with $c=-1.3<-1$; see \eqref{3sum} and subsection \ref{sumc<-1}.
	The compactified horizontal $\lambda$-axis $\{s=0\}$ (black) represents the one-cluster case of total synchrony $\xi_3=\xi_1$.
	All 2-cluster bifurcation curves coexist, in the same phase space $\mathbf{x}\in X_0$\,, for realizable ratios $\alpha=N_1/N_3\in (0,1]$.
	The color shading indicates fixed values of $\alpha$ increasing from $\alpha=0$ (yellow) to $\alpha=1$ (blue), along each bifurcation curve in the $(\lambda,s)$ plane.
	The quadratic coefficient $q$ in \eqref{lamas} changes sign at $\alpha=\alpha_c \in (0,1)$; see \eqref{ac}.
	Specifically, $\alpha_c=0.5821...$ for $c=-1.3$.
	The redundant cases $1/\alpha=N_3/N_1\in (0,1)$ are omitted.
	Red: the two branches of extreme saddle-node values $(\lambda, s)=(\lambda_\mathrm{minmax}(\alpha),s_\mathrm{minmax}(\alpha))$ on each bifurcation curve; see \eqref{sminmax}, \eqref{lamminmax}.
	Positive $q$, for $0\leq\alpha<\alpha_c$, imply positive $s_\mathrm{minmax}$.
	Negative $q$, for $\alpha_c<\alpha \leq 1$, imply $s_\mathrm{minmax}<0$.
	The size ratio $\alpha=\alpha_c$ is realized in the limit of infinite $|\lambda|$ and $|s|$.
	In-/stability of each equilibrium $s_*$, within 2-cluster dynamics, can easily be derived from exchange of stability, at $\lambda=0$ and the saddle-nodes, or explicitly from \eqref{ODEs}.
}
	\label{fig_4_1} 
\centering
	\includegraphics[width=0.9\textwidth]{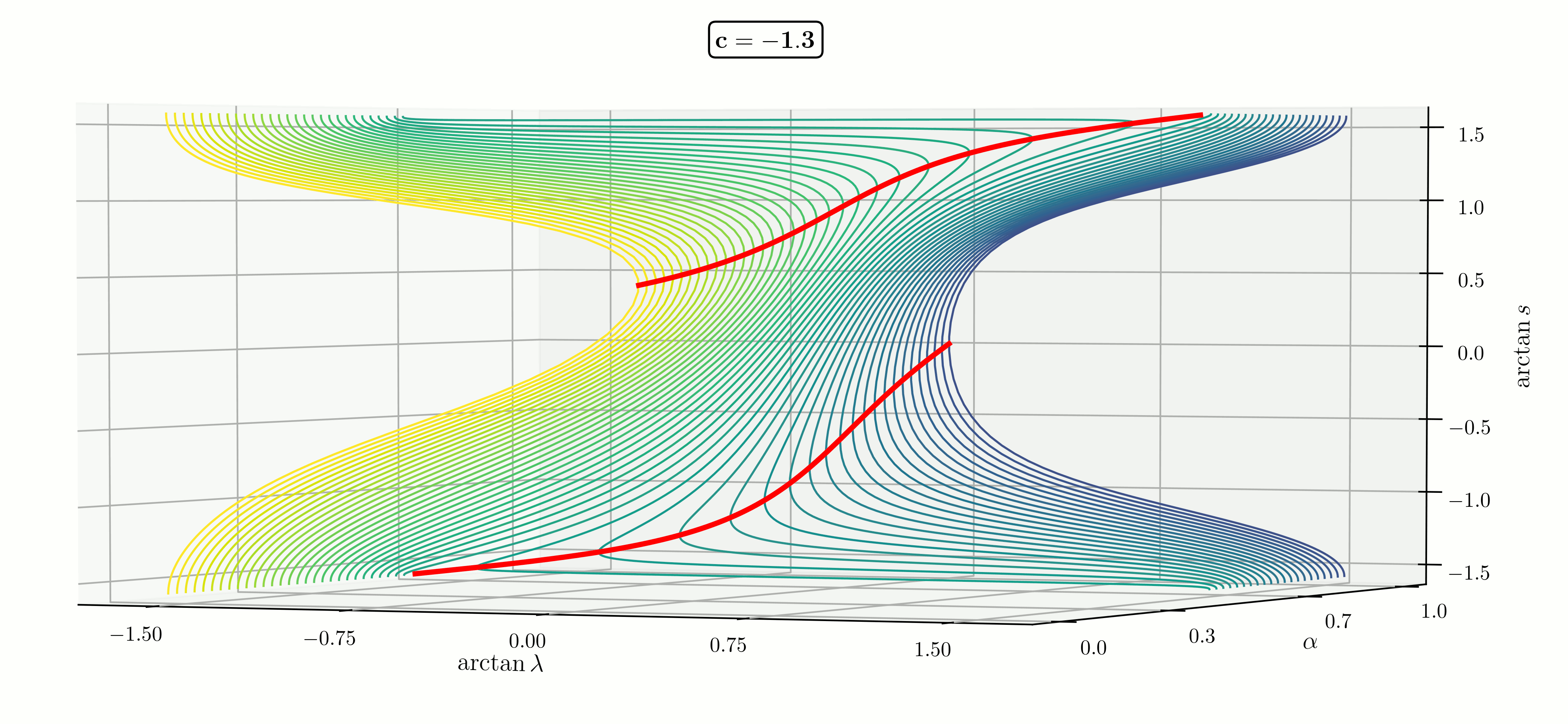}
	\caption{
	The bifurcation diagram of figure \ref{fig_4_1}, rotated such that the cluster size ratios $\alpha\in (0,1]$ can be visualized as a second ``parameter''.
	Color coding as before, but with yellow in front and blue in the background. Note the red fold curves, for the projections into the horizontal plane $(\arctan\lambda,\,\alpha)$ of figure \ref{fig_4_3}.
}
	\label{fig_4_2} 
\end{figure}
%%%%%%%%%%%%%%%%%%%%%%%%%%%%%%

In this case, the quadratic coefficient $q=(\alpha-1)^2$ is nonnegative and vanishes at $\alpha=1$, only.
Note how the $s$-dynamics becomes linear, $\dot{s}=\lambda s$, at $\alpha=1$; see \eqref{a=1}.
For $0\leq\alpha<1$ we my rescale $s$ to $\widetilde{s}:=(1-\alpha)s$ and obtain the $\alpha$ independent ODE
\begin{equation}
	\label{ODEstilde}
	\dot{\widetilde{s}} = \widetilde{s}\left(\lambda-\widetilde{s}+\widetilde{s}^2 \right)\,.
\end{equation}
which coincides with the case $\alpha=0$ of \eqref{a=0}.

Stability of the equilibria $s_*=\widetilde{s}_*/(1-\alpha)$ for any $0\leq\alpha<1$ is easily determined.
For $\lambda<1/4$, we have three equilibria $\widetilde{s}_*$.
Since 
\begin{equation}
\label{qs3}
\dot{s} = q s^3+\ldots
\end{equation} 
with $q=(\alpha-1)^2>0$, the top and bottom equilibrium are unstable, while the intermediate equilibrium is stable.
At $\lambda=1/4$, of course, we obtain a saddle-node equilibrium $s_*=\tfrac{1}{2}(1-\alpha)^{-1}$.
For $\lambda>1/4$ only the trivial equilibrium $s_*=0$ of total synchrony remains, which is unstable for all $\lambda>0$.

%%%%%%%%%%%%%%%%%%%%%%%%%%%%%%
\begin{figure}[p]
\centering
	\includegraphics[width=0.7\textwidth]{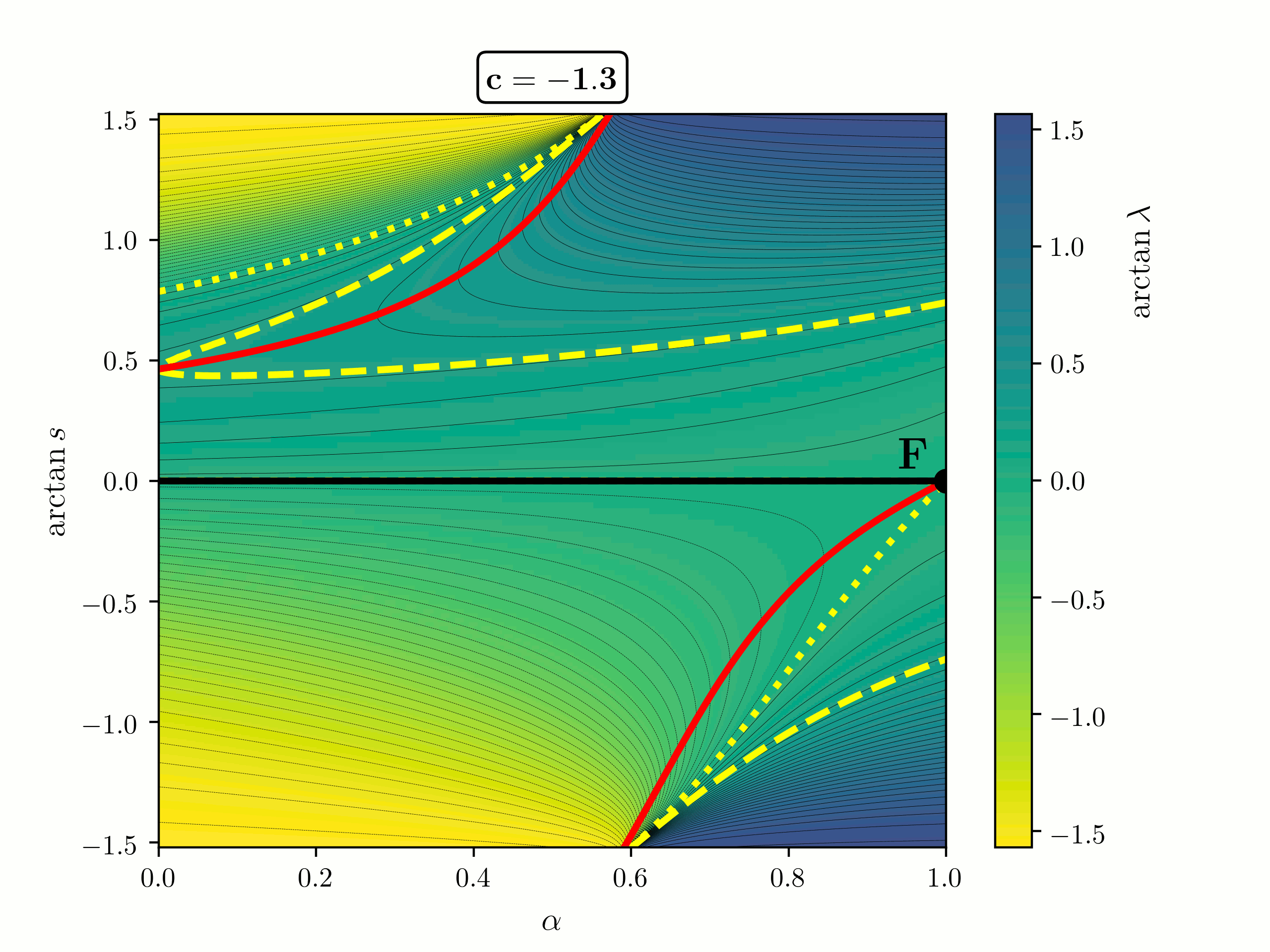}
	\caption{
	Level curves of $\lambda=\lambda(\alpha,s)$ for 2-cluster equilibria $s=s_*>0$ of the ODE flow \eqref{ODEs} with $c=-1.3<-1$. 
	Here $\alpha= N_1/N_3$ is horizontal, and $\arctan s$ is plotted vertically.
	Colors from yellow to blue indicate increasing values of $-\infty<\lambda<+\infty$, this time.
	Note the black and the two dotted yellow level curves of $\lambda=0$ which intersect at the only critical point $\mathbf{F}$ of $\lambda(\alpha,s)$.
	In particular, any level curve begins and terminates at the boundary, as described in the text.
	Another example is the dashed yellow level curve of the value $\lambda=1/4$.
	Restricted to the left vertical $s$-axis, at $\alpha=0$, this is the maximal value of $\lambda$.
	As in figure \ref{fig_4_1}, the two red curves indicate the values $s=s_\mathrm{minmax}(\alpha)$ where saddle-node bifurcations occur at the levels $\lambda=\lambda_\mathrm{minmax}(\alpha)$.
	Equivalently, they indicate extremal values of $\alpha$, on level curves of $\lambda$ in that region.
	Note the pole $s=\pm\infty$ at critical size ratio $\alpha=\alpha_c=0.5821...$ .
	The region of stable equilibria $s=s_*\,$, within 2-cluster dynamics \eqref{ODEs}, is located between the two red curves.}
	\label{fig_4_3} 
\centering
	\includegraphics[width = 0.90\textwidth]{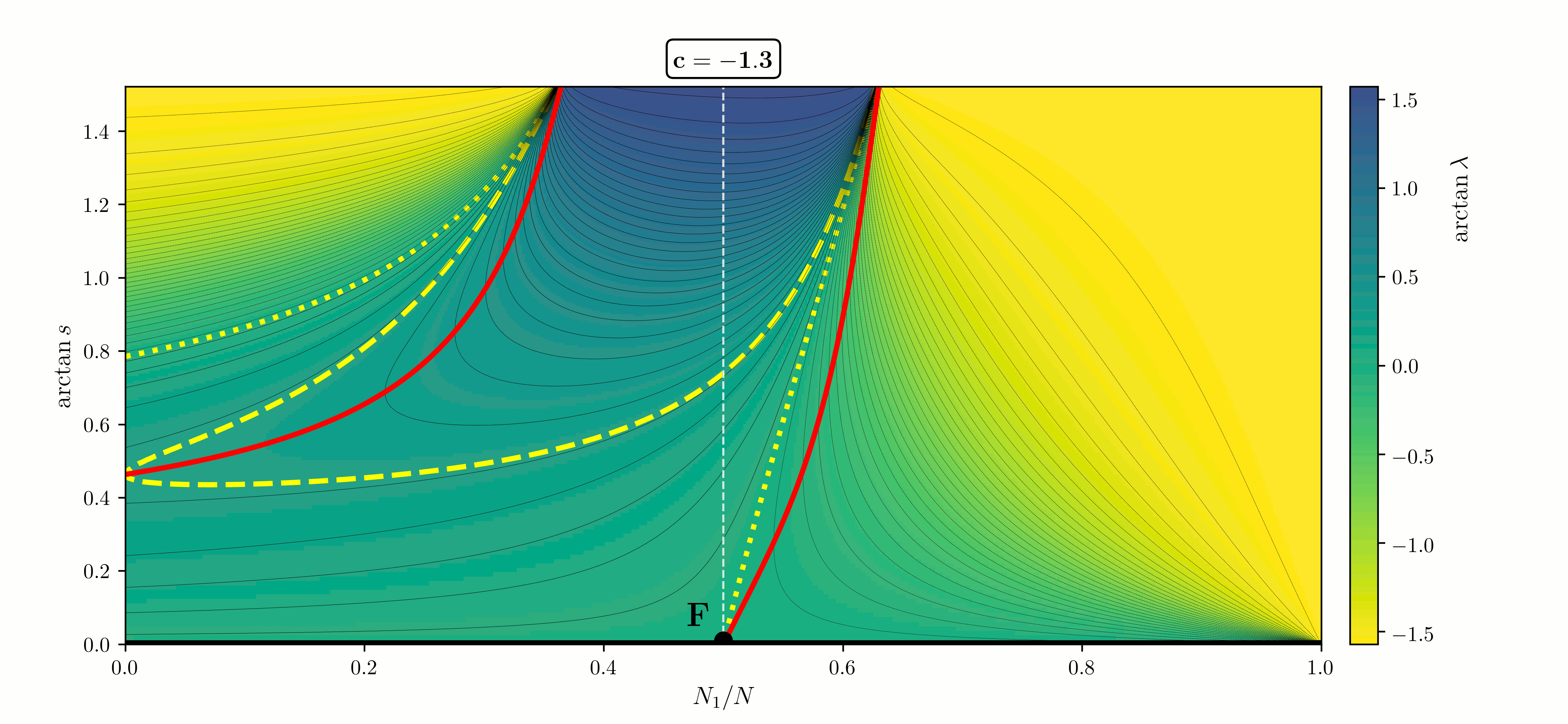}
	\caption{
	The substitution $N_1 \leftrightarrow N_3,\ y_1 \leftrightarrow -y_2$ allows for a gluing identification $s \leftrightarrow -s$ at the right boundary $\alpha=N_1/N_3=1$ of figure \ref{fig_4_3}.
	The new horizontal axis $N_1/N=\alpha/(\alpha+1)\in[0,1]$ therefore compactifies cluster ratios $\alpha\in[0,\infty]$ and allows us to omit $s<0$ as redundant.
	The break-even point $N_1/N=1/2$, alias $\alpha=1$, of equal cluster size $N_1=N_3$ is marked by a dashed white vertical line. 
	Again, the region of stable equilibria $s=s_*\,$, within 2-cluster dynamics \eqref{ODEs}, is located between the two red curves.
	Note the poles $s=+\infty$ at the critical size ratio $N_1/N=\alpha_c/(\alpha_c+1)=0.3679...$ and its complement $1/(\alpha_c+1)=0.6321...$ .
}
	\label{fig_4_4} 
\end{figure}
%%%%%%%%%%%%%%%%%%%%%%%%%%%%%%

\subsection{The case $\mathbf{c<-1}$}\label{sumc<-1}

In this case, the quadratic coefficient $q=q(\alpha)$ in \eqref{ODEs}, \eqref{ODEq} changes sign strictly, at
\begin{equation}
	\label{ac}
	\alpha = \alpha_c := \tfrac{1}{2}(1-c-\sqrt{(1-c)^2-4})\ \in\ (0,1)\,.
\end{equation}
Specifically $q>0$, for $0\leq\alpha<\alpha_c$, and $q<0$, for $\alpha_c<\alpha\leq1$.
Interchanging $N_1$ with $N_3\,,$ we omit the redundant reciprocal cases $\alpha= N_1/N_3>1$, for now.

For a specific, but not quite arbitrary, example we fix the cubic coefficient $c=-1.3$\,.
See figure \ref{fig_4_1} for the resulting bifurcation diagrams of \eqref{ODEs}.
The 2-cluster equilibria $s=s_*\neq 0$, at any fixed $\lambda=\lambda_*$ and size ratio $\alpha=\alpha_*\,$, appear as the intersections $s=s_*$ of the bifurcation curve for size ratio $\alpha$ with the vertical line $\lambda=\lambda_*\,$, in this plot.
The size ratio $\alpha=N_1/N_3$ may be considered as a fixed ``parameter'', in any of the invariant cluster subspaces \eqref{2Mcl}.
We therefore plot the bifurcation diagrams as a family of curves, parametrized over discrete values $\alpha$.
Color coding is from yellow, at $\alpha=0$, to blue, at $\alpha=1$.
Since all these bifurcation diagrams coexist, in the large $(N-1)$-dimensional phase space $X_0\,$, we superimpose all bifurcation curves in figure \ref{fig_4_1}.
Figure \ref{fig_4_2} unfolds this superposition with respect to the size ratios $\alpha$.

The less standard contour plot of figure \ref{fig_4_3} tracks the level contours of the parameter $\lambda=\lambda(\alpha,s)$, as a function of $0 < \alpha =N_1/N_3 \leq1$ (horizontal) and $-\pi/2<\arctan s<\pi/2$ (vertical), at which 2-cluster equilibria $s=s_*\neq 0$ occur; see \eqref{lamas}.
We use $\arctan s$ again, rather than $s$ itself, for compactification of the unbounded range of $s\in\mathbb{R}$.
The level contours are in fact level curves because the only critical point $\mathbf{F}$ of $\lambda$, located at $(N_1/N_3=1\, , s=0)$, is a nondegenerate saddle.
This accounts for the two level curves of the level contour $\lambda = 0$, one black and one dotted yellow, which intersect at $\mathbf{F}$.
The third level curve  of $\lambda = 0$ emanates from the left boundary as a dotted yellow curve.

Next we drop the assumption $\alpha= N_1/N_3 \leq 1$ and allow arbitrary sizes $N_1, N_3$ of the two clusters.
Without loss of generality, we may then label the large clusters $(N_1, \xi_1)$ and $(N_3, \xi_3)$ such that the asynchrony
\begin{equation}
	\label{s>0}
	s=(\xi_3- \xi_1)/ (\alpha + 1)>0
\end{equation}
is strictly positive.
This allows us to discard the redundantly symmetric case $s<0$, a priori.
Caution is required because our choice admits any cluster ratio $\alpha = N_1/N_3 \in (0, \infty)$.
To represent $\alpha$, we therefore use the percentage $N_1/N= \alpha/(\alpha + 1) \in [0,1]$ as a compactification of the horizontal axis, in figure \ref{fig_4_4} and all subsequent level plots of the same style.
The important break-even point $\alpha = N_1/N_3 = 1$ of equal cluster parity $N_1= N_3 \, $, alias $N_1/N=1/2$, is marked by a dashed white vertical line.

Each level curve of $\lambda (\alpha, s) = \lambda$ terminates at two points on the boundary of figure \ref{fig_4_3}.
Any termination at the upper or lower boundary $s=\pm\infty$ must occur at $\alpha=\alpha_c \,$, where $q=0$.
Indeed, $\lambda=-qs^2+\ldots$ in \eqref{lamas} implies limits $\lambda=-(\mathrm{sign}\, q)\cdot \infty$, for $s=\pm\infty$ and $q\neq0$\,. 
At the left and right boundaries $\alpha=0$ and $\alpha=1$ we encounter the values $\lambda(0,s)=s(1-s)$ and $\lambda(1,s)=-(c+1)s^2$\,, respectively. See \eqref{a=0}, \eqref{a=1}.

Along each level curve $\lambda(\alpha,s)=\lambda_* \, $, we may also determine the local extrema of $\alpha$, i.e.~the vertical tangents of the level curves.
Equivalently, these are the local extrema of $\lambda(\alpha,s)$, for any fixed $\alpha=\alpha_* \, $.
An elementary calculation shows that these curves are given by the level contours of $0=\partial_s \lambda(\alpha,s)=1-\alpha-2\,qs$, i.e.
\begin{align}
\label{sminmax}
s=s_\mathrm{minmax}(\alpha) :=&\ \tfrac{1}{2} (1-\alpha)/q>0 \,,\\
%\end{equation}
%\begin{equation}
\label{lamminmax}
\lambda=\lambda_\mathrm{minmax}(\alpha):=&\ \lambda(\alpha,s_\mathrm{minmax}(\alpha))= \tfrac{1}{4} (1-\alpha)^2/q\,.
\end{align}
These locations are marked as two red curves in figures \ref{fig_4_1}, \ref{fig_4_3}, \ref{fig_4_4}.
Comparing \eqref{lamas} and \eqref{sminmax}, the red curves of saddle-nodes occur at half the $s$-value of the nontrivial dotted yellow level curve $\lambda=0$, for each $\alpha$.

In-/stability of each equilibrium $s=s_*$ can be derived easily from exchange of stability, at $\lambda=0$, or explicitly from \eqref{ODEs}.
We caution the alert reader, however, that stability is asserted only within \eqref{ODEs}, i.e., within the restricted phase space of 2-cluster dynamics.
Dynamics transverse to that invariant subspace, e.g.~by rebellions, is not yet accounted for at this stage.
As in the previous subsection \ref{sumc=-1}, positive $q=q(\alpha)>0$ implies instability of the largest and smallest equilibria $s_* \, $, and stability of any intermediate $s_* \,, $ on each level curve $\lambda$ and for each fixed $\alpha$.
This identifies the region of $s$ between the two red saddle-node curves $s=s_\mathrm{minmax}(\alpha)$ as the only region of stable equilibria $s=s_* \, $.
We call such regions of $\alpha, s_*$ where the equilibrium $s \equiv s_*$ is stable an $s$\emph{-stable region}.
The \emph{$s$-unstable region} consists of the two parts below and above the two red saddle-node curves. 

Negative sign $q(\alpha,c)<0$ in contrast, which only occurs for $\alpha_c\leq \alpha \leq 1/ \alpha_c $, indicates stability of the largest and smallest equilibria $s=s_* \,$, and instability of any intermediate $s_* \, $, there.
In particular, this also identifies $\alpha_c < \alpha < 1/ \alpha_c$ as the region where the dynamics of \ref{ODEs} is \emph{dissipative}, i.e.~where solutions $s(t)$ are attracted to a bounded region in forward time.

%%%%%%%%%%%%%%%%%%%%%%%%%%%%%%
\begin{figure}[t]
\centering
	\includegraphics[width=0.9\textwidth]{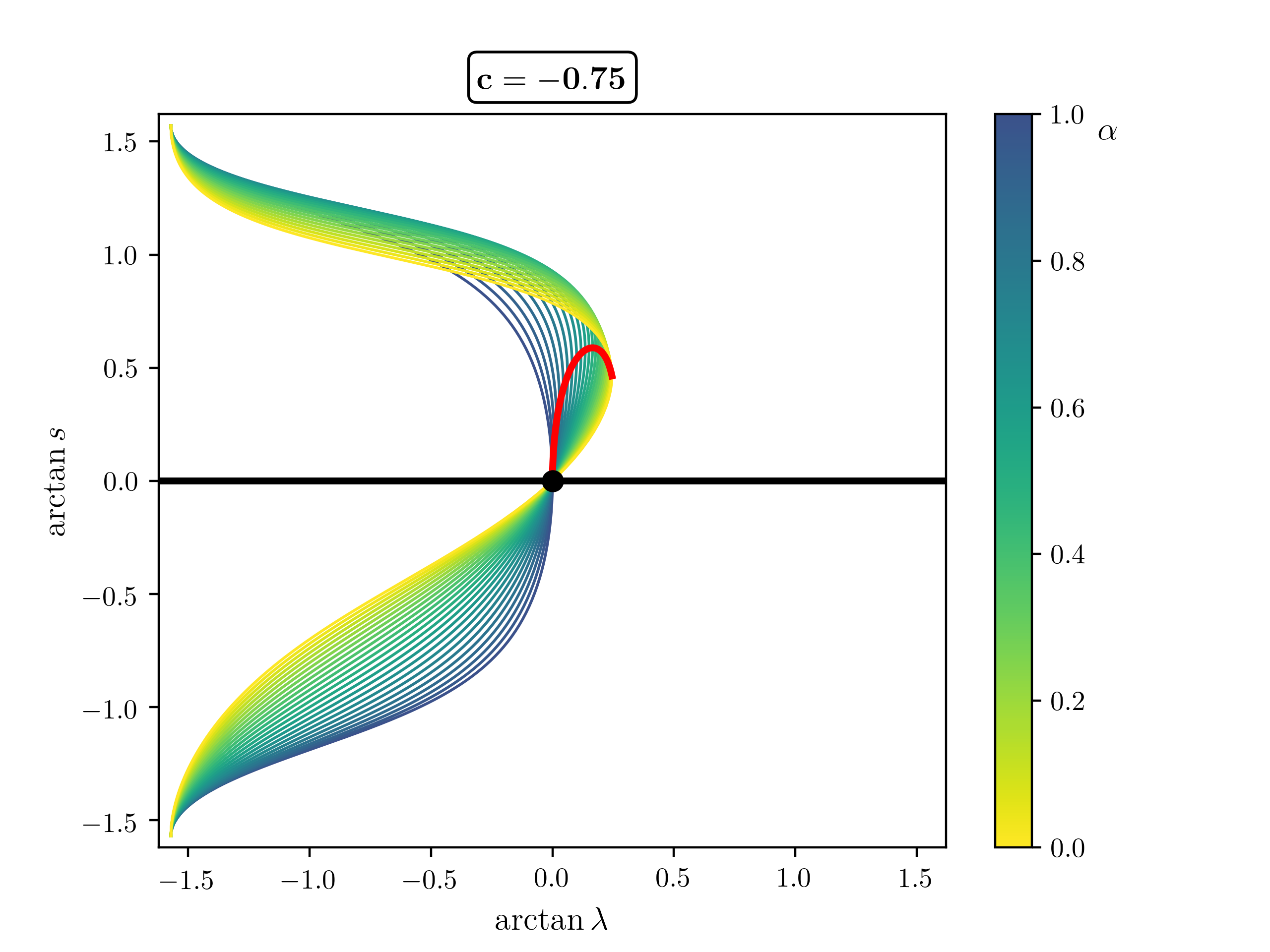}
	\caption{
	Global bifurcation diagrams for the compactified asynchrony $s$ of 2-cluster equilibria in the ODE flow \eqref{ODEs}, at $c=-0.75>-1$; see subsection \ref{sumc>-1}.	
	The quadratic coefficient $q$ remains positive for all $0\leq \alpha \leq 1$.
	See figure \ref{fig_4_1} for colors and in-/stability of each equilibrium $s=s_* \,$, within 2-cluster dynamics \eqref{ODEs}.
	Note the single red branch of saddle-nodes at extremals $\lambda=\lambda_\mathrm{minmax}(\alpha) $, according to \eqref{sminmax},\eqref{lamminmax}. Indeed $q>0$ implies $\lambda_{\mathrm{minmax}}>0$ and $s_\mathrm{minmax}>0$.
}
	\label{fig_4_5}
\end{figure}
%%%%%%%%%%%%%%%%%%%%%%%%%%%%%%
%%%%%%%%%%%%%%%%%%%%%%%%%%%%%%
\begin{figure}[p]
\centering
	\includegraphics[width=0.8\textwidth]{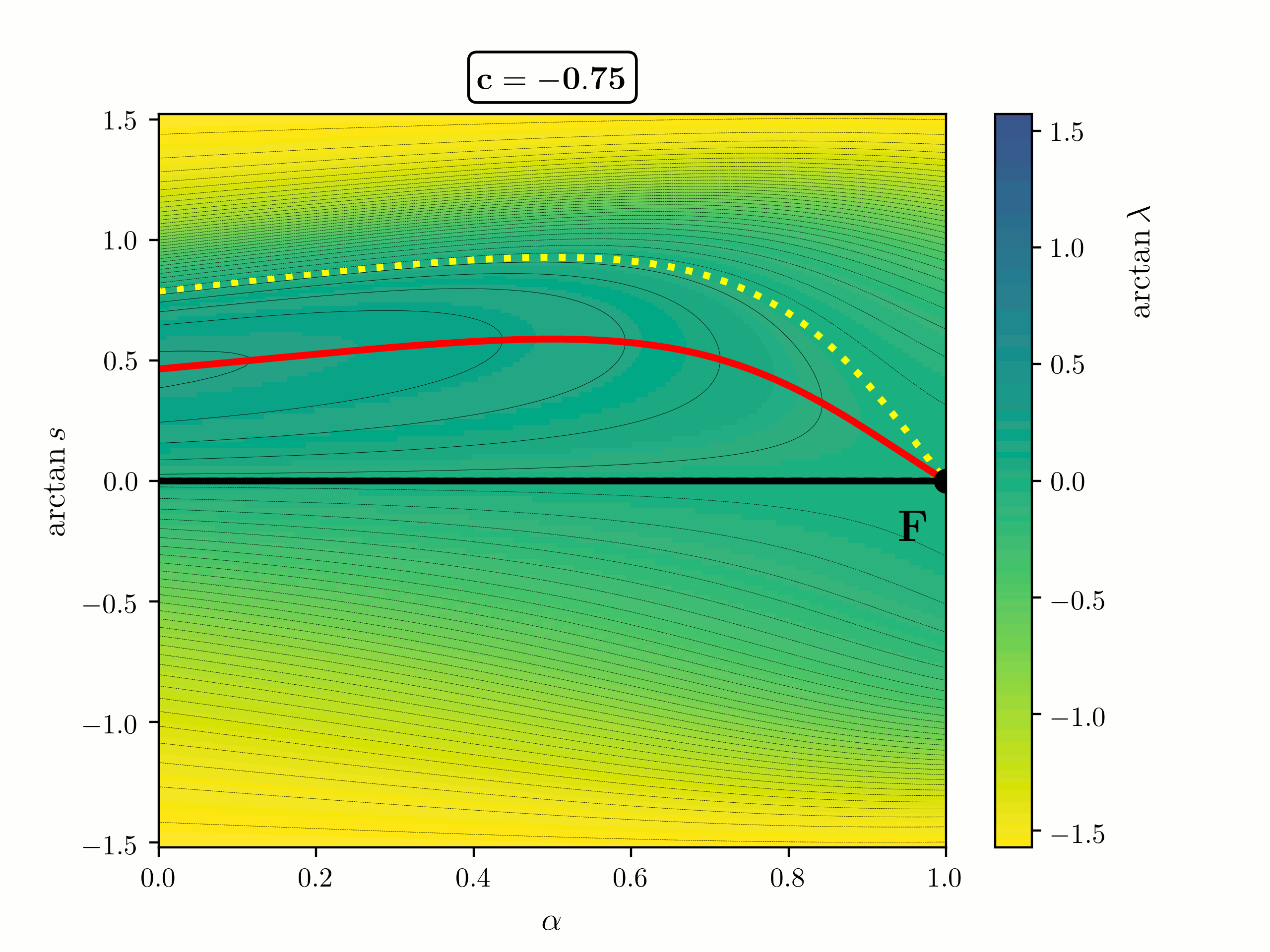}
	\caption{
	Level curves of $\lambda=\lambda(\alpha,s)$ for 2-cluster equilibria $s=s_*>0$ of the ODE flow \eqref{ODEs} with $c=-0.75>-1$, in analogy to the case $c=-1.3$ of figure \ref{fig_4_3}. 
	See the legend there.
	Note that $-\infty<\lambda \leq 1/4$ is now bounded above.
	There are only two level curves for $\lambda=0$, one black and one dotted yellow. 
	Again they intersect at the only critical point $\mathbf{F}$ of $\lambda(\alpha,s)$. 
	The maximal values of $\alpha$, on level curves of $\lambda$, form a single red curve of saddle-node bifurcations, this time. 
	See also figure \ref{fig_4_5}.
	All level curves of $\lambda$ still begin and terminate at the boundary, as described in the text.
	The only region of stable equilibria $s=s_*\,$, within 2-cluster dynamics \eqref{ODEs}, is located between the black horizontal axis and the red saddle-node curve.}
\bigskip
	\label{fig_4_6}
	\includegraphics[width = 1.00\textwidth]{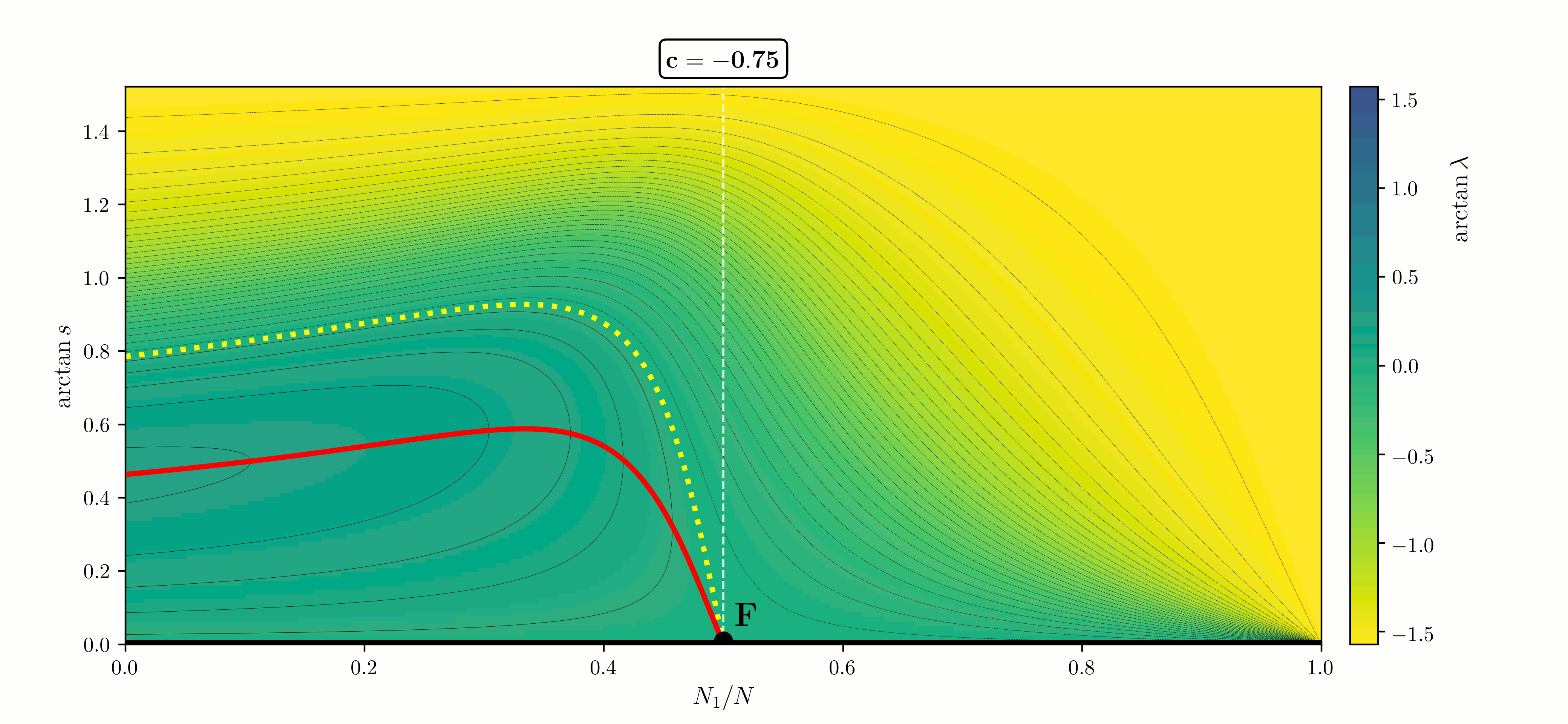}
	\caption{
	Glued version of figure \ref{fig_4_6}.
	The horizontal axis is $N_1/N= \alpha/(\alpha + 1)$, and $s<0$ has been omitted as redundant, analogously to the derivation of figure \ref{fig_4_4} from figure \ref{fig_4_3}.
	The only region of stable equilibria $s=s_*>0$, within 2-cluster dynamics \eqref{ODEs}, is still confined between the black horizontal $\alpha$-axis $s=0$, and the red saddle-node curve $s=s_{\mathrm{minimax}}(\alpha)$.
	}
	\label{fig_4_7}
\end{figure}
%%%%%%%%%%%%%%%%%%%%%%%%%%%%%%

\subsection{The case $\mathbf{c>-1}$}\label{sumc>-1}

In this case, the quadratic coefficient $q=q(\alpha)$ in \eqref{ODEs}, \eqref{ODEq} is strictly positive, for all $0 < \alpha < \infty$.
Therefore our discussion follows the part of the previous subsection \ref{sumc<-1} for the case $q>0$.
For an explicit, but relevant, example we fix the cubic coefficient $c=-0.75$\,.
See figure \ref{fig_4_5} for the resulting bifurcation diagrams of \eqref{ODEs}.

In the less standard contour plots of figures \ref{fig_4_6}, \ref{fig_4_7}, analogously to figures \ref{fig_4_3}, \ref{fig_4_4}, we present the level curves of the parameter $\lambda=\lambda(\alpha,s)$.
The only critical point of $\lambda$ is still the nondegenerate saddle $\mathbf{F}$, with two associated level curves $\lambda=0$ (black and dotted yellow).
This time, $-\infty<\lambda \leq 1/4$ is bounded above, globally, as is already visible from the bifurcation diagrams of figure \ref{fig_4_5}.
The maximal $\lambda$ is attained on the left boundary $\alpha=0$, at $s=1/2$.

Each level curve terminates at two points on the boundary of figures \ref{fig_4_6}, \ref{fig_4_7}, as before.
Since $\lambda=-\infty$ at $s=+\infty$, this time, all terminations occur at the right and left boundaries.
In the left region, delimited by the black and yellow level curves of $\lambda=0$, both terminations are located on the left boundary.
The maximal value of $\alpha$, along each of the interior level curves of $0<\lambda < 1/4$, occurs on the red curve of saddle-nodes, of course. Again that red curve is located at half the $s$-value of the dotted yellow level curve $\lambda=0$, for each $\alpha$.
Above the region delimited by the dotted yellow curve $\lambda=0$, in figure \ref{fig_4_7}, level curves connect the two vertical boundaries $\alpha = 0$ and $\alpha= \infty $. 

In-/stability of each equilibrium $s=s_*>0$ can be derived as before.
Because $q>0$, the region of $s$ between the  black horizontal $\alpha$-axis and the red  saddle-node curve $s=s_\mathrm{minmax}(\alpha)$ is the only $s$-stable region, within 2-cluster dynamics \eqref{ODEs}, for any $0<\alpha<\infty$.
\FloatBarrier

\section{Rebellions, rebel flows, and blocking}\label{rebdy}

In the previous section \ref{sumdy} we have seen how the asynchrony variable $s=(\xi_3-\xi_1)/(\alpha+1)$ either tends to total 1-cluster synchrony $s\equiv 0$ of the two large clusters, or to a nontrivial equilibrium $s\equiv s_*\neq 0$, for $t\rightarrow\infty$.  
In the present section, we study the remaining heteroclinic dynamics of a small rebel cluster $(N_2,\xi_2)$, e.g.~for $N_2=1$, when the two large clusters have already equilibrated.
At fixed $s\equiv s_*\neq 0$, the two large clusters $(N_1,\xi_1)$ and $(N_3,\xi_3)$ then compete for the rebels $\xi_2$ in size.
In fact, we only have to address the remaining ODE for $y:=\xi_2- \xi_1= (\alpha +1)y_1\,$; see \eqref{xi2y}, \eqref{ODEy1s}.
Taking a more global view point, we also explain how to encode and represent the individual heteroclinic rebellions by a rebel flow.

Total synchrony $s\equiv0$ of large clusters leads to
\begin{equation}
\label{y1dot_s=0}
\dot{y}=y(\lambda + y + y^2)
\end{equation}
for $y=(\alpha + 1)y_1 \, $. 
To derive \eqref{y1dot_s=0} we directly insert $y_2=s-y_1=-y_1$ in \eqref{ODEy1y2}, or we formally replace $\dot{s}/s$ by $\lambda$ in \eqref{ODEy1s} due to \eqref{ODEsq}.

For $\lambda>1/4$, we obtain global instability of the fully synchronous 1-cluster equilibrium $0 \equiv s= (\xi_3-\xi_1)/ (\alpha + 1)$ towards rebels $y=\xi_2 - \xi_1\,$, which escape to $\pm \infty$.
For $0 \neq \lambda < 1/4$, in contrast, we obtain a unique stable equilibrium $y \equiv y_* \,$.
The domain of attraction is delimited by the remaining two linearly unstable equilibria, beyond which rebels $y$ escape to $\pm \infty$, respectively, as before.
Only for  $\lambda < 0$ we have stability of $y_*=0$ against rebellion, in this sense.
This reflects the local stability of the trivial 1-cluster $\mathbf{x}=0$ of total synchrony in the full system \eqref{ODExn}, of course.
For $0< \lambda < 1/4$, where $0>y_*> - 1/2$, rebellion can lead to the gradual formation of a tiny stable rebel cluster at $y_*= - \tfrac{1}{2}(1-\sqrt{1-4\lambda})$, at least as long as its size $N_2$ remains small compared to $N \approx N_1+ N_3$.
Also note the presence of a linearly unstable rebel cluster at $y \equiv - \tfrac{1}{2}(1+ \sqrt{1-4 \lambda})$, for all $\lambda< 1/4$.

The 2-cluster case $s \equiv s_*\neq 0$, where the two large clusters are distinct and compete for the rebels $(N_2,\xi_2) \,$, is much more interesting.
From \eqref{s>0} we recall $s>0$, without loss of generality.

Scaling \eqref{ODEy1s}, \eqref{y1bar} to $y:=(\alpha+1)y_1=\xi_2-\xi_1$ again, we obtain the cubic ODE
\begin{align}
	\label{ODEys}
\dot{y} =&\ y\,(y-(\alpha+1)s)\,(y-\bar{y}(s))\,,\\
	\label{ybar}
	\bar{y}(s) :=&\ (2-\alpha)s-1\,.
\end{align}

We repeat that $s \equiv s_*> 0$ is constant here.
In particular, the term $\dot{s}/s$ from \eqref{ODEy1s} drops out in \eqref{ODEys}.
The equilibrium $y=0$ indicates $\xi_2=\xi_1 \,$: the rebels $\xi_2$ are at the large cluster $(N_1,\xi_1)$.
The equilibrium $y_1=s$, i.e.~$y=(\alpha+1)s$, in contrast, indicates $\xi_2=\xi_3 \, $: the rebels are with the competing large cluster $(N_3, \xi_3)$.
Indeed, $y_1=s$ is equivalent to $y_2=0$, by \eqref{3sum}, and hence to $\xi_2=\xi_3\,$, by \eqref{xi2y}.
The third equilibrium $y=\bar{y}(s)$ denotes a 3-cluster equilibrium where, in general, the tiny rebel cluster establishes its own equilibrium balance, holding out against both large clusters.

We can now introduce and explain the central concept of a \emph{rebel flow}.
Suppose a nonstationary solution $y=y(t)$ of the scalar ODE \eqref{ODEys} remains bounded for all positive and negative times $t\in\mathbb{R}$.
Then $y(t)$ is heteroclinic.
First consider a heteroclinic rebel migration in \eqref{ODEys} from $y=0$ to $y=(\alpha+1)s$, as $t$ increases from $t=-\infty$ to $t=+\infty$. 
This means that rebels leave the cluster $(N_1,\xi_1)$ in favor of the cluster $(N_3,\xi_3)$.
As we have explained in the introduction, the minimal case $N_2=1$ of single rebel heteroclinicity then leads to the neighboring cluster configuration $(N_1-1,N_3+1)$.
For large $N<\infty$, therefore, each such heteroclinic orbit amounts to a small discrete step decreasing the rational value of $\alpha\,$: from $N_1/N_3$ to $(N_1-1)/(N_3+1)$.
The parameter $\lambda = \lambda(\alpha,s)$, on the other hand, remains constant along heteroclinic orbits; see \eqref{lamas} for the limiting case $N=\infty$.
The discrete-valued size ratio $\alpha=N_1/N_3\,$, along with $N_1/N$, becomes continuous and real-valued in our asymptotics of large $N \rightarrow\infty$.
Even though $\alpha$ actually remains constant, for $N=\infty$, we therefore denote the heteroclinic decrease of $\alpha$, of infinitesimal order $1/N$, by a magenta arrow towards smaller $\alpha=N_1/N_3$ along the level curve of constant $\lambda$.
Normalizing the magenta tangent vectors to unit length, this defines the rebel flow for rebels which favor the cluster $N_3$ over $N_1$.
Following the rebel flow lines indicates the total effect of concatenated infinitesimal rebellions.
See the illustrations for the cases $c=-1.3$ and $c=-0.75$ in the contour plots \ref{fig_5_3} and \ref{fig_5_5} of $ \lambda(\alpha,s)$, further below.

In the opposite direction, a heteroclinic rebel migration of \eqref{ODEys} from $y=(\alpha+1)s$ to $y=0$ indicates how rebels leave the cluster $(N_3,\xi_3)$ in favor of the cluster $(N_1,\xi_1)$.
We indicate this heteroclinic rebellion towards infinitesimally larger $\alpha$ by a magenta unit tangent towards larger $\alpha$ and $N_1/N=\alpha/(\alpha+1)$, accordingly.
See the illustrations in the contour plots \ref{fig_5_2} -- \ref{fig_5_5} below.

As we have remarked in the introduction, the rebel flow proceeds along the level curves of the fixed parameter $\lambda=\lambda(\alpha,s)$, in either case.
Therefore the planar rebel flow is integrable, with preserved first integral $\lambda$.

Let us briefly adapt the stability discussion in \cite{elm,stelm} to draw some consequences for the rebel flow, in terms of 2-cluster instability.
By \eqref{2Mcl}, 2-clusters are invariant under the permutation subgroup $S_\mathbf{N} = S_{N_1}\times S_{N_3}\,$.
Therefore, the linearization at stationary 2-clusters is $S_\mathbf{N}$-equivariant.
The standard irreducible representation of $S_N$ on $X_0$ decomposes, accordingly, into three irreducible representations $X_0 = X_s \oplus X_{01} \oplus X_{03}$ of $S_\mathbf{N}\,$, each invariant under the linearization.
The subspace $X_{01}$ of dimension $N_1-1$ features $\mathbf{x} \in X_0$ with $N_3=N-N_1$ vanishing components $x_{N_1+1},\ldots,x_N\,$.
Similarly, the subspace $X_{03}$ of dimension $N_3-1$ features $\mathbf{x} \in X_0$ with vanishing components $x_{1},\ldots,x_{N_1}\,$.
The one-dimensional subspace $X_s\,$ of $X_0\,$, finally, addresses 2-cluster perturbations with $x_1=\ldots=x_{N_1}=\xi_1$ and $x_{N_1+1}=\ldots=x_N=\xi_3\,$.

In section \ref{sumdy} we have distinguished regions of $s$-unstable and $s$-stable stationary 2-clusters.
This refers to one-dimensional in-/stability in $X_s\,$, based on the autonomous 2-cluster dynamics of $s$ in \eqref{ODEs}, only.
Rebel flow dynamics with decreasing $\alpha$ refers to single rebel heteroclinics $N_2=1$ from cluster $N_1$ to cluster $N_3$ initiated by linear instability in $X_{01}\,$; see \eqref{ODEys}.
This implies linear instability in $X_{01}\,$, of dimension $N_1-1$, but complementary stability in $X_{03}$\,.
Conversely, rebel dynamics with increasing $\alpha$ implies $N_3-1$ additional unstable dimensions, in $X_{03}\,$, but complementary stability in $X_{01}\,$.
It is this dichotomy of opposite stabilities in $X_{01}$ and $X_{03}\,$, essentially, which enables us to describe the global rebel dynamics by a rebel flow.

%%%%%%%%%%%%%%%%%%%%%%%%%%%%%%
\begin{figure}[p]
\centering
	\includegraphics[width=1.0\textwidth]{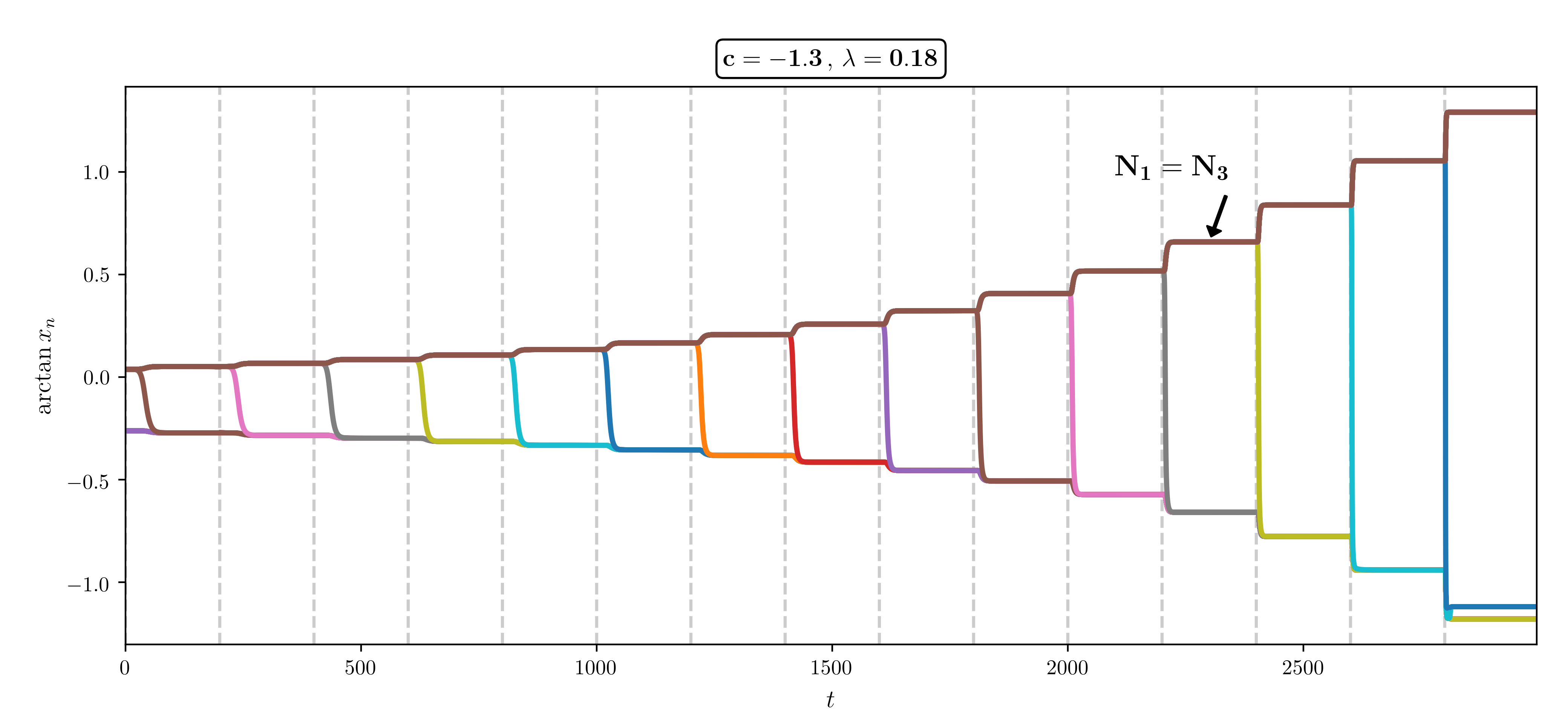}
	\caption{
	Heteroclinic rebel transients obtained from numerical simulations of \eqref{ODExn} for $c=-1.3$, $\lambda=0.18$ and $N=32$ units. 
	We consider 2-cluster solutions $x_1= \dots = x_{N_1} = \xi_1, \ x_{N_1+1} = \dots = x_N = \xi_3$ with $\xi_1$ and $\xi_3$ as in section \ref{large}. 
	Starting from $N_1=4$ and $n=N_1+1$, we apply a small random
perturbation to $x_n=\xi_3\,$ at a time indicated by the dashed vertical lines, only.
	We then integrate the system \eqref{ODExn} with rebel  $x_n\neq \xi_1\,, \xi_3\,$, until the dynamics settles again.
	See the colored rebel transients of $x_n$ from $\xi_3$ (top) to $\xi_1$ (bottom), along which the rebel $x_n$ changes its cluster affiliation.
	We repeat this process until the system enters a blocking region because a stationary 3-cluster state is initiated by the rebel at $x_n=\xi_2$ (blue) between $\xi_1$ and $\xi_3$; see the last state shown in the figure.
	See \cite{FieTatra, Ash07, KuGiOtt2015} for related ideas.}
	\label{fig_5_1}
	\includegraphics[width = 1.00\textwidth]{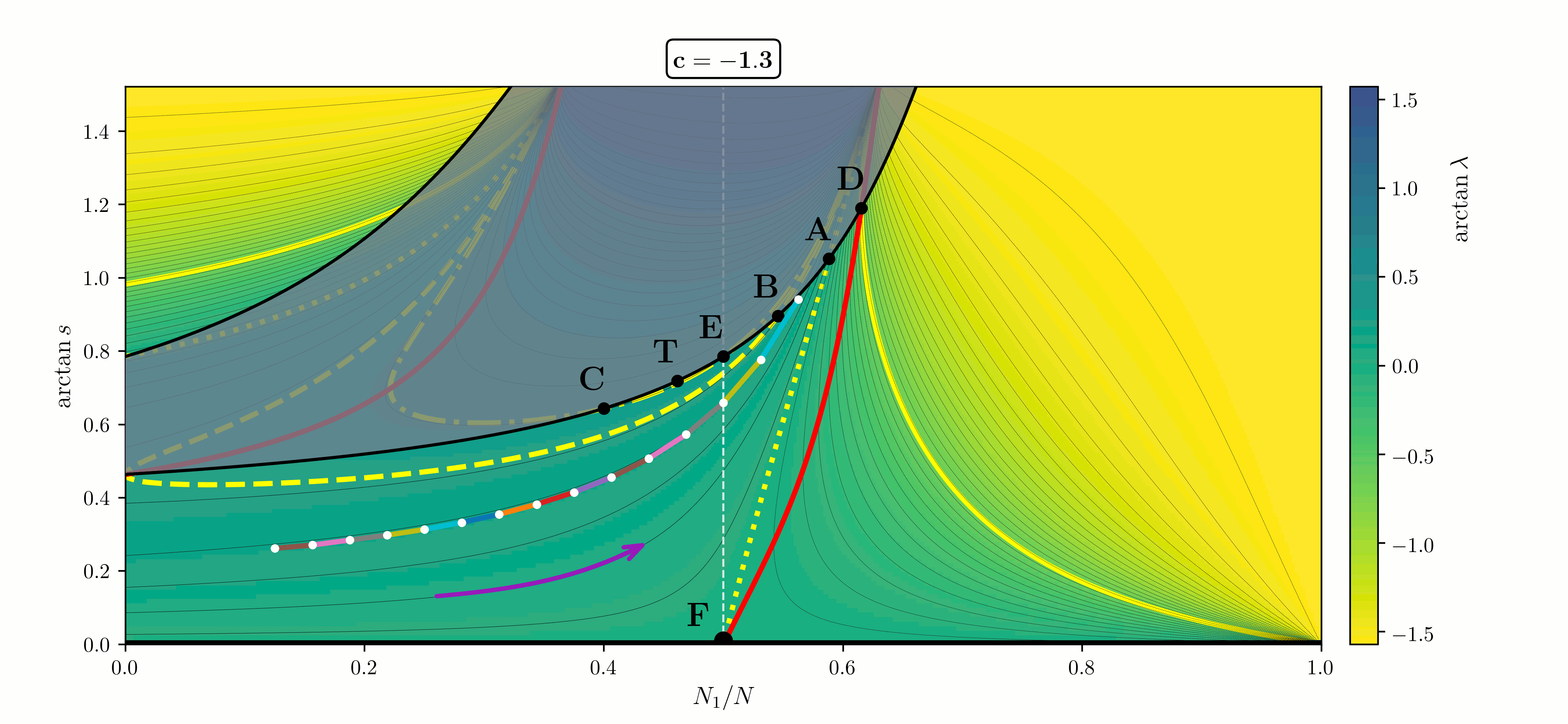}
	\caption{
	The rebel transients of figure \ref{fig_5_1} are inserted into the diagram of $\lambda$-levels from figure \ref{fig_4_4}, within the level curve of $\lambda=0.18$.
	The color coding of the heteroclinic rebel transients is the same as before.
	A magenta arrow indicates the rebel flow induced by the rebel transients.
	For further discussion of the yellow curves, and of the dark shaded blocking region where stationary rebel 3-clusters bifurcate and persist, we refer to figure \ref{fig_5_3} below.}
	\label{fig_5_2}
\end{figure}
%%%%%%%%%%%%%%%%%%%%%%%%%%%%%%

In figure \ref{fig_5_1} we illustrate the rebel flow by numerical integration of \eqref{ODExn} for $c=-1.3$, $\lambda=0.18$ and $N=32$ units.
As initial condition, a two cluster solution $x_1= \dots = x_{N_1} = \xi_1, \  x_{N_1+1} = \dots = x_N = \xi_3$ was chosen, with
$\xi_1$ and $\xi_3$ as in section \ref{large}.
For $N_1=4$, initially, this corresponds to an initial 2-cluster proportion of $N_1/N = 0.125$.
We then perturb a single unit $x_n\,, \ n=N_1+1,$ in cluster $\xi_3\,$, and integrate forward in time until the dynamics no longer changes.
As a result, we observe heteroclinic rebel dynamics, that is, the perturbed unit $x_n$ changes its cluster affiliation from $\xi_3$ to $\xi_1$.
In other words, $N_1=5$, after the rebel transient.
We repeat this process, for ever increasing cluster sizes $N_1\,$.
Note the successive heteroclinic transients of the rebels $x_n\,$, from $\xi_3$ down to $\xi_1<\xi_3\,$.
After 12 transients, of course, equal cluster parity $N_1=N_2=16$ is reached.
After 15 transients, the dynamics enters a blocking region and finally settles on a three cluster solution; see the bottom right part of figure \ref{fig_5_1} and further explanations below.
At this stage, the third coexisting cluster $y=\bar{y}(s)>0$ of \eqref{ODEys}, at $\xi_2<\xi_3$ near $\xi_1<\xi_2\,$, consists of just one single rebel element.
The trajectories of figure \ref{fig_5_1} are also visualized as a discretized rebel flow in the $(N_1/N, s)$ plane of contour plot \ref{fig_5_2},
with corresponding  color coding of the rebel transients.

For numerical integration, we employed the implicit Adams method provided by SciPy; see \cite{2020SciPy-NMeth}.
After each perturbation, we subtracted the mean of the ensemble to ensure the perturbations are contained in the above representation subspace $X_s\oplus X_{03}$ of the phase space $X_0$ for \eqref{ODExn}.
Note that by choosing initial conditions in the 2-cluster subspace with just a single unit perturbed,
we suppress transitions in which multiple units might change their cluster affiliation.
We also suppress instabilities that might break up the 2-cluster altogether, in favor of a 3-cluster.

In the language of \cite{FieTatra} or \cite{Ash07}, figures \ref{fig_5_1}, \ref{fig_5_2} indicate a path in the connection graph or the heteroclinic web of cluster transitions.
See also \cite{KuGiOtt2015}, fig.~6, for a related simulation in the Stuart-Landau setting of our section \ref{sl} with $N=1000$ oscillators.

The direction of the heteroclinic transients partially determines the ordering of the two asymptotic large 2-clusters, by the decreasing energy or Lyapunov function $I(\mathbf{x})$ in \eqref{gracub}. 
This would require a nontrivial calculation, otherwise.
The transitivity of that order, simply following the level curves of $\lambda$ along our arrows, possesses a dynamic counterpart.
Assuming transversality of the stable and unstable manifolds of the target and source 2-cluster equilibria, respectively, along heteroclinic orbits, there also exists a direct heteroclinic connection between any two equilibria connected by a directed sequence of heteroclinic orbits, for the same parameter $\lambda$.
This \emph{dynamic transitivity} is a consequence of the so-called $\lambda$-Lemma; see for example \cite{padm}.
The useful property of transversality of invariant manifolds, often called the Morse or Morse-Smale property, is generic for general vector fields, by the Kupka-Smale theorem.
In PDE settings like \cite{firo18}, such transversality is long known to hold automatically; see \cite{ang86}.
For our present class of equivariant vector fields \eqref{ODExn}, however, transversality is a much more delicate assumption -- somewhat beyond the scope of our present paper. 
As long as the size $N_2$ of the rebel cluster remains small compared to $N$, however, we may still concatenate a small number $N_2$ of single rebel transients to obtain limited transitivity.
Following the rebel flow indicates the effect of concatenated infinitesimal rebellions, by dynamic transitivity or due to successive perturbations as in figure \ref{fig_5_1}.

Heteroclinic orbits between $y=0$ and $y=(\alpha+1)s$ are blocked, a priori, when the third rebel equilibrium $y=\bar{y}(s)$ of \eqref{ODEys} is located strictly between $y=0$ and $y=(\alpha+1)s$.
Therefore we call the equilibrium $\bar{y}(s)$ in \eqref{ODEys}, \eqref{ybar} \emph{blocking}, if $0< \overline{y}(s)<(\alpha+1)s$.
The \emph{blocking regions}, in contour plots  \ref{fig_5_2}--\ref{fig_5_5}, consist of those $(N_1/N,s)$ for which the equilibrium $\bar{y}(s)$ blocks rebel heteroclinic orbits between the two large competing clusters.
Instead the rebels are ready to form a tiny third cluster between the large ones, which may turn out stable, destabilizing the larger competitors, or unstable, stabilizing the 2-cluster status quo.

The \emph{blocking boundaries} of the blocking region are characterized by those values of $(\alpha,s)$ for which $\bar{y}(s)=0$ or $\bar{y}(s)=(\alpha+1)s$, respectively.
For the blocking boundary $\bar{y}(s)=0$  we obtain the graphs
\begin{align}
	\label{s0}
	s=s_0(\alpha):=&\quad\ \frac{1}{2-\alpha}\ >0 \,,\\
	\label{lam0}
	\lambda=\lambda_0(\alpha):=&\ \frac{1-(c+2)\alpha}{(2-\alpha)^2}\,.
\end{align}
Indeed \eqref{lam0} follows from \eqref{s0} and \eqref{lamas}.
The blocking boundary $\bar{y}(s)=(\alpha+1)s$ is analogously characterized by
\begin{align}
	\label{s1}
	s=s_1(\alpha):=&\qquad\frac{1}{1-2\alpha}\ > 0\,,\\
	\label{lam1}
	\lambda=\lambda_1(\alpha):=&\ \alpha\,\frac{\alpha-(c+2)}{(1-2\alpha)^2}\,.
\end{align}
See figures \ref{fig_5_2}--\ref{fig_5_5}, where we have added the two black blocking boundaries to the corresponding previous contour plots \ref{fig_4_4}, \ref{fig_4_7}, for $c=-1.3, -0.75$. The blocking boundaries are easily distinguished by their values at $\alpha=0:\ s_0=1/2,\ \lambda_0=1/4$ versus $s_1=1,\ \lambda_1=0$.
Also note the poles at $\alpha=2,\ N_1/N=2/3$ and at $\alpha= 1/2,\ N_1/N=1/3$, respectively.
See \cite{kf20} for a more detailed discussion of these poles.

%%%%%%%%%%%%%%%%%%%%%%%%%%%%%%
\begin{figure}[p] 
\centering
	\includegraphics[width = 0.95\textwidth]{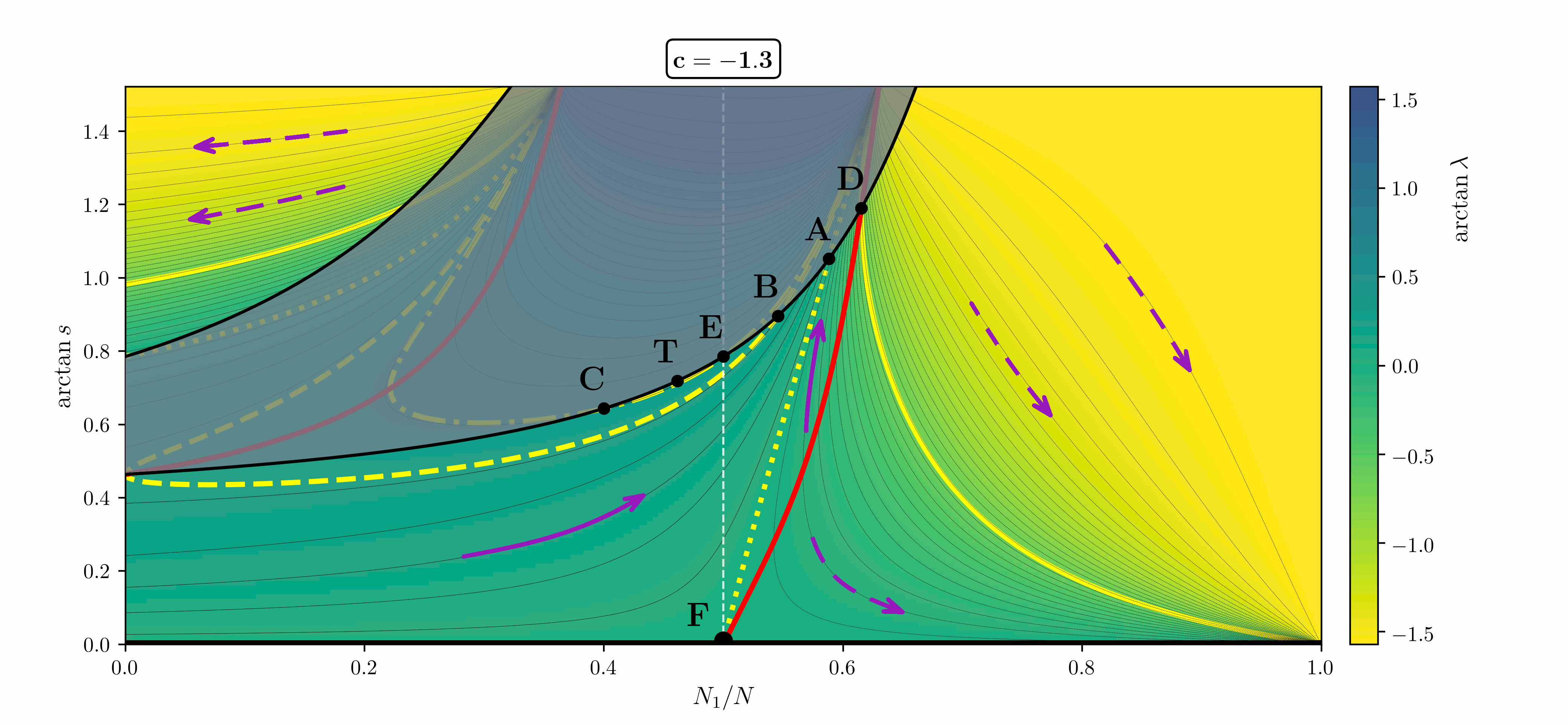}
	\caption{
	Contour plot of $\lambda=\lambda(\alpha,s)$ for 2-cluster equilibria $s=s_*>0$ of the ODE flow \eqref{ODEs} with $c=-1.3<-1$, in coordinates $(N_1/N,\arctan s)$.
	See figure \ref{fig_4_4} for axes and color codings.
	The two black curves mark the boundaries of the blocking region.
	The new dot-dashed yellow curve between $\mathbf{C}$ and $\mathbf{E}$ marks the level $\lambda= \lambda(\mathbf{E})= -(c+1)$.
	See figure \ref{fig_5_4} for zooms into that region and a discussion of the tangency point $\mathbf{T}$.
	The dashed yellow curve  $\lambda=1/4$ indicates the $\lambda$-level where one black blocking boundary terminates at $N_1/N=0$. 
	The other black blocking boundary left terminates at the level $\lambda=0$ indicated by the previous dotted yellow curve.
	The blocking region is located between the two black boundaries and is indicated by a darker shading.
	Outside the shaded blocking region, magenta arrows indicate the rebel flow along the level curves of $\lambda=\lambda(\alpha,s)$.
	Arrows to the right, i.e.~towards larger cluster fractions $N_1/N_3 \,$, indicate heteroclinic rebel orbits from the cluster $N_3$ to the cluster $N_1 \,$.
	Similarly, arrows to the left, i.e.~towards smaller fractions $N_1/N \,$, indicate heteroclinic rebel orbits in the opposite direction, favoring the cluster $N_3 \,$.
	Note how directions change across the blocking region and across the red saddle-node curves.
	Magenta arrows are drawn solid, in the 2-cluster $s$-stable region, and are drawn dashed in the $s$-unstable region; see also figures \ref{fig_4_4} and \ref{fig_4_7}.
	In the $s$-stable region, for example, rebellions from $N_3$ to any $N_1<N_3$ will cause $N_1$ to grow beyond equal parity $N_1=N_3 \,,$ across the dashed white line $N_1/N=1/2$ between $\mathbf{E}$ and $\mathbf{F}$: from minority to majority.
	Growth of $N_1$ only terminates at the black blocking boundary, between $\mathbf{A}$ and $\mathbf{B}$.
	See text and figure \ref{fig_6_5} for further details.
	}	
	\label{fig_5_3} 
\centering
	\includegraphics[width = 0.95\textwidth]{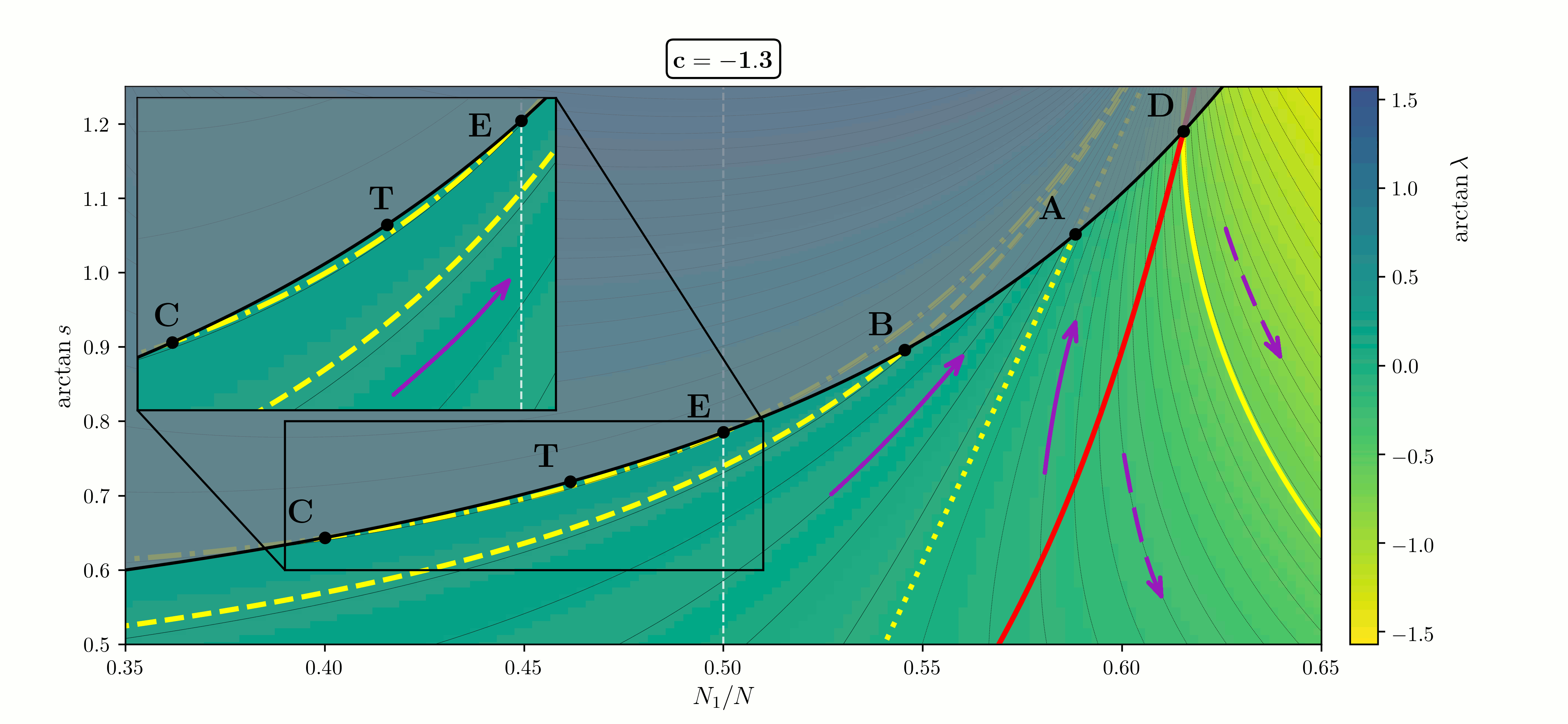}
	\caption{
	Zooms of figure \ref{fig_5_3}.
	Note the tangency point $\mathbf{T}$, further enlarged in the insert.
	Level contours of $\lambda$ are tangent to the blocking boundary at $\mathbf{T}$, from inside the dark shaded blocking region.
	In particular, level curves which emanate from the blocking boundary, between $\mathbf{C}$ and $\mathbf{T}$, terminate on the blocking boundary, between the break-even point $\mathbf{E}$ and the tangency $\mathbf{T}$.
	}
	\label{fig_5_4} 
\end{figure}
%%%%%%%%%%%%%%%%%%%%%%%%%%%%%%

It is remarkable that the rebel dynamics \eqref{ODEys} do not depend on the cubic coefficient $c$, at all. 
In particular the blocking regions in the $(\alpha,s)$-plane, and their black boundaries \eqref{s0}, \eqref{s1}, coincide in the contour plots \ref{fig_5_2}--\ref{fig_5_5}. 
Any differences arise from the configuration of the level contours $\lambda=\lambda(\alpha,s)$, which certainly depend on $c$ via \eqref{lamas}; see also \eqref{lam0} and \eqref{lam1}.

We determine the blocking region and the directions of the rebel flows in \eqref{ODEys} next.
Off the black blocking boundaries \eqref{s0}, \eqref{s1}, we sort the three equilibria $y=0,\ (\alpha+1)s,\ \bar{y}(s)$ as $\eta_1<\eta_2<\eta_3\,$, i.e.
\begin{equation}
\label{ysort}
\{0,(\alpha+1)s,\bar{y}(s)\} = \{\eta_1,\eta_2,\eta_3\}\,.
\end{equation}
Then $\dot{y}=(y-\eta_1)(y-\eta_2)(y-\eta_3)$ implies instability of the smallest and largest equilibria $y=\eta_1,\eta_3\,$, and stability of the intermediate equilibrium $y=\eta_2$.
The two heteroclinic orbits run from $\eta_1$ and $\eta_3$ to $\eta_2\,$, respectively.

%%%%%%%%%%%%%%%%%%%%%%%%%%%%%%
\begin{figure}[t] 
\centering
	\includegraphics[width = 1.00\textwidth]{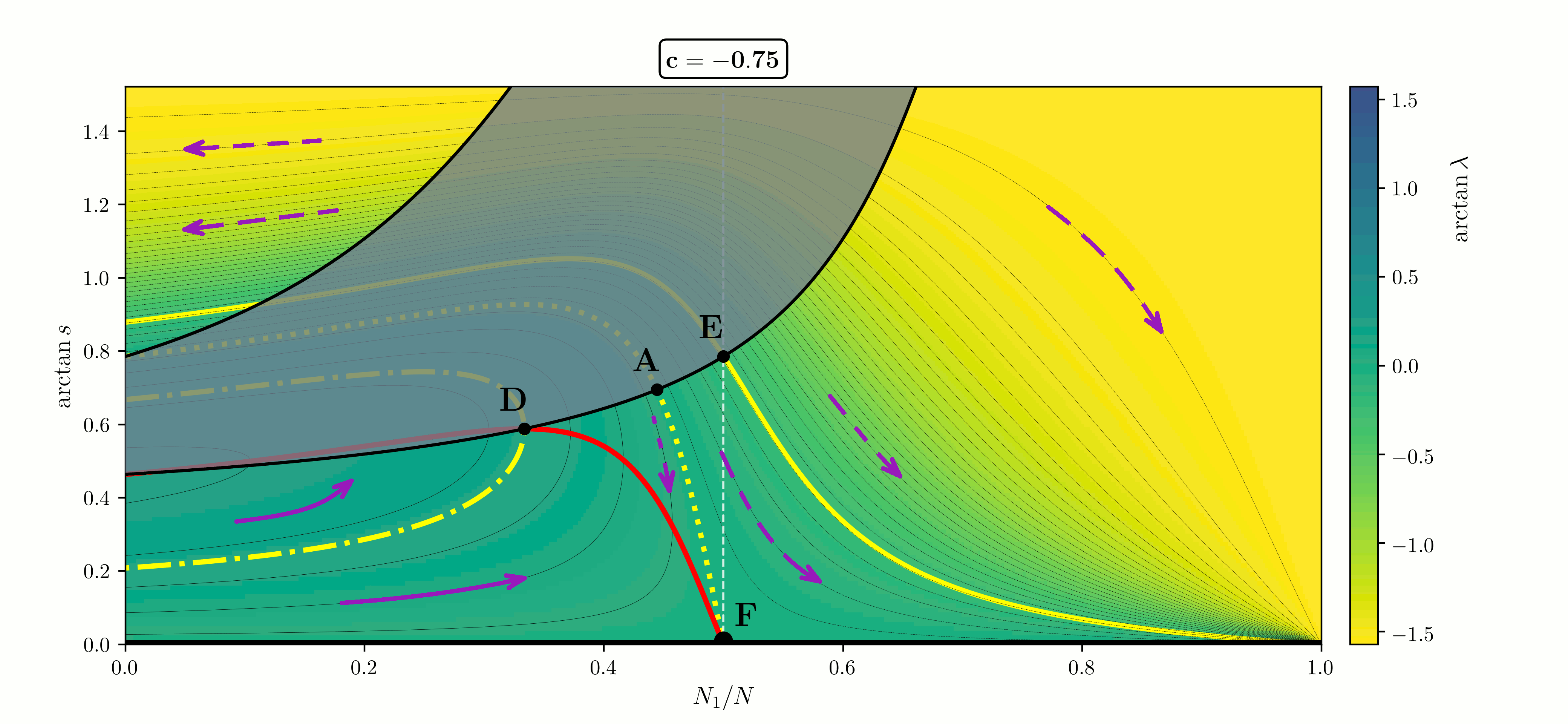}
	\caption{
	Level curves of $\lambda=\lambda(\alpha,s)$ as in figure \ref{fig_5_3}, but for $c=-0.75>-1$.
	The basic locations of the $s$-stable and the shaded $y$-blocking regions look similar, at first sight, but there are subtle differences in detail. See text and figure \ref{fig_6_7} for further details.
	}	
	\label{fig_5_5} 
\end{figure}
%%%%%%%%%%%%%%%%%%%%%%%%%%%%%%

For $s>0$, i.e.~$0<(\alpha+1)s$, this leaves us with the following three cases for the third equilibrium $\bar{y}(s)$.

\begin{description}
  \item[Region 1 (increasing $\alpha$):] $\bar{y}(s)=\eta_1$. \\
  This case is equivalent to $(2-\alpha)s-1=\bar{y}(s)=\eta_1<0=\eta_2<\eta_3=(\alpha+1)s$, i.e.~$s>0$ is between the horizontal axis and the lower black blocking curve $s_0$ of \eqref{s0}.
  Then blocking does not occur, and heteroclinic rebel migration $y= \xi_2- \xi_1$ runs from $y=(\alpha+1)s=\eta_3$ down to $y=0=\eta_2\,$, i.e.~from the cluster $(N_3,\xi_3)$ towards the cluster $(N_1,\xi_1) \, $.
We therefore indicate the rebel flow by a \emph{magenta arrow towards larger} $\alpha$ and $N_1/N$, in figures \ref{fig_5_3}-- \ref{fig_5_5}.
  \item[Region 2 (blocking):] $\bar{y}(s)=\eta_2$. \\
  Then $\eta_1=0<\bar{y}(s)=(2-\alpha)s-1=\eta_2<\eta_3=(\alpha+1)s$, i.e.~$s$ is between the two black blocking curves.
  Blocking occurs, and heteroclinic rebel migration from either large cluster gets stuck at the intermediate equilibrium $y=\bar{y}(s)=\eta_2$. 
  The resulting tiny new stationary rebel cluster at that 3-cluster equilibrium may in fact grow, at the expense of both large clusters, and with \emph{indefinite effects on their proportion} $\alpha$.
  Figures \ref{fig_5_2}--\ref{fig_5_5} indicate this blocking region by a darker shading.
  \item[Region 3 (decreasing $\alpha$):] $\bar{y}(s)=\eta_3$. \\
  Then $(2-\alpha)s-1=\bar{y}(s)=\eta_3> \eta_2=(\alpha+1)s>\eta_1=0$, i.e.~$s$ is above the upper red curve, and hence $0 \leq\alpha<1/2,\ 0\leq N_1/N<1/3$.
  Blocking does not occur, and heteroclinic rebel migration runs from $y=0=\eta_1$ upwards to $y=(\alpha+1)s=\eta_2\,$, i.e.~from the smaller cluster $(N_1,\xi_1)$ towards the larger cluster $(N_3,\xi_3)$: from minority to majority.
We indicate this rebel flow by a \emph{magenta arrow towards smaller} $\alpha$, in figures \ref{fig_5_3} and \ref{fig_5_5}.
\end{description}

For example consider the s-stable 2-cluster region in figure \ref{fig_5_3}, i.e.~for $c=-1.3$.
The region is located in the wedge between the lower black blocking boundary $s_0$ and the right red saddle-node curve.
All level contours $\lambda=\lambda(\alpha,s)$ in that region are oriented, along the solid magenta arrows, towards their termination at the black  blocking boundary $s_0$ to the left of $\mathbf{D}$.
Rebel heteroclinic migration towards the cluster $N_1$  erodes the cluster $N_3 \,$, until rebel flow termination of the $2$-cluster regime at the blocking boundary $s_0 \,$.

In fact, consider the $s$-stable 2-cluster states, which start out below the dashed yellow level curve $\lambda=1/4$ from a minority cluster $N_1<N_3 \,$, i.e.~from the left of the dashed white line $N_1/N=\tfrac{1}{2}$ of equal parity $N_1=N_3 \,$.
All these initial conditions will be prone to heteroclinic rebellions from the cluster $N_3$ to $N_1 \,$.
The rebel flow of concatenated rebellions drives them across the dashed white line and well into the region $N_1>N_3\,$: from minority to majority, across equal cluster size.

In figure \ref{fig_5_5} in contrast, at $c=-0.75$, the red saddle-node boundary confines the $s$-stable subregion of region $1$ to the left of the dashed white line $N_1/N=\tfrac{1}{2}$, i.e.~to $N_1<N_3 \,$.
Therefore heteroclinic rebel orbits starting in the $s$-stable region cannot achieve equal parity, anymore -- not even upon patient concatenation.
Instead, they face one of two possibilities: 
\begin{enumerate}[label=\alph*)]
  \item termination by blocking at the black blocking curve $s_0 \,$, or 
  \item termination at the red saddle-node curve.
\end{enumerate}
Migration from the larger cluster $N_3$ to $N_1$ gets stuck by an emerging, tiny but stationary, third rebel cluster, in case (a).
The blocking 3-cluster equilibrium $y=\bar{y}(s)\gtrsim 0$ emerges near the smaller cluster $y=0$, across the black blocking boundary $s=s_0(\alpha)$.
In case (b), any further increase of $N_1$ prevents any stationary status quo 2-cluster configurations, a priori.
The cause is the saddle-node termination of the 2-clusters.
Indeed, the value of $\alpha= N_1/N_3$ at the saddle-node intersection is the maximal available value of $\alpha$ for any 2-cluster equilibrium, at that particular level of $\lambda$.

For later reference we also determine the regions of the cubic coefficient $c$ for which the black blocking boundaries $s=s_0(\alpha)$ and $s=s_1(\alpha)$, respectively, intersect with specific relevant dotted or dashed level curves of $\lambda$, or with the red saddle-node curves $s=s_\textrm{minmax}(\alpha)$.

Specifically we claim the following four intersection points $\mathbf{A}$--$\mathbf{D}$ of the lower black blocking boundary $s_0(\alpha)=1/(2-\alpha)$:
\begin{align}
\label{s00}
    \mathbf{A}:=s_0\cap\{\lambda=0\},\qquad\qquad\,  0<\alpha<2\qquad \Longleftrightarrow &\qquad -\tfrac{3}{2}<c<+\infty ;   \\
    \label{s0c}
    \mathbf{B}:=s_0\cap\{\lambda=\tfrac{1}{4}\},\qquad\qquad\,  0<\alpha<2\qquad \Longleftrightarrow &\qquad -\tfrac{3}{2}<c<-1 ;   \\
    \label{s025}
\mathbf{C}:=s_0\cap\{\lambda=-(c+1)\},\quad 0<\alpha<2\qquad \Longleftrightarrow &\qquad -\tfrac{3}{2}<c<-\tfrac{5}{4} ;   \\
    \label{s0mm}
    \mathbf{D}:=s_0\cap s_\mathrm{minmax}\,,\qquad\qquad\  0<\alpha<2\qquad \Longleftrightarrow &\qquad -\tfrac{3}{2}<c<-\tfrac{1}{2}.   
\end{align}
We also claim the following intersection $\mathbf{D}'$ of the upper black blocking boundary $s_1(\alpha)=1/(1-2\alpha)>0$ :
\begin{equation}
    \label{s1mm0}
    \mathbf{D}':=s_1\cap s_\mathrm{minmax}\,,\qquad \qquad \, \,  0< \alpha< \tfrac{1}{2}\qquad \Longleftrightarrow \qquad -\infty<c<-\tfrac{3}{2}\,.   \\ 
\end{equation}
In addition, we mark the following two intersections with the dashed white line $N_1/N= \alpha/ (\alpha+1)= 1/2$ of equal parity $\alpha= N_1/N_3=1$:
\begin{equation}
\label{ef}
\begin{aligned}
\mathbf{E}& =  s_0 \cap \{\alpha =1\} \,,  &\quad &  \lambda=-(c+1) ; \\
\mathbf{F}& =  (\alpha=1, \ s=0)\,,  &\quad &   \lambda=0 \,  .
\end{aligned}
\end{equation}

The elementary proofs all follow the same pattern.
We first insert $s_0=1/(2-\alpha)>0$ from \eqref{s0} and the values of $\lambda$ in
\eqref{lamas} or the expression \eqref{sminmax} for $s_\mathrm{minmax} \,,$ as required. 
For the specified $\lambda$-values, we may alternatively invoke \eqref{lam0}, \eqref{lam1}.
The resulting linear equation for $c$ provides the following explicit expressions:
\begin{align}
\label{cs00}
    \mathbf{A}&=s_0\cap\{\lambda=0\}: &c&=\phantom{-}(1-2\alpha)/\alpha\,,    & \alpha= 1/(c+2)\,; \phantom{(c+1)}
     \\
\label{cs025}
    \mathbf{B}&=s_0\cap\{\lambda=\tfrac{1}{4}\}: &c&=-\tfrac{1}{4}\alpha-1\,,     & \alpha=-4(c+1) \,; \phantom{/(c5))}  \\
\label{cs0c}
    \mathbf{C}&=s_0\cap\{\lambda=-(c+1)\}:\  &c&=-(\alpha-5)/(\alpha-4)\,, \ & \alpha=(4c+5)/(c+1) \,;   \\
\label{cs0mm}
    \mathbf{D}&=s_0\cap s_\mathrm{minmax}:\    &c&=-(\alpha+1)/2\,, \quad &
     \begin{aligned}
     \alpha&=-2c-1 \,,\\
     \lambda&=(1+c)/(3+2c)\, .
     \end{aligned}
\end{align}
This proves the four claims \eqref{s00}--\eqref{s0mm} on $s_0$.
For $\ s_1=1/(1-2\alpha)$ from \eqref{s1}, we obtain analogously
\begin{equation}
\label{cs1mm}
    \ \mathbf{D'}=s_1\cap s_\mathrm{minmax}:\qquad\qquad\  c=-\tfrac{1}{2}(\alpha+1)/\alpha\,, \qquad\qquad 
    \begin{aligned}
    \alpha &= -1/(2c+1) \,, \\
    \lambda &= (1+c)/(3+2c)\, .
    \end{aligned}
\end{equation}
This proves the remaining claim \eqref{s1mm0}.
We have omitted variants $\mathbf{A', B', C'} \in s_1$ which will be irrelevant for our subsequent discussion.

It remains to address possible tangencies between level curves $\lambda = \lambda (\alpha, s)$ and the black boundaries of the blocking region, in the $(N_1/N, \arctan s)$-plane.
At such tangencies, the emanation/termination behavior of the rebel flow changes, as we will illustrate in the next section.
For now, we note that such tangencies $\mathbf{T, T'}$ are characterized by unique extrema of $\lambda_\iota (\alpha):= \lambda(\alpha, s_\iota(\alpha))$ along the black blocking boundaries $s_\iota(\alpha),\  \iota = 0,1$.
Elementary calculations of high school type for the rational expressions \eqref{lam0}, \eqref{lam1} of $\lambda_\iota(\alpha) $ provide the explicit expressions
\begin{align}
\label{t0}
&\mathbf{T}: \quad &c&=-2(\alpha+1)/(\alpha+2) \in (-\tfrac{3}{2}, -1) \,, & 
	\begin{aligned}
	\alpha&=-2(c+1)/(c+2) \,, \\
 	\lambda&= \tfrac{1}{4}(c+2)^2/(2c+3) \,;
 	\end{aligned}\\
\label{t1}
&\mathbf{T'}: \quad &c&=-2(\alpha + 1)/(2 \alpha +1) \in (-2, -1)\,, &
	\begin{aligned} 
	\alpha &= -\tfrac{1}{2}(c+2)/(c+1) \,, \\
 	\lambda&= \tfrac{1}{4}(c+2)^2/(2c+3)\, .
 	\end{aligned}
\end{align}

See contour plot \ref{fig_5_4} for an illustration of the tangency point $\mathbf{T}$, at $c=-1.3$.

Since we are democratically interested in minority/majority transitions across the dashed white line $N_1/N=\tfrac{1}{2}$, below the black blocking curve $s_0\,$, we also determine the values of $c$ where $\mathbf{A}, \ldots , \mathbf{D}$, and $\mathbf{T}$ cross $\alpha=1$:
\begin{center}
\begin{tabular}{ |c||c|c|c|c|c| } 
 \hline
$ \alpha=1 $& $\mathbf{A}$ & $\mathbf{B}$ & $\mathbf{C}$ & $\mathbf{D}$ & $\mathbf{T}$ \\ 
 \hline 
 $c$ & $-1$ & $-4/3$ & $-5/4$ & $-1$ & $-4/3$ \\ 
 \hline
\end{tabular}
\end{center}

In conclusion we observe crucial changes in the above intersection behavior at the six critical cubic coefficients $c= -2, -\tfrac{3}{2}, -\tfrac{4}{3}, -\tfrac{5}{4}, -1, -\tfrac{1}{2}$, as  announced in \eqref{cri} and as exemplified in the next section.

\section{Results}\label{resul}

This section presents a concise summary of our main theoretical results on the global heteroclinic rebel dynamics by rebel migrations among stationary 2-clusters, in the limit of large dimension $N\rightarrow+\infty$.
We recall that our results are based on the skew product structure \eqref{ODEsq}, \eqref{ODEy1s}; 
see also the scaled version \eqref{ODEys}.
Our presentation is based on the central concept of rebel flows; see section \ref{rebdy}. 
We explicitly caution the reader to proceed with care: thorough familiarity with the peculiarities of our approach, as carefully outlined in the introduction and as substantiated and exemplified, particularly, in sections \ref{sumdy} and \ref{rebdy} is now required.
This includes a clear understanding of our explanations of the rebel flows \ref{fig_5_1} -- \ref{fig_5_5}.

As announced in \eqref{cri}, we will illustrate the rebel flows of the 3-cluster system \eqref{ODExi}.
Each rebel flow represents the global heteroclinic rebel dynamics in the positive quadrant of cluster fractions $N_1/N$, horizontally, and compactified asynchrony $s$ of stationary 2-clusters, vertically.
The flow lines follow the conserved level contours $\lambda=\lambda(\alpha,s)$ of the integrable planar rebel flow, for all feasible real parameters $\lambda$.
See \eqref{lamas}.
We recall the relation $N_1/N=\alpha/(\alpha+1)$ between the size ratio $\alpha=N_1/N_3$ of the large clusters and the compactified horizontal axis, in all rebel flow diagrams.

We only address the non-blocking regions, where rebel heteroclinics between 2-cluster equilibria are not blocked by stationary 3-clusters; see section \ref{rebdy}. 
In all figures we have shaded the region in the plane $(N_1/N,\arctan(s))$ where  heteroclinic rebel orbits between the two large clusters are blocked, according to section \ref{rebdy}.

\begin{figure}[t!] 
\centering
	\includegraphics[width = 1.00\textwidth]{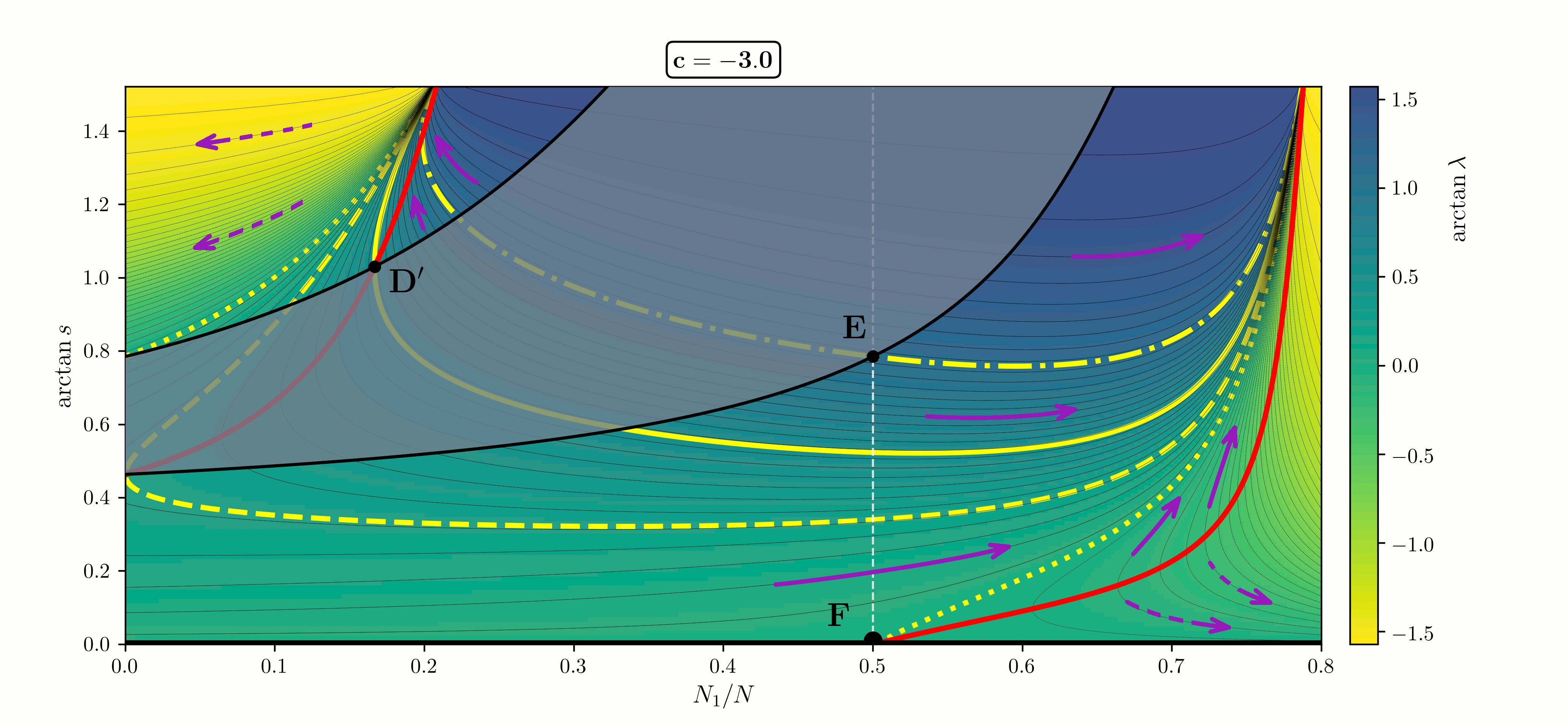}
	\caption{
	Rebel flows in the plane $(N_1/N, \arctan(s))$ for case \ref{resul1}, $-\infty<c=-3<-2$. Color coding and legends as in figures \ref{fig_5_3}, \ref{fig_5_5}. 
	For definition of the intersection points $\mathbf{D',E, F}$, see \eqref{cs1mm}, \eqref{ef}. 
	The shaded region marks blocking of rebel heteroclinic migration between the two large clusters of size ratio $\alpha=N_1/N_3 \,$. 
	Magenta arrows indicate the integrable rebel flow on the level curves of the preserved first integral $\lambda= \lambda(\alpha, s)$. 
	Solid magenta arrows are used in the $s$-stable region of the asynchronous $2$-cluster equilibrium $s=s_*>0$. 		Dashed magenta arrows account for  the two $s$-unstable regions. See text for a detailed interpretation.}	
	\label{fig_6_1} 
\centering
	\includegraphics[width = 1.00\textwidth]{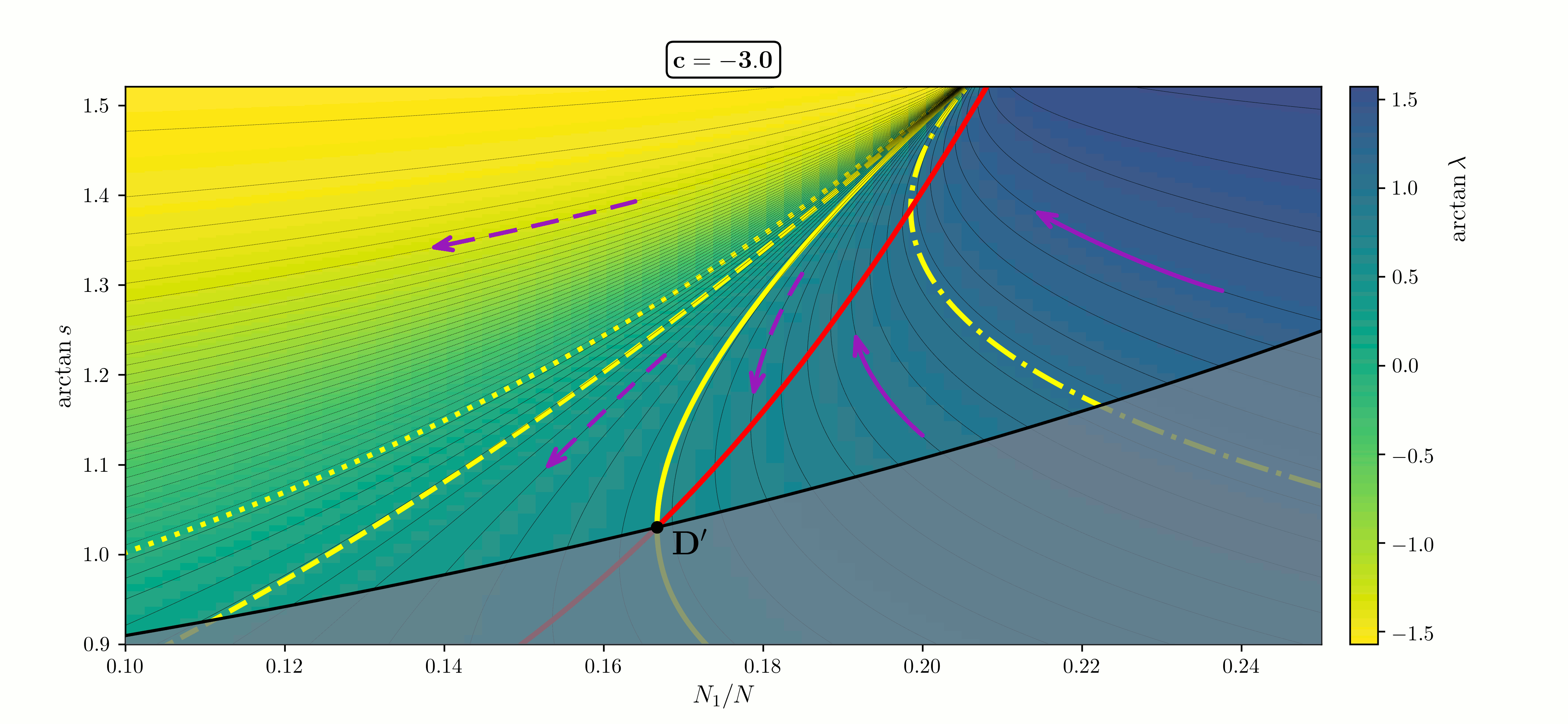}
	\caption{
	Zoom into the upper left $s$-unstable and $s$-stable regions of figure \ref{fig_6_1}.
	}	
	\label{fig_6_2}
\end{figure}

The remaining cubic coefficient $c$ in the reference ODE \eqref{ODExn}, \eqref{ODExi} is fixed, in each rebel flow diagram.
The seven relevant, and distinct, intervals of $c$ are separated by the six critical cubic coefficients $c=-2$, $-\tfrac{3}{2}$, $-\tfrac{4}{3}$, $-\tfrac{5}{4}$, $-1$, $-\tfrac{1}{2}$ of \eqref{cri}, as identified in sections \ref{sumc=-1} and \ref{rebdy} above.
We illustrate the rebel flows, in the seven intervals, by the representatively chosen noncritical coefficients
\begin{equation}
\label{ncri}
	c\quad =\quad -3,\ -1.77,\ -1.37,\ -1.3,\ -1.12,\ -0.75,\ +1\,;
\end{equation}
see figures \ref{fig_6_1}--\ref{fig_6_8} in sections \ref{resul1} -- \ref{resul7}.

%%%%%%%%%%%%%%%%%%%%%%%%%%%%%%
%%%%%%%%%%%%%%%%%%%%%%%%%%%%%%

The driving 2-cluster dynamics $s>0$ of \eqref{ODEsq} is assumed to have reached an $s$-unstable or $s$-stable 2-cluster equilibrium $s=s_*>0$, according to section \ref{sumdy}.
We recall from figures \ref{fig_5_3}--\ref{fig_5_5} how solid magenta arrows along level curves of the conserved first integral $\lambda=(\lambda, s)$ indicate heteroclinic rebellions in $s$-stable regions.
Dashed magenta arrows indicate $s$-unstable regions.
This leaves two dashed magenta regions in each of the figures \ref{fig_6_1}--\ref{fig_6_8} below.

To enforce $s$-stability, in regions which are not $s$-stable according to section \ref{sumdy}, originally, we may reverse time in all ODEs.
For the coefficients $A, B, C$ in \eqref{ODEgen} this amounts to a reversal of all signs.
In \eqref{ODExn} and the following sections, we just replace $\dot{x}_n=\ldots, \  \dot{s}=\ldots,\ \dot{y}= \ldots$ by $-\dot{x}_n= \ldots, \ -\dot{s}= \ldots,\ -\dot{y}= \ldots\ $.
This would extend rebel flows through saddle-node curves, consistently.
In the following, however, we will refrain from such partial time reversals.
This will emphasize the forward or backward destruction of large 2-cluster equilibria near saddle nodes, by small rebellions, and will avoid confusion in the global interpretation of our rebel flow diagrams.

\subsection{The rebel flow for $\mathbf{-\infty < c < -2}$}
\label{resul1}

We begin with the rebel flow for $-\infty<c=-3<-2$ of figure \ref{fig_6_1}.
There are two $s$-stable non-blocked regions, indicated by solid magenta arrows.
The dashed magenta arrows indicate the two $s$-unstable regions.

The lower $s$-stable region of solid magenta arrows is located between the lower black blocking boundary $s_0$ and the right red saddle-node curve.
It is split in four by three  separating non-solid yellow level curves of the bifurcation parameter $\lambda$.
In all four subregions, the cluster $N_1$ wins at the expense of $N_3 \,$.
The rebel flow of successive heteroclinic rebellions leads to infinite growth of the 2-cluster asynchrony, 
\begin{equation}
\label{s-infty}
s=(\xi_3-\xi_1)/(\alpha+1) \ \longrightarrow\  +\infty\,,
\end{equation}
via a size ratio $\alpha=N_1/N_3$ which increases asymptotically to $1/\alpha_c\,,$ given by \eqref{ac}.

Between the dotted and the dashed yellow separatrices, i.e.~for $0<\lambda < 1/4$, all directed level curves of $\lambda$ emanate from the left boundary $N_1/N=0$ and terminate at $\alpha=1/\alpha_c \,,\ s= +\infty$.
This means that successive heteroclinic rebel migrations favor the growth of arbitrarily small clusters $(N_1,\xi_1)$, ``out of the blue", over the  cluster $(N_3,\xi_3)$, until the cluster asynchrony blows up, $s\nearrow +\infty$, at the maximal sustainable size ratio $\alpha= N_1/N_3= 1/\alpha_c>1$: from minority $N_1<N_3$ to majority $N_1>N_3$.

Between the dashed and the dot-dashed yellow separatrices, i.e.~for $1/4<\lambda<\lambda(\mathbf{E})=-(c+1)$, a minority cluster of size $N_1<N_3$ can still reach majority, by the rebel flow, until the cluster asynchrony $s$ blows up.
This time, however, at least a critical minimal size $N_1$ of the smaller cluster is required, which depends on the parameter value $\lambda$.
Indeed that critical size is determined by the realizable value of $\alpha = N_1/N_3$ at the intersection of the level curve of $\lambda$ with the black blocking boundary $s_0 \,$.

Above the dot-dashed separatrix, i.e.~for $\lambda>-(c+1) = \lambda(\mathbf{E})$, the rebel growth of $N_1$ does not cross the dashed white line $N_1/N=\tfrac{1}{2}$\,.
The cluster size $N_1 \,,$ initiating to the right of $\mathbf{E}$ on the blocking boundary $s_0\,,$ must exceed $N_3$ from the start.
To the right of the dotted separatrix from $\mathbf{F}$, i.e.~for given $\lambda<0$, the minimally required cluster size of $N_1$ is determined by the value $\alpha$ of the cluster ratio $N_1/N_3$ on the right red saddle node curve $s_\mathrm{minmax}$ corresponding to $\lambda_\mathrm{minmax}= \lambda$.

The upper $s$-stable region of solid magenta arrows is located in the triangular wedge above $\mathbf{D'}$, between the left red saddle-node curve $s_\mathrm{minmax}$ and the upper black blocking boundary $s_1 \,$.
Rebellions there originate from $s_1$ and decrease $\alpha= N_1/N_3<1$, until they terminate at the left red curve $s_\mathrm{minmax}>0$, where 2-cluster solutions disappear into saddle-node bifurcations. See also the zoom \ref{fig_6_2} of figure \ref{fig_6_1}.

Similar remarks apply to the remaining two non-blocking regions, which are $s$-unstable.
The dashed magenta arrows indicate the resulting rebel flow.
The upper left $s$-unstable region is bounded below by the black blocking boundary $s=s_1(\alpha)$ and, on the right, by the left red saddle-node curve $s=s_\mathrm{minmax}(\alpha)$; see \eqref{s1} and \eqref{sminmax}.
In \eqref{s1mm0} and \eqref{cs1mm} we have denoted their intersection by $\mathbf{D'}$.
The two yellow separatrix levels $\lambda(\alpha, s)= 0$, dotted, and $\lambda (\alpha, s)= \lambda (\mathbf{D'})$, solid, define three subregions, which are distinguished by the eventual fate of successive heteroclinic rebel migrations under the rebel flow.
This leads to the following trichotomy, depending on the parameter $\lambda$ in the region of the initial 2-cluster configuration.
First, the rebel flow may drive $N_1$ to extinction, at the left boundary $\alpha=0$.
Second, successive rebellions towards the black blocking boundary $s_1(\alpha)$ to the left of $\mathbf{D'}$ may ultimately position a small third cluster in-between the two large clusters.
This will stop further rebellions between them.
Or, third, successive rebellions may eventually disrupt the stationary 2-cluster configuration altogether, at the left red saddle-node cluster configuration $s_\mathrm{minmax}(\alpha)$ to the right of $\mathbf{D'}$.
In all three cases, the ongoing decay of $\alpha=N_1/N_3$ may originate from asynchrony up to $s=+\infty$, at finite size ratios $\alpha$ up to $\alpha_c<1$.

The lower right region of dashed magenta rebel dynamics does not involve unbounded asynchrony, for fixed parameter $\lambda$.
All rebellions favor $N_1$ over $N_3 \,,$ this time, and terminate at $\alpha= N_1/N_3= \infty$ alias $N_3=0,\ N_1=N$.
Heteroclinic rebels defect from $N_3$ to the larger cluster $N_1 \,$.
Defection originates from the red saddle-node boundary $s_\mathrm{minmax}(\alpha)>0$, to the right of $\mathbf{F}$, for some $\lambda$-dependent minimal $\alpha= N_1/N_3>1$.
Note that the majority $N_1>N_3$ prevails, because the dashed white line $N_1/N=\tfrac{1}{2}$ is not crossed.

\subsection{The rebel flow for $\mathbf{-2<c< -3/2}$}
\label{resul2}

\begin{figure}[t] 
\centering
	\includegraphics[width = 1.00\textwidth]{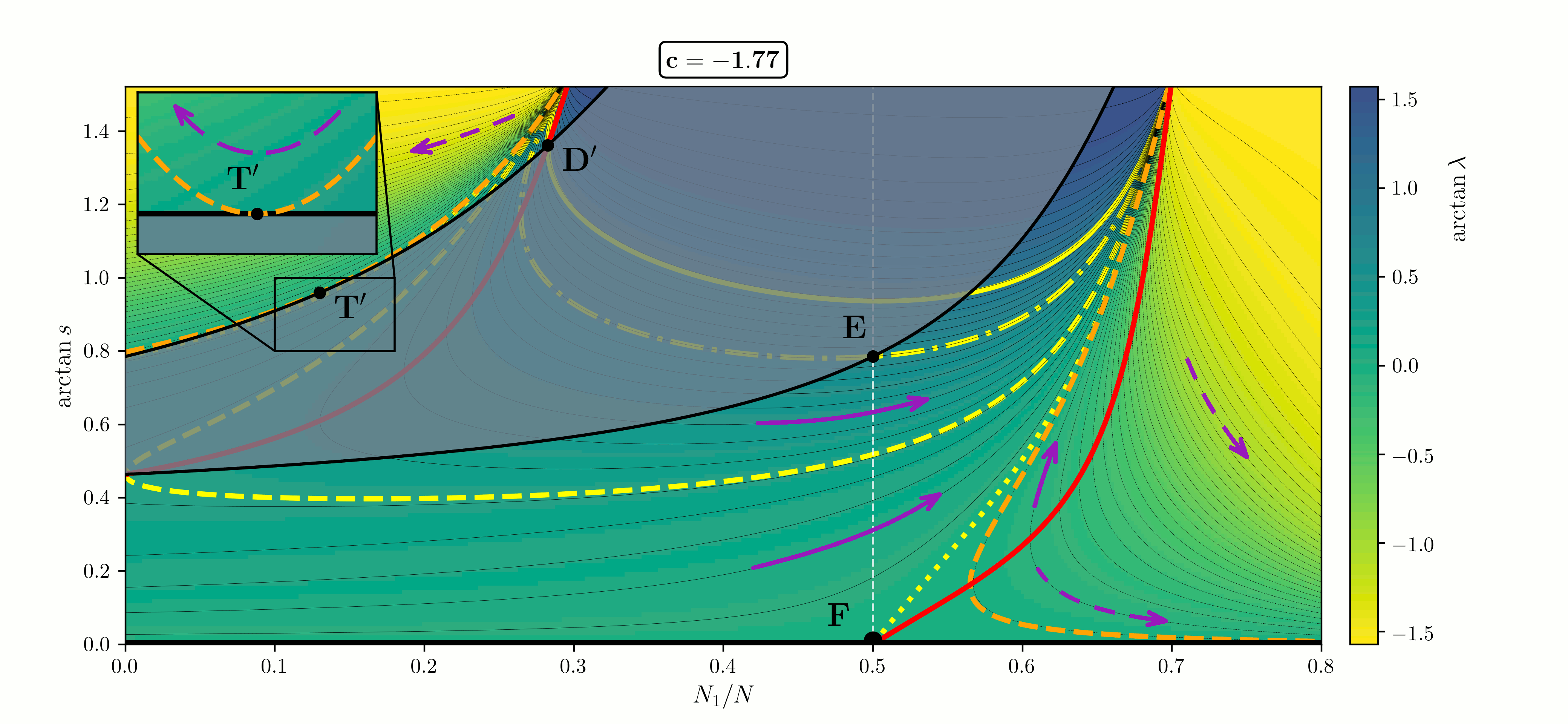}
	\caption{
	Rebel flows in the plane $(N_1/N, \arctan(s))$ for case \ref{resul2}, $-2<c=-1.77<-3/2$. For definition of the tangency $\mathbf{T'}$ between the upper black blocking boundary and the level contour $\lambda(\alpha, s)= \lambda(\mathbf{T'})$; see \eqref{t1}.}	
	\label{fig_6_3} 
\end{figure}

We address the rebel flow for  $-2<c=-1.77<-3/2$ of figure \ref{fig_6_3} next.
The description is identical to the previous case $c=-3$, in the original $s$-stable regions with solid magenta arrows, and in the lower right $s$-unstable region with dashed arrows.
Note however the intersection point $\mathbf{D'}$ and the new tangency point $\mathbf{T'}$ on the upper black blocking boundary $s_1 \,$; see \eqref{t1}.
These points only affect level contours in the upper left $s$-unstable region of dashed magenta rebel arrows.
The corner point $\mathbf{D'}$ and its level contour $\lambda(\alpha, s)= \lambda(\mathbf{D'})$ retain their previous significance.
See in particular the previous zoom in figure \ref{fig_6_2}.
However, the level contour $\lambda (\alpha, s)= \lambda(\mathbf{T'})$  of the new tangency point $\mathbf{T'}$ consists of two branches.
Only above $\lambda= \lambda(\mathbf{T'})$, the rebel flow still terminates at $N_1=0$, originating from asynchrony $s =+\infty$ at $\alpha=\alpha_c \,$.
Below the left branch of $\lambda= \lambda (\mathbf{T'})$, the rebel flow originates from the blocking boundary, instead.
Below the right branch, the rebel flow still originates from $s = +\infty, \ \alpha=\alpha_c\,$, as before, but reaches the black blocking boundary at a minimal cluster ratio $\alpha=\alpha (\lambda)>0$.

%%%%%%%%%%%%%%%%%%%%%%%%%%%%%%
%%%%%%%%%%%%%%%%%%%%%%%%%%%%%%

\subsection{The rebel flow for $\mathbf{-3/2<c<-4/3}$}
\label{resul3}

\begin{figure}[t!] 
\centering
	\includegraphics[width = 1.00\textwidth]{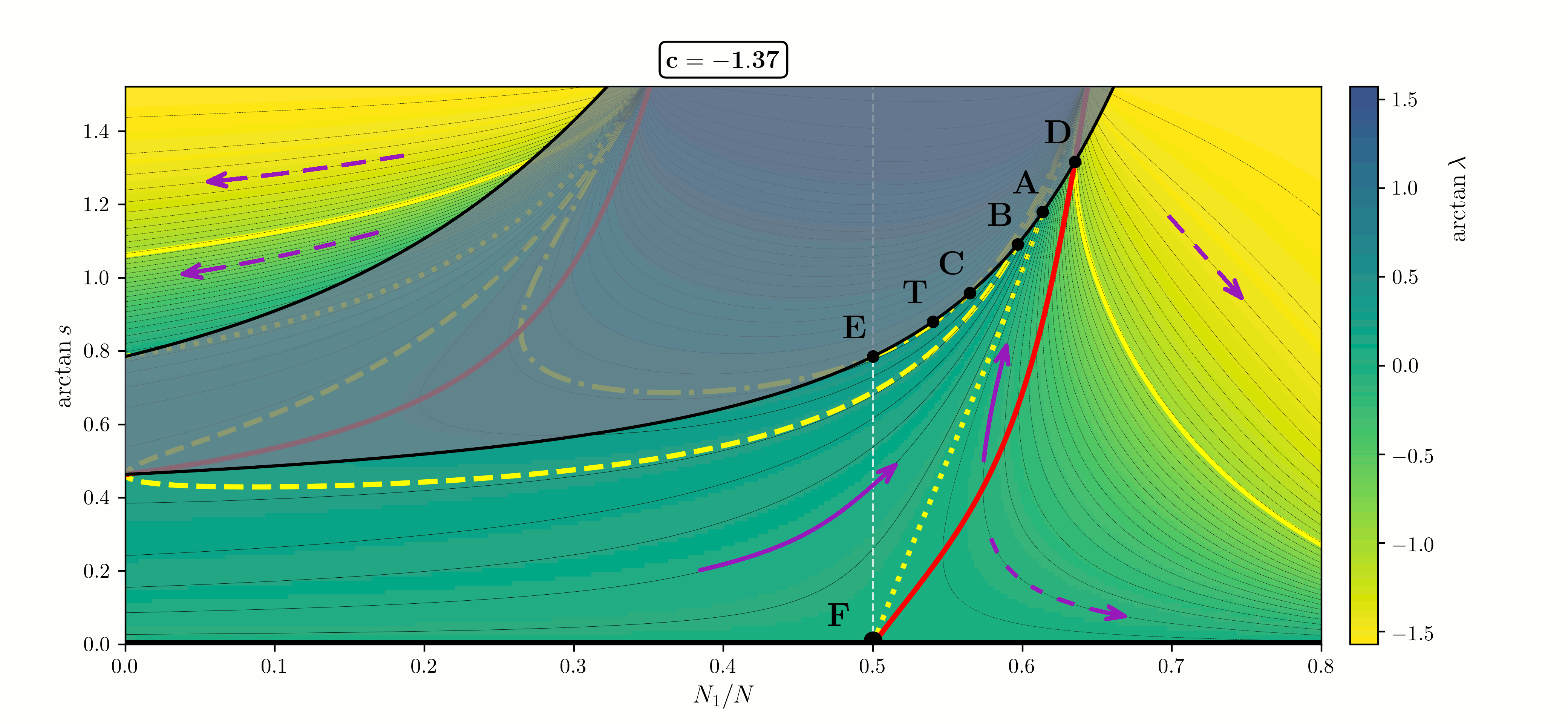}
	\caption{
	Rebel flows in the plane $(N_1/N , \arctan(s))$ for case \ref{resul3}, $-3/2<c=-1.37<-4/3$.}		
	\label{fig_6_4} 
	\includegraphics[width = 1.00\textwidth]{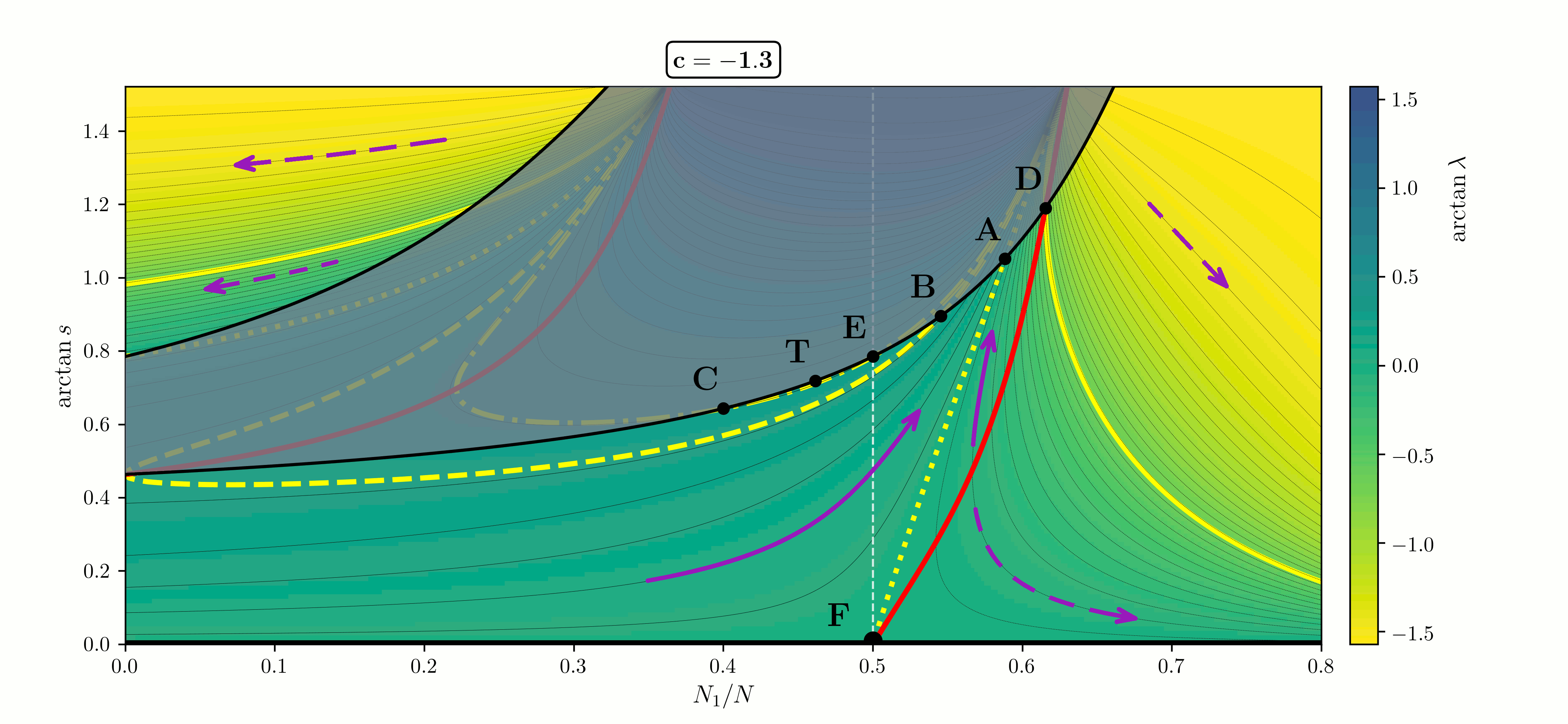}
	\caption{
	Rebel flows in the plane $(N_1/N, \arctan(s))$ for case \ref{resul4}, $-4/3<c=-1.3<-5/4$. See also figure \ref{fig_5_3}.}	
	\label{fig_6_5} 
\end{figure}

The rebel flow for $-3/2<c=-1.37<-4/3$ of figure \ref{fig_6_4} features only a single $s$-stable region of solid magenta arrows, and two $s$-unstable regions of dashed magenta arrows.
The only $s$-stable region, lower triangular between the lower black blocking boundary $s_0$ and the right red saddle-node curve, has now detached from the singular tip $s= + \infty$ at $\alpha = 1/\alpha_c \,$.
The new tip is located at $\mathbf{D}$; see \eqref{s0mm} and \eqref{cs0mm}.
Along $s_0 \,,$ the new intersection points $\mathbf{A, B, C}$ have appeared, with the yellow $\lambda$-levels $\lambda=\lambda(\mathbf{A})=0=\lambda (\mathbf{F})$(dotted), $\lambda=\lambda(\mathbf{B})= 1/4$ (dashed), and $\lambda=\lambda(\mathbf{C})= -(c+1)= \lambda(\mathbf{E})$ (dot-dashed), respectively.
See \eqref{s00}--\eqref{s0c} and \eqref{cs00}--\eqref{cs0c}.
The three yellow separatrices define four subregions.

%%%%%%%%%%%%%%%%%%%%%%%%%%%%%%
%%%%%%%%%%%%%%%%%%%%%%%%%%%%%%	

For $\lambda<0$, i.e.~in the triangular subregion $\mathbf{ADF}$, the rebel flow starts from the red saddle-node curve $\mathbf{DF}$ and terminates at the black blocking boundary segment $\mathbf{AD}$.
In the pentagonal subregion $0<\lambda< \tfrac{1}{4} \,,$ the rebel flow starts from ``blue sky", at $\alpha= N_1/N_3=0$, with tiny $N_1\,$.
The successive rebellions gain majority as they cross the dashed white break-even line $N_1=N_3\,,$ and terminate at the black blocking boundary segment $\mathbf{AB}$.
For $\tfrac{1}{4}< \lambda< -(c+1)= \lambda(\mathbf{E})$, rebellions still gain majority across the dashed white line, but they start on the black blocking boundary $s_0$ to the left of $\mathbf{E}$ and terminate on the segment $\mathbf{BC}$ of $s_0 \,$.
The black blocking segment $\mathbf{EC}$ to the right of $\mathbf{E}$, finally, exhibits a new tangency $\mathbf{T}$ with the level contours of $\lambda$; see \eqref{t0}.
For $-(c+1)= \lambda(\mathbf{E})< \lambda< \lambda(\mathbf{T})$ this leads to rebellions, from $N_3$ to increasing $N_1>N_3$ already in majority, which start and terminate at the black blocking boundary: from $\mathbf{ET}$ to $\mathbf{CT}$. 
Except for the location to the right of the dashed white break-even line, the dynamics follows the zoom in figure \ref{fig_5_4}.

The upper region of dashed magenta arrows has simplified: the rebel flow now originates from the black blocking boundary $s_1 \,,$ with finite $\alpha$ and $s$, and terminates at $\alpha=0$.
The lower region of dashed magenta arrows, likewise, terminates at $\alpha= +\infty \,, \  N_3=0$.
For $\lambda< \lambda (\mathbf{D})$, the rebel flow lines originate from black blocking $s_0$ and, for $\lambda(\mathbf{D})< \lambda <0$, at saddle-nodes (red).

\subsection{The rebel flow for $\mathbf{-4/3<c<-5/4}$}
\label{resul4}

The rebel flow for $-4/3<c=-1.3<-5/4$ of figure \ref{fig_6_5} has been prepared in section \ref{rebdy}; see figure \ref{fig_5_3}.
The two regions of dashed magenta arrows correspond to the previous case \ref{resul3}, verbatim.

The $s$-stable triangular region of solid magenta arrows looks quite similar to figure \ref{fig_6_4}, except for the position of the dashed white line $\mathbf{EF}$ of equal parity $N_1=N_3\,$.
The segment $\mathbf{CT}$ on the black blocking boundary has in fact moved from the right of $\mathbf{E}$ to the left of $\mathbf{E}$, i.e.~from size ratios $\alpha >1$ to $\alpha <1$.
Of the four $s$-stable regions separated by the three yellow level curves $\lambda=0, \tfrac{1}{4}, -(c+1)$, this only effects the region $-c+1<\lambda< \lambda(\mathbf{T})$ which now features a minority $N_1 \,,$ still growing, rather than a majority.
Rebellions lead from $\mathbf{CT}$ to $\mathbf{ET}$, this time.

%%%%%%%%%%%%%%%%%%%%%%%%%%%%%%
\begin{figure}[t!] 
\centering
	\includegraphics[width=\textwidth, height=0.5\textwidth]{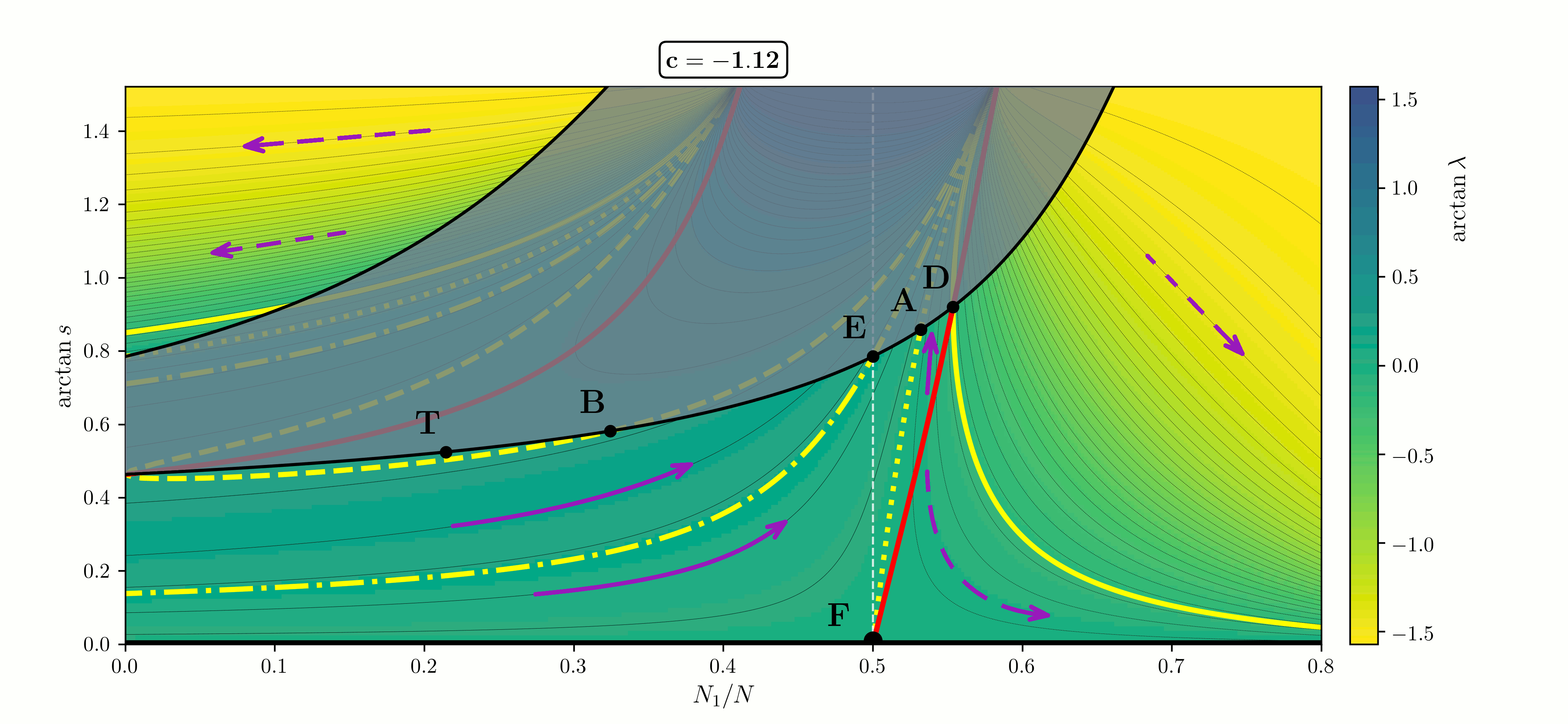}
	\caption{
	Rebel flows in the plane $(N_1/N, \arctan(s))$ for case \ref{resul5}, $-5/4<c=-1.12<-1$.}	
	\label{fig_6_6} 
\end{figure}
%%%%%%%%%%%%%%%%%%%%%%%%%%%%%%

\subsection{The rebel flow for $\mathbf{-5/4<c<-1}$}
\label{resul5}

For $-5/4<c=-1.12<-1$, as in figure \ref{fig_6_6}, the rebel flow in the two regions of dashed magenta arrows remains the same, qualitatively, as in the two previous figures \ref{fig_6_4} and \ref{fig_6_5}.
In the remaining unique $s$-stable region of solid magenta arrows, the dot-dashed yellow level $\lambda=-(c+1)= \lambda(\mathbf{E})$ has dropped below the dashed yellow level $\lambda= \tfrac{1}{4}$, as $c$ increased through $-5/4$.
The region of rebellion from ``blue sky" minority $N_1=0$ to majority $N_1>N_3 \,,$ across the dashed white line $\mathbf{EF}$, therefore requires $0<\lambda< -(c+1)$, now.
Rebel flow termination occurs at the black blocking segment $\mathbf{EA}$ of $s_0 \,$.
The second intersection point $\mathbf{C}$ of $s_0$ with the yellow level $\lambda= -(c+1)= \lambda(\mathbf{E})$ has disappeared.
The rebel flow in the region $\tfrac{1}{4}< \lambda<\lambda(\mathbf{T})$ now features growth of the minority $N_1$ from the black blocking segment of $s_0$ on the left of $\mathbf{T}$ to $\mathbf{TB}$.

%%%%%%%%%%%%%%%%%%%%%%%%%%%%%%
\begin{figure}[t!] 
\centering
	\includegraphics[width = 1.00\textwidth]{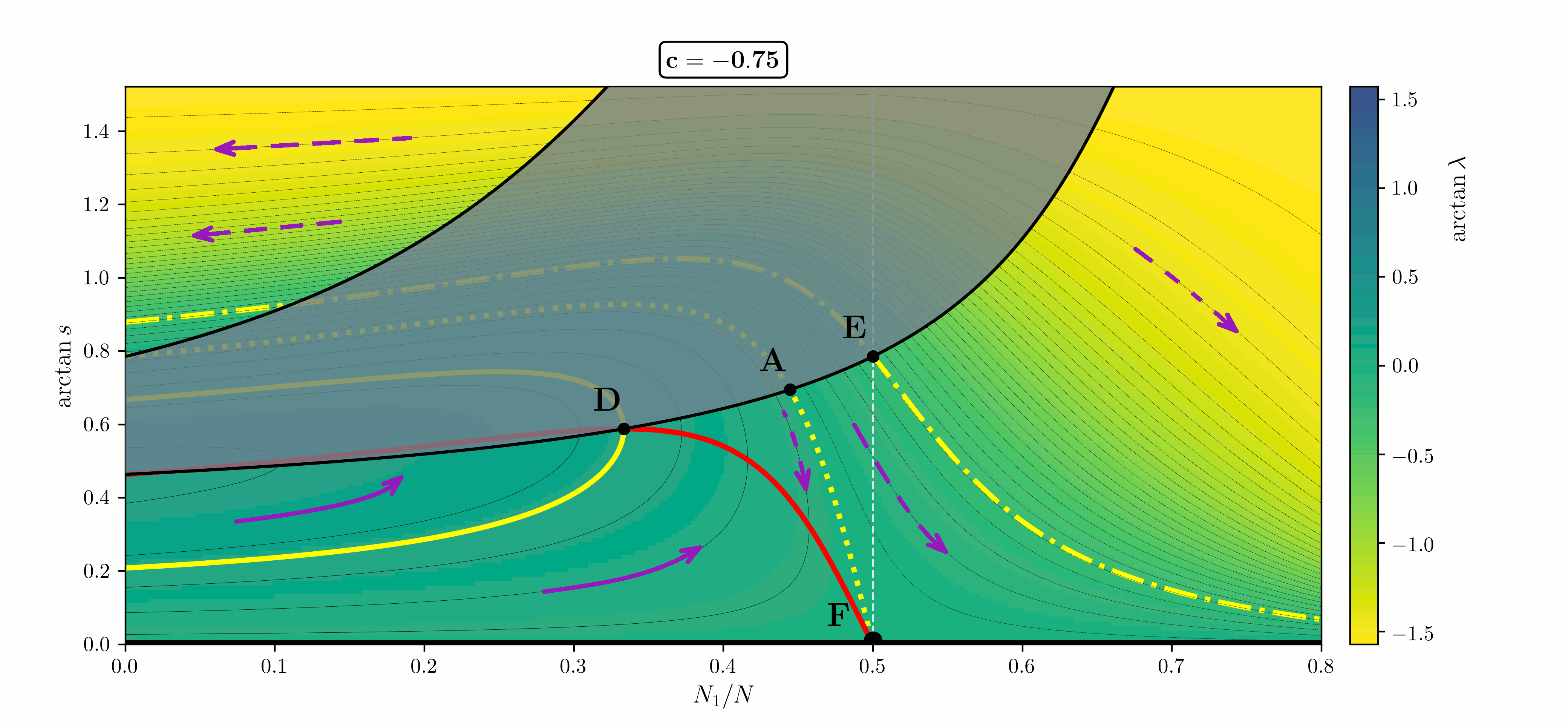}
	\caption{
	Rebel flows in the plane $(N_1/N\alpha, \arctan(s))$ for case \ref{resul6}, $-1<c=-0.75<-1/2$.}	
	\label{fig_6_7} 
\end{figure}
%%%%%%%%%%%%%%%%%%%%%%%%%%%%%%

\subsection{The rebel flow for $\mathbf{-1<c<-1/2}$}
\label{resul6}

The rebel flow for $-1<c=-0.75<-1/2$ of figure \ref{fig_6_7} has also been prepared in section \ref{rebdy}; see figure \ref{fig_5_5}. 
The upper left $s$-unstable region of dashed magenta arrows remains the same, qualitatively, as in figures \ref{fig_6_4}--\ref{fig_6_6}.
The original $s$-stable region of solid magenta arrows is now contained in the region $N_1/N$ to the left of the dashed white line $N_1=N_3$ of equal parity.
Thus $N_1$ is, and remains, in minority $N_1<N_3$ there.
The only yellow separatrix $\lambda= \lambda(\mathbf{D})>0$ (solid) highlights the intersection $\mathbf{D}$ of the lower black blocking boundary $s_0$ with the only remaining red saddle-node curve $s_\mathrm{minmax} \,$; see \eqref{s0mm} and \eqref{cs0mm}.
All rebel flow lines start from ``blue sky", i.e.~at vanishing $N_1 \,$.
In the subregion $0<\lambda< \lambda(\mathbf{D})$ they terminate at the red saddle-nodes.
In the complementary subregion $\lambda(\mathbf{D})< \lambda < \tfrac{1}{4}$, they terminate at the black blocking boundary.

The second $s$-unstable region of dashed magenta arrows, on the right, is now subdivided into three subregions by the two yellow separatrices $\mathbf{AF}$ of $\lambda=0$ (dotted) and $\lambda=-(c+1)= \lambda(\mathbf{E})$ (dot-dashed).
For $\lambda(\mathbf{D})>\lambda>\lambda(\mathbf{A})=0$, i.e.~in the left subregion, the rebel flow diminishes the larger cluster $N_3$ from the segment $\mathbf{AD}$ of the black blocking boundary $s_0 \,$ to the red saddle-nodes, where $N_3$ is still in majority.
For $\lambda<\lambda(\mathbf{A})=0$, in contrast, rebel flow lines lead to extinction of $N_3$.
In the middle subregion $\lambda(\mathbf{A})=0>c>-(c+1)=\lambda(\mathbf{E})$, the majority cluster $N_3$ from the black blocking segment $\mathbf{AE}$ of $s_0 \,$ crosses the dashed white line of equal parity, under the rebel flow, until it goes extinct.
In the right subregion $\lambda(\mathbf{E})=-(c+1)>\lambda $ of the rebel flow, the cluster $N_3$ remains a minority on its path to extinction, originating from the black blocking segment of $s_0 \,,$ to the right of $\mathbf{E}$, at a finite value of $s$.

%%%%%%%%%%%%%%%%%%%%%%%%%%%%%%
\begin{figure}[t] 
\centering
	\includegraphics[width = 1.00\textwidth]{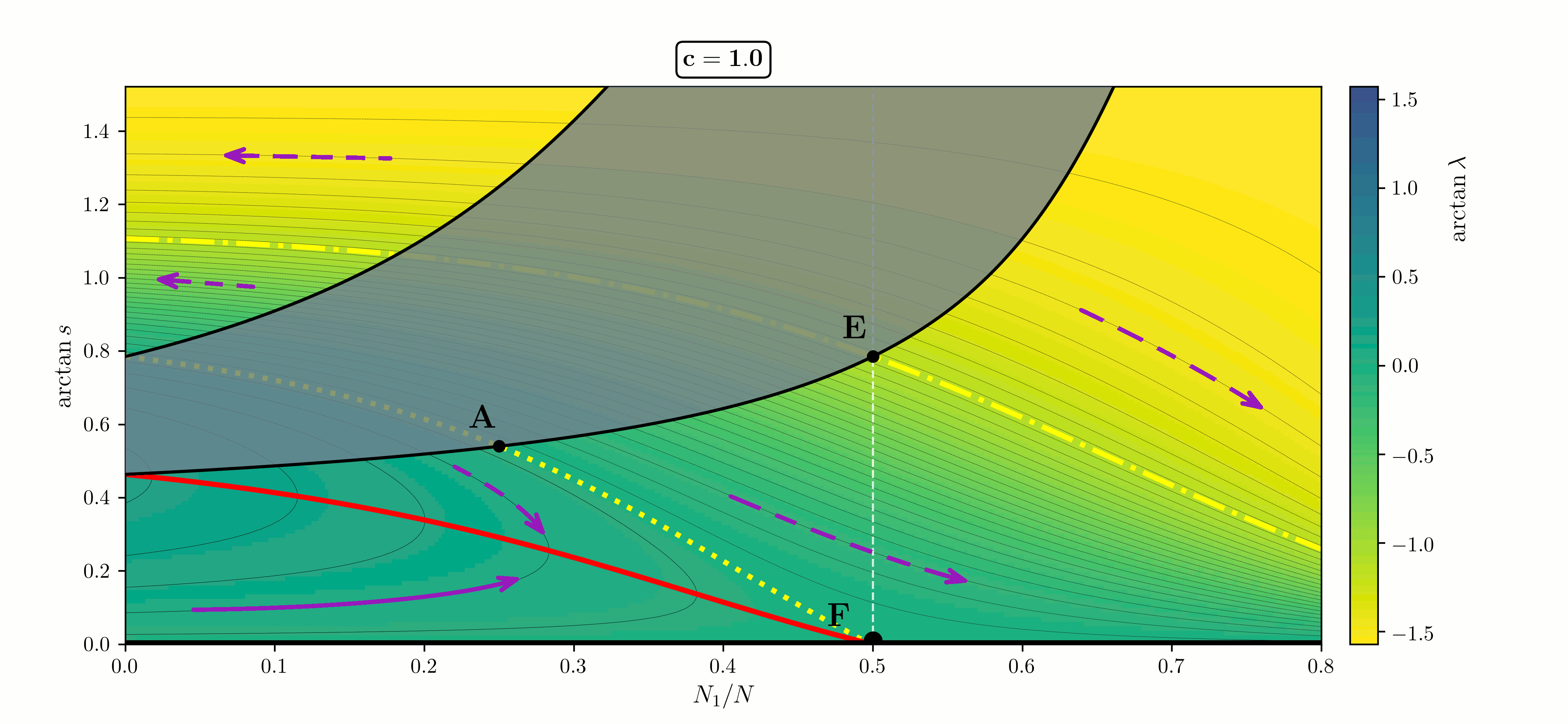}
	\caption{
	Rebel flows in the plane $(N_1/N, \arctan(s))$ for case \ref{resul7}, $-1/2<c=+1<+\infty$.}	
	\label{fig_6_8} 
\end{figure}
%%%%%%%%%%%%%%%%%%%%%%%%%%%%%%

\subsection{The rebel flow for $\mathbf{-1/2<c< +\infty}$}
\label{resul7}
The final rebel flow is $-1/2<c=1<+\infty$,  as in figure \ref{fig_6_8}.
As for all $c>-3/2$ we obtain a single $s$-stable region, with solid magenta arrows, and two $s$-unstable regions with dashed magenta arrows; see figures \ref{fig_6_4}--\ref{fig_6_7}.
The upper left dashed magenta region remains unchanged.
The solid magenta region has lost $\mathbf{D}$ from its boundary: all rebel flow lines originate from red saddle-nodes and terminate at $N_1=0 \,, \alpha=0$, with $N_1$ remaining in minority.

The wedge of the lower dashed magenta region, between the red saddle nodes and the lower black blocking boundary $s_0$, now reaches all the way to the left tip at $\alpha=0, s=\tfrac{1}{2}$ where $\lambda=\tfrac{1}{4}$.
The two yellow level curves $\lambda=0=\lambda(\mathbf{A})$ (dotted) and $\lambda= -(c+1)= \lambda (\mathbf{E})< -\tfrac{1}{2}$ (dot-dashed) divide the region into three subregions, just as in the previous case \ref{resul6} of $-1<c<-1/2$; see figure \ref{fig_6_7}.

The only difference, now, is that rebellions for $0<\lambda< \tfrac{1}{4}$ originate from anywhere on the black blocking boundary, to the left of $\mathbf{A}$, i.e.~at any size ratio $0<\alpha=N_1/N_3< \alpha(\mathbf{A})=1/(c+2)<2/3$.
This is in marked contrast to the previous case \ref{resul6}, where the size ratio $\alpha=N_1/N_3$  remained bounded away from $\alpha=0$ by $0<-2c-1= \alpha(\mathbf{D})< \alpha < \alpha (\mathbf{A})=1/(c+2)<1$.
The two other subregions of $\lambda<0$, as before, show how the rebel flow leads the cluster $N_3$ to extinction from the maximal value of $N_3/N=1-\alpha/(\alpha+1)$ on the black blocking boundary, which is sustainable at the given level of $\lambda < 0$.

If we reverse time, to make this $s$-unstable region $s$-stable, then the growth of $N_3=0$ to the maximally sustainable $N_3$ reveals the limitations of rebel dynamics defecting to an emerging minority.

\section{Example: Stuart-Landau oscillators with global coupling}\label{sl}

In this section, we study $N$ globally coupled, identical Stuart-Landau oscillators
\begin{equation}
\label{sln}
\dot{W}_n=(1-(1+\mathrm{i}\gamma)|W_n|^2)W_n+ \beta \cdot (\langle W \rangle- W_n)\, .
\end{equation}
Here $W_n \in \mathbb{C}$ indicate phase and amplitude of the $n$-th oscillator, $n=1, \ldots, N$.
We consider real amplitude dependence $\gamma$ of individual periods, complex coupling $\beta \in \mathbb{C}$. 
As before, $\langle W \rangle := \tfrac{1}{N} \sum W_n$ abbreviates the average or mean field.
Note $S_N\,$-equivariance of \eqref{sln} under the action analogous to \eqref{SN}.

For a background and motivation we recall how \eqref{sln} often serves, in physics, as a ``normal form'' for oscillatory systems close to the onset of oscillation and under the influence of a linear coupling through the mean field \cite{Kura84, VGM2008}. 
This normal form has been established to be a good approximation in a multitude of contexts from various disciplines, whether it be in physics, chemistry, biology, neuroscience, social dynamics, or engineering. 
For an overview see e.g.~\cite{Pik2003, Pik2015} or references 1-15 in \cite{KuGiOtt2015}.

Our motivation to study \eqref{sln} is to gain a deeper understanding of the dynamics of oscillatory electrochemical systems.
Indeed global, linear coupling often controls the evolution of the electrostatic potential of the working electrode, a crucial dynamic variable in electrochemical systems \cite{ WaKiHu2000, Kr2001, PlLiKr2004, VaBeBoKr2005, MiGMKr2009, KoKuKiss2014, SchoZenHeKr2014, SchKr2015, PaHuToKr2017, LiKi2018, NoKrVa2019, HaGAKi2019}. 
The global coupling originates from the electric control of the device: any potential drop in the electrolyte or the external electric circuit is fed back to the evolution of the electrode potential at any location. 
Yet, there are many other situations where the dynamics of the electric potential is governed in almost the same manner as in electrochemical systems. 
Examples include semiconductor devices \cite{scholl2001}, gas discharge tubes \cite{PuBoAM2010}, or arrays of Josephson junctions \cite{BeVeBe1997}. 
Along with these numerous applications go various theoretical studies of the globally coupled Stuart-Landau ensemble  \cite{NaKu1993, Naku1994, NaKu1995, HaRap1992, ShiFran1989, MatStro1990, MMSt1993, DaNa2004, DaNa2006, KuGiOtt2015, KeHaKr2019}.

\begin{figure}[t] 
\centering
	\includegraphics[width = 1.00\textwidth]{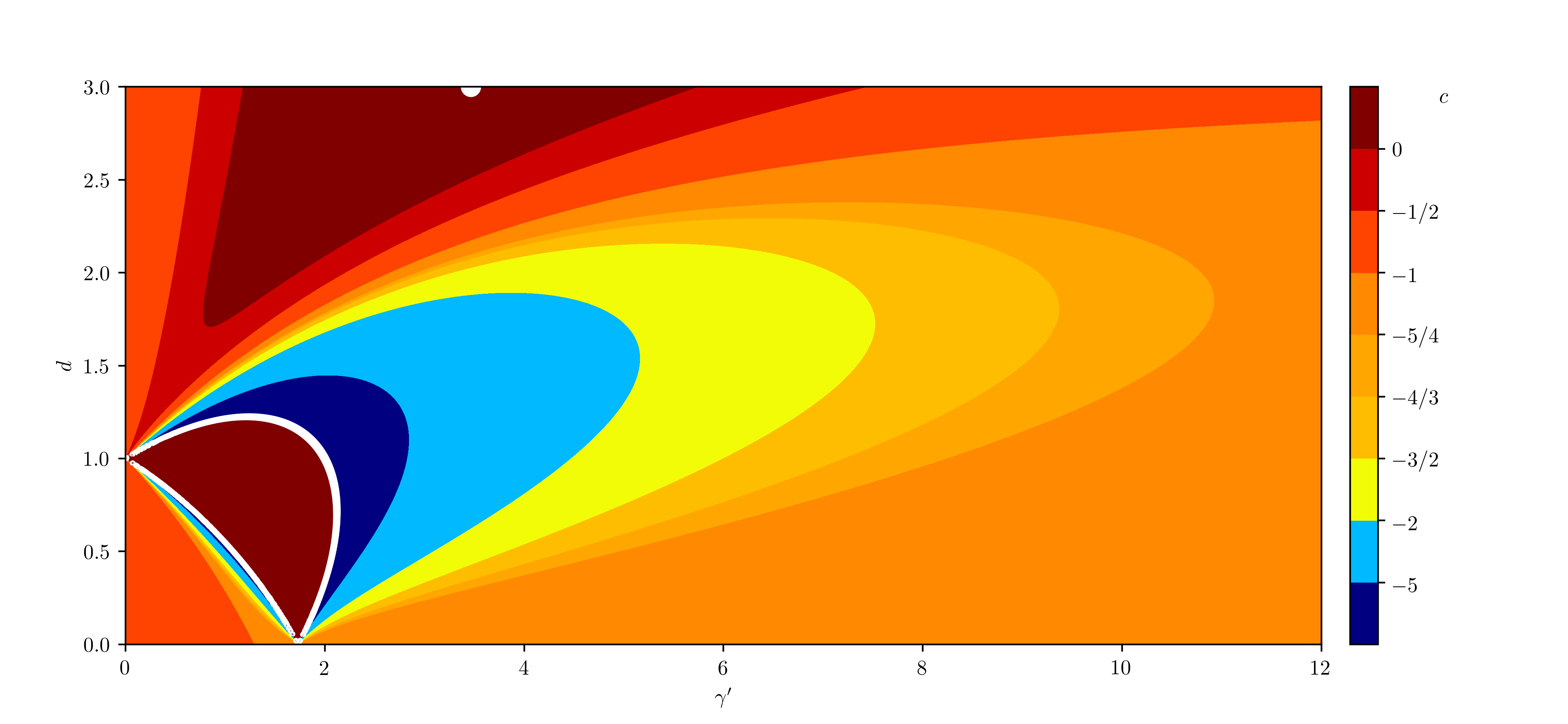}
	\caption{
	Level sets of the cubic coefficient $c=c(\gamma', d)$ in the cubic $S_N$ normal form \eqref{ODExn}, as a function of the positive parameters $\gamma'$ and $d$. 
	See \eqref{c=}. 
	Since $c(\gamma', d)=c(-\gamma', d)$ is quadratic in $\gamma'$, we only plot positive $\gamma', d$. 
	See \eqref{d} and \eqref{beta} for expressions of $d$ and $\gamma'$ in terms of the original coefficients $\gamma \in \mathbb{R}$, of period-amplitude dependence, and $\beta \in \mathbb{C}$, of complex linear coupling, in the Stuart-Landau setting \eqref{sln}. 
	The singular set $c=\pm \infty$, alias $B=0$, is indicated by the white crescent. 
	For the seven resulting rebel flows in the union of the two lowest, blue intervals of $c$, the union of the two uppermost red intervals, and the five remaining intermediate $c$-intervals, respectively, see the representative figures \ref{fig_6_1}-- \ref{fig_6_8} of sections \ref{resul1} -- \ref{resul7}. 
	The white dot at $\gamma'=2\sqrt{3}$ on the upper boundary $d=3$ indicates the 2-cluster singularity $c=1$ of fig.~\ref{fig_6_8}.}	\label{fig_7_1} 
\end{figure}

As \cite{Ash16} point out, $S_N\,$-equivariant coupling of phase oscillators allows for significantly more general nonlinear coupling terms than just a mean field average.
It is therefore perhaps surprising, that all seven rebel flows of sections \ref{resul1} -- \ref{resul7} will appear under linear mean field coupling \eqref{sln}.
On the other hand, we will cover all nondegenerate regions of the real three-dimensional parameter space in \eqref{sln} by our complete list of just seven one-parameter rebel flows.
We refer to figure \ref{fig_7_1} for a summary of these results.

Specifically, we consider bifurcations from the globally synchronous periodic solution
\begin{equation}
\label{sync}
W_n(t)= \exp(-\mathrm{i}\gamma t)
\end{equation}
of amplitude 1 and minimal period $2\pi/ \gamma$.
We consider the \emph{Benjamin-Feir instability} of total synchrony, i.e.~bifurcation due to a nontrivially vanishing critical Floquet exponent of high geometric multiplicity; see \cite{BenFeir, HaRap1992}.
This differs, from the start, from approaches like \cite{DiRo09, Ash16} which focus on Hopf bifurcation in system \eqref{sln}, on invariant $N$-tori, and on their phase field equations.

Somewhat unconventionally, we rewrite \eqref{sln} in complex log-polar coordinates $Z_n=R_n+\mathrm{i}\Psi_n$ of $W_n= \exp(Z_n)$ as
\begin{equation}
\label{Zndot}
\dot{Z}_n=\dot{W}_n/W_n= 1-(1+ \mathrm{i}\gamma)|W_n|^2+ \beta(-1+\tfrac{1}{N} \sum^N_{k=1} W_k/W_n) \, .
\end{equation}
We now invoke the notation \eqref{avg} and define 
\begin{equation}
\label{R_n}
\begin{aligned}
R:  =  \tfrac{1}{N}\sum R_n\,, \qquad  & r_n:= \widetilde{R_n}= R_n-R, \\
\Phi:= \tfrac{1}{N} \sum \Phi_n\,, \qquad & \varphi_n: = \widetilde{\Phi_n}= \Phi_n - \Phi,  \\
Z:=R + \mathrm{i}\Phi,  \quad \qquad & z_n:=r_n+ \mathrm{i} \varphi_n\,, \ \mathbf{z}=(z_n)^N_{n=1} \,, \\
\end{aligned}
\end{equation}
to derive
\begin{align}
	\label{phindot}
	\dot{\varphi}_n &= -\gamma \ e^{2R} \widetilde{e^{2r_n}} \, + \mathrm{Im}(\beta \langle e^z \, \rangle \widetilde{e^{-z_n}} \,), \\
	\label{rndot}
	\dot{r}_n \, &=  -e^{2R} \, \widetilde{e^{2r_n}} \, + \mathrm{Re}(\beta \langle e^z \,  \rangle \widetilde{e^{-z_n}} \,), \\
	\label{Rdot} 
	\dot{R} \phantom{n}&= 1-e^{2R} \, \langle e^{2r} \, \rangle + \mathrm{Re}(\beta(\langle e^{z} \, \rangle \langle e^{-z}\, \rangle -1))\,, 
\end{align}
and the average phase
\begin{equation}
\label{Phidot}
\dot{\Phi}= -\gamma e^{2R} \langle e^{2r} \rangle + \mathrm{Im}(\beta (\langle e^z \, \rangle \langle e^{-z} \, \rangle -1)) \,.
\end{equation}
Here we have slightly extended the notation \eqref{avg} to include the complex exponential average and deviation
\begin{equation}
\label{avge}
\begin{aligned}
\langle e^z \rangle &:= \tfrac{1}{N} \sum^N_{n=1} e^{z_n}= \tfrac{1}{N} \sum^N_{n=1} \sum^\infty_{m=0} \tfrac{1}{m!}z^m_n = \sum^\infty_{m=0} \tfrac{1}{m!} \langle z^m \rangle \,, \qquad \textrm{and}\\
\widetilde{e^{z_n}} &:= e^{z_n}- \langle e^z \rangle = \sum^\infty_{m=0} \tfrac{1}{m!}(z^m_n- \langle z^m \rangle ) = \sum^\infty_{m=0} \tfrac{1}{m!} \widetilde{z^m_n} \,. 
\end{aligned}
\end{equation}
The globally synchronous solution \eqref{sync} becomes the \emph{trivial equilibrium} $\mathbf{z}=0, R=0$ of \eqref{phindot}--\eqref{Rdot}, in this notation.
The average phase $\Phi (t)$ does not appear in these ODEs, due to $S^1$-equivariance of the original Stuart-Landau system \eqref{sln} under uniform phase shifts.
We will therefore ignore the average phase $\Phi(t)$, henceforth.
We only keep in mind how equilibria of $\mathbf{z}, R$, and heteroclinic orbits between them, actually indicate periodic orbits and their heteroclinic connections, via the skew product structure of $\dot{\Phi}= \ldots$, driven by the $\Phi$ independent dynamics of $\mathbf{z}, R$, only.

Our task, in the present section, is the derivation of the reduced flow \eqref{ODEgen}, i.e.
\begin{equation}
\label{ODEgensl}
\dot{x}_n= \mu_+x_n+ A \widetilde{x^2_n}+ B \widetilde{x^3_n} + C \langle x^2 \rangle x_n + \ldots
\end{equation}
in a center manifold of the trivial equilibrium $\mathbf{z}=0, \ R=0$ of the system \eqref{phindot}-- \eqref{Rdot}, at a zero Benjamin-Feir eigenvalue $\lambda:=\mu_+$ of the linearization.
See for example \cite{carr, chha, vancmf} for a background on center manifolds.

An outline of this standard procedure is as follows.
We replace $z_n=r_n+\mathrm{i}\varphi_n \in \mathbb{C}$ by suitable linear real coordinates $(x_n\,, y_n)$ such that the eigenspace of the mandatory eigenvalue $\mu_+=0$ is given by $\mathbf{y}=0, \ R=0$.
The remaining eigenvalues will be $\mu_-<0$, for $\mathbf{x}=0, \ R=0$, and $\mu_0=-2$, for $\mathbf{x}= \mathbf{y}=0$.
Since $\langle z \rangle =0$, by construction of $z_n= \widetilde{R_n}+ \mathrm{i} \widetilde{\Phi_n} \,,$ we will inherit $\langle x \rangle =0 = \langle y \rangle$, i.e.~$\mathbf{x, y} \in X_0$ will each realize the standard representation of $S_N \,$; see \eqref{SN}, \eqref{X0}.
Since the $S_N\,$-invariant center manifold can be written as a graph of $(\mathbf{y},R)$ over $\mathbf{x}$, tangent to the eigenspace of $\mu_+=0$ at the trivial equilibrium, truncation to second order yields
\begin{align}
	\label{yncmf}
y_n&=a \widetilde{x^2_n}+ \ldots \ , \\
	\label{Rcmf}
R&=b \langle x^2 \rangle + \ldots \ ,
\end{align}
with suitable real coefficients $a,b$ calculated below.
Substitution of \eqref{yncmf}, \eqref{Rcmf} into the ODE $\ \dot{x}_n = \ldots$ with vanishing linear part then allows us to determine the coefficients $A, B, C$ of the reduced flow \eqref{ODEgensl} in the center manifold, up to third order in $\mathbf{x}$, as required for our analysis of \eqref{ODEgen}, \eqref{ODExn}.
We can then invoke the results of sections \ref{intro}--\ref{resul} to detect rebel heteroclinic migration between periodic 2-cluster solutions of the globally coupled Stuart-Landau system \eqref{sln}.
See \cite{elm, stelm} for another example in a Darwinian evolution setting of sympatric speciation.

To substantiate the above outline we start from the linear change of coordinates
\begin{equation}
\begin{aligned}
\label{cmftrf}
2dx_n:&=-r_n+\tfrac{d+1}{\gamma '} \varphi_n \,,& \qquad & r_n=(1-d)x_n+(1+d)y_n \,, \\
2dy_n:&=+r_n+ \tfrac{d-1}{\gamma '} \varphi_n \,, & \qquad & \varphi_n = \gamma'  x_n + \gamma' y_n \,,
\end{aligned}
\end{equation}
for $n=1, \ldots, N $.
The system on the right defines the inverse of the system on the left.
Here $d$ abbreviates the discriminant root
\begin{equation}
\label{d}
d:= \sqrt{1- \beta^2_I- 2\gamma \beta_I} >0 \,,
\end{equation}
writing the real and imaginary parts of the complex linear coupling as $\beta= \beta_R + \mathrm{i}\beta_I \,$.
Of course we assume \emph{positive discriminant}, i.e.
\begin{equation}
\label{dis}
\beta^2_I+ 2 \gamma \beta_I< 1 \,,
\end{equation}
because we address real eigenvalues $\lambda=\mu_+$. 
For complex instabilities $\lambda$ see for example \cite{Bury}.
The coefficient $\gamma'$ in \eqref{cmftrf} is defined as
\begin{equation}
\label{g'}
\gamma':= \beta_I\,+ 2 \gamma \ . 
\end{equation}

The two real eigenvalues of the linearization of \eqref{phindot}, \eqref{rndot} at the trivial equilibrium $\mathbf{z}=0, R=0$ are
\begin{equation}
\label{mu}
\mu_\pm= -(\beta_R+1)\pm d \,.
\end{equation}
Note that each of the real eigenvalues $\mu_-< \mu_+$ is of algebraic and geometric multiplicity $N-1$.
Indeed the eigenspaces $\mathbf{x}=0, \  R=0$ and $\mathbf{y}=0, \ R=0$ are each isomorphic to the standard irreducible representation $X_0$ of $S_N \,$.
The requisite Benjamin-Feir center manifold eigenvalue $\lambda=\mu_+= 0$, at bifurcation, is picked such that $\mu_-< 0= \mu_+$ and the algebraically simple eigenvalue $\mu_0=-2$, in the synchrony direction of $R$, ensure exponential stability towards the reduced flow on the center manifold of $\mu_+ \,$.
We collect some relations among the available coefficients:
\begin{equation}
\begin{aligned}
\label{beta}
\beta_R &=d-1 \,, \\
\gamma' \beta_I &= (\beta_I + 2 \gamma)  \beta_I= 1-d^2=-(\beta_R+2)\beta_R \,, \\
\beta&=(d-1)(1-\mathrm{i}(d+1)/\gamma') \, .
\end{aligned}
\end{equation}
Indeed, the first line follows from $\mu_+=0$ and \eqref{mu}.
The second line uses definition \eqref{g'} of $\gamma ' \,,$ the definition \eqref{d} of $d$, and the first line.
The third line follows from the first and the second.
In summary, \eqref{g'} and \eqref{beta} allow us to express the three free real parameters $\gamma, \beta_R$, $\beta_I$ of \eqref{sln} by the two real parameters $\gamma'$ and $d$, at $\mu_+=0$, \emph{with the only remaining constraint} $d>0 \neq \gamma '$.
We will therefore express the remaining coefficients $a,b$ of \eqref{yncmf}, \eqref{Rcmf}, and $A, B, C$ of \eqref{ODEgensl} in terms of $\gamma'$ and $d$.

To calculate $a,b$ we use existence and $C^k$ differentiability of the center manifold para\-metrizations $R=R(\mathbf{x}),\ r=r(\mathbf{x}),\ y=y(\mathbf{x})$, for any $k>0$.
See \cite{vancmf}.
We first expand the transformed ODE
\begin{equation}
\begin{aligned}
\label{yndot}
0 + \ldots =2d y'_n( \mathbf{x}) \dot{\mathbf{x}} &= 2d\dot{y}_n= \dot{r}_n+ \tfrac{d-1}{\gamma'}\dot{\varphi}_n= \\
&=-(1+(\lambda-1)\tfrac{\gamma}{\gamma '})e^{2R} \widetilde{e^{2r_n}}+ \mathrm{Re} \left((1- \mathrm{i} \tfrac{d-1}{\gamma'}) \beta (\langle e^z \rangle \widetilde{e^{-z_n}})\right)= \\
&= \mu_-y_n+ \ldots \ .
\end{aligned}
\end{equation}
Of course $y'_n(\mathbf{x})$ denotes the derivative of the center manifold parametrization $y_n(\mathbf{x})$ with respect to $\mathbf{x}$, here.
We have substituted \eqref{rndot}, \eqref{phindot} on the right, after the transformation \eqref{cmftrf}.
On the left, we have inserted the quadratic expansion \eqref{yncmf}.
Note that $\dot{\mathbf{x}}= \mu_+\mathbf{x}+ \ldots$ with $\mu_+ = 0$ is at least quadratic.
Moreover, tangency of the center manifold to the eigenspace $y_n=R=0$ implies $y_n'(0)=0$.
Therefore, the left hand side of \eqref{yndot} starts at (omitted) cubic order.
Substitution  of \eqref{cmftrf}, \eqref{beta}, and the expansion \eqref{yncmf} on the right side of \eqref{yndot}, yield the desired result
\begin{equation}
\label{a=}
a= \tfrac{1-d}{8 \gamma'^2 d^2}\left( \gamma'^2+ (d-1)^2\right) \left(\gamma'^2+ 3 (d^2-1) \right) \,,
\end{equation}
by comparison of quadratic coefficients. 
For $R=R(\mathbf{x})$, we analogously obtain
\begin{equation}
\label{Rprime}
0 + \ldots= R'(\mathbf{x}) \dot{\mathbf{x}}= \dot{R}= 1-e^{2R} \langle e^{2r} \rangle + \mathrm{Re} \left( \beta ( \langle e^z \rangle \langle e^{-z} \rangle -1 ) \right) \,,
\end{equation}
with a left hand side of at least cubic order.
Substitutions and comparison of second order coefficients yield
\begin{equation}
\label{b=}
b=\tfrac{1}{2}(1-d)\left(\gamma'^2+(d-1)(d+5)\right) \,.
\end{equation}

To calculate the reduced flow $\dot{x}_n =f_n(\mathbf{x})$ in the center manifold, to order $k \geq 2$, it is always sufficient to expand the center manifold itself to order $k-1$.
To determine the quadratic coefficient $A$ and the cubic coefficients $B, C$ in \eqref{ODEgen}, we therefore expand
\begin{equation}
\label{xndot}
\begin{aligned}
2d\dot{x}_n =& - \dot{r}_n + \tfrac{d+1}{\gamma'} \dot{\varphi}_n= \\
=&- \left(-1+(d+1)\tfrac{\gamma}{\gamma '}\right) e^{2R}\widetilde{e^{2r_k}}+ \mathrm{Re} \left( -1-\mathrm{i}\tfrac{d+1}{\gamma '} \beta (\langle e^z \rangle \widetilde{e^{-z_k}})\right)= \\
=&\mu_+x_n+A \widetilde{x^2_n}+ B \widetilde{x^3_n}+ C \langle x^2 \rangle x_n + \ldots
\end{aligned}
\end{equation}
to cubic order.
We use the substitutions \eqref{cmftrf} and \eqref{beta} and insert the quadratic expansions \eqref{yncmf}, \eqref{Rcmf} to finally obtain, with the prerequisite stamina,
\begin{align}
\label{A=}
%&\resizebox{.5\hsize}{!}{$\displaystyle{
&A= \tfrac{d-1}{4 \gamma'^{2}d} \left(\gamma'^{2}+(d+1)^2 \right) \left( \gamma'^2-3(d-1)^2 \right) \,; \\%}$}\\
\label{B=}%&\resizebox{.9\hsize}{!}{$\displaystyle{
&\begin{aligned}
B=-\tfrac{1}{d} \left( \tfrac{d-1}{4 \gamma'^2 d} \right)^2 &\left( \gamma'^2+(d+1)^2 \right) \left(\gamma'^2+(d-1)^2 \right)  \cdot \\
\cdot &  \left( (\gamma'+ d)^2 + 2d^2 - 3 \right) \left( (\gamma'-d)^2 + 2d^2-3 \right) \,; %}$}
\end{aligned}\\
\label{C=} %&\resizebox{.8\hsize}{!}{$\displaystyle{
&\begin{aligned}
C=\tfrac{1}{d} \left( \tfrac{d-1}{4 \gamma'^2d} \right)^2 
\bigg( \gamma'^8-&4(2d^3-7d^2+1) \gamma'^6 -2(8d^5+ d^4- 56 d^3+22 d^2 +1) \gamma'^4- \\
-&4(d+1)^3(d-1)^2(2d^2+3d-3) \gamma'^2 + 9 (d^2-1)^4 \bigg) \, .
\end{aligned}%}$}
\end{align}
(As is always recommended for such calculations, these results were obtained, independently, by the authors K.F. and B.F.; see also our companion paper \cite{kf20}.)

In particular, scaling \eqref{scale} for nonzero $A, B$ and truncation to cubic order lead to the cubic normal form \eqref{ODExn} studied in the previous sections.
The one remaining cubic coefficient $c=C/B$, according to \eqref{cdef}, then becomes
\begin{equation}
\label{c=}
c= \tfrac{\gamma'^8-4(2d^3-7d^2+1) \gamma'^6-2(8d^5+d^4-56d^3 +22d^2+1)\gamma'^4-4(d+1)^3(d-1)^2(2d^2+3d-3)\gamma'^2+9(d^2-1)^4}{- \left( \gamma'^2+(d+1)^2 \right) \left(\gamma'^2+(d-1)^2 \right)  \left( (\gamma'+d)^2 + 2d^2 -3 \right) \left( (\gamma'-d)^2+ 2d^2-3 \right)} \ .
\end{equation}

%%%%%%%%%%%%%%%%%%%%%%%%%%%%%%
%%%%%%%%%%%%%%%%%%%%%%%%%%%%%%

Our results are summarized in the contour plot \ref{fig_7_1} of $c \in \mathbb{R}$.
First we note that the rational function $c=c(\gamma', d)$ of \eqref{c=} is even in $\gamma'$.
This follows from the parameter symmetry $(\gamma,\beta_I) \mapsto (-\gamma,-\beta_I)$ under complex conjugation of all $W_n$ in the Stuart-Landau system \eqref{sln}; see \cite{KeDiss}.
We can therefore omit negative $\gamma'$ and only consider $d, \gamma'>0$.
We recall the expressions \eqref{d} and \eqref{beta} for $d$ and $\gamma'$, in terms of the original coefficients $\gamma \in \mathbb{R}$ and $\beta \in \mathbb{C}$ of the coupled Stuart-Landau system \eqref{sln}.
The coefficient $\gamma$ regulates the soft-/hard-spring  characteristic of the individual Stuart-Landau oscillator, i.e.~the monotone dependence of period on amplitude.
Complex linear mean field coupling is regulated by $\beta$.
Colors in the contour plot \ref{fig_7_1} indicate the seven intervals of $c$ which are complementary to the six critical levels 
\begin{equation}
\label{ccri}
	c\quad =\quad -2,\ -\tfrac{3}{2},\ -\tfrac{4}{3},\ -\tfrac{5}{4},\ -1,\ -\tfrac{1}{2}\,,
\end{equation}
as identified in section \ref{rebdy}.
We have further split the blue intervals $c<-2$ and the red intervals $c>-1/2$, for clarity of the contour plot.
Note how values $c$ in all intervals do occur, for suitable parameters $\gamma$ and $\beta$.
The associated seven rebel flows with parameter $\lambda=\mu_+$ have been established and discussed in sections \ref{resul1} -- \ref{resul7}.

For further illustration we relate our present results to the detailed discussion of the \emph{2-cluster singularity} in \cite{KeDiss, KeHaKr2019} and in the companion paper \cite{kf20}.
By definition, the 2-cluster singularity refers to the bifurcation at the Benjamin-Feir instability $\lambda=0$ of an odd nonlinearity $A=0$ in the dynamics \eqref{ODEgen} on the center manifold.
% This specifies a point of codimension two in figure \ref{fig_7_2}.

From the outset, we note that any analysis of 2-cluster equilibria is subsumed as $N_2=0$ in our present setting. 
Therefore such results hold for all $N$, and are not restricted to any asymptotics of large $N$.
This extends to the bifurcation curves of rebel 3-cluster equilibria, at the blocking curves.
Indeed, the defining kernels of the linearization are independent of the size of the bifurcating cluster.
See (4.39) and (4.40) in \cite{elm}, \cite{stelm}, and our discussion of 2-cluster instability in section \ref{rebdy}.
The very value $A=0$, however, is oddly absent in our scaled asymmetric version \eqref{ODExn}, due to the singular scaling \eqref{scale} with $\tau=B/A^2$.

We can easily determine the 2-cluster singularities in the parameters $\gamma',d$ of figure \ref{fig_7_1}. 
Indeed, $A=0$ in our derivation \eqref{A=} is equivalent to the pair of straight lines 
\begin{equation}
\label{A=0}
\gamma'^2=3(d-1)^2\,.
\end{equation} 
Quite remarkably, insertion of \eqref{A=0}, to eliminate $\gamma'$, collapses the formidable expression \eqref{c=} of the cubic coefficient $c$ in the scaled center manifold dynamics \eqref{ODExn}, along these lines, to become
\begin{equation}
\label{c=d-2}
c=d-2\,.
\end{equation}

%%%%%%%%%%%%%%%%%%%%%%%%%%%%%%
%\begin{figure}[t] 
%\centering
%	\includegraphics[width = 1.00\textwidth]{t2_numeric}%{fig_7_2}
%	\caption{
%	The 2-cluster singularities with size ratios $\alpha = N_1/N_3= 1:15,\ldots, 6:10$  for $N=16$ oscillators in the all-to-all coupled Stuart-Landau setting \eqref{sln}.
%	In the center manifold formulation \eqref{ODEgen} the 2-cluster singularities correspond to $A=0$, where the nonlinearity is odd.
%	FUTURE: In the case $c=1$ of this figure, parameters are $\beta= 2\,-\,4/\sqrt{3}\,\textrm{i},\ \gamma=5/\sqrt{3}$.
%	CURRENTLY: In the case $\gamma=2$ of this figure, parameters are $c=-\tfrac{3}{2}+\sqrt{3},\ \beta= 2\,-\,4/\sqrt{3}\,\textrm{i}$.
%	The two bifurcation curves $\lambda_\iota(A)$, for $\iota=0,1$, are parabolas, asymptotically, with quadratic coefficients given by the blocking coefficients $\lambda_\iota(\alpha)/B$, respectively. 
%	See \eqref{lABC}.
%	Quite analogously, the third bifurcation curve $\lambda_\textrm{sn}(A)$, of 2-cluster saddle-node bifurcations, is characterized by the quadratic coefficient $\lambda_\textrm{minmax}/B$. 
%	See table 7.1 for the predicted quadratic coefficients.
%	Numerical simulations agree, with a relative error of order \textbf{ ?????????????}.	}
%	\label{fig_7_2} 
%\end{figure}
%%%%%%%%%%%%%%%%%%%%%%%%%%%%%%

Conversely, for given $c>-2$, we can now invoke \eqref{A=0}, \eqref{beta}, and \eqref{g'}, successively, to determine the parameters of the 2-cluster singularity as
\begin{equation}
\label{dgamma'}
d=c+2,\qquad \gamma'  =  \sqrt{3}\, (c+1)\,. 
\end{equation}
Since \eqref{A=0} is purely quadratic, we may in fact replace any occurrence of $\sqrt{3}$, here and below, by $-\sqrt{3}$.
For brevity, we will only address the positive sign.

At $\lambda=0$, relations \eqref{beta} then determine the original parameters $\beta,\gamma$ as
\begin{align}
\label{betac}
\beta &=(c+1) - \textrm{i}\,\tfrac{1}{\sqrt{3}}\, (c+3)\,,  \\
\label{gammac}
 \gamma &= \tfrac{1}{\sqrt{3}}( 2c+3) \,.
\end{align}
Insertion of \eqref{dgamma'} in \eqref{B=} and \eqref{C=}, respectively, determines the modest expressions
\begin{align}
\label{Bc=}
&B=-\tfrac{8}{3} (c+1)^2 (c^2+3c+3)/ (2+c)\,,\\
\label{Cc=} 
&C = Bc\,.
\end{align}
Of course we may just as well invoke \eqref{gammac}, anytime, to alternatively express all other parameters in terms of the soft/hard spring constant $\gamma$ of \eqref{sln}, at the 2-cluster singularity.
In fact, even case \ref{resul1}, $c<-2$, of 2-cluster singularity discriminants $-d=c+2$ occurs, albeit at the expense of a repelling center manifold associated to $\lambda=\mu_-=0$, with an unstable transverse eigenvalue $\mu_+=-2(c+2)>0$; see \eqref{mu}.

In the language of section \ref{resul}, each size ratio $\alpha=N_1:N_3$ gives rise to up to three particular nonzero bifurcation values of the parameter $\lambda$ in the scaled center manifold dynamics \eqref{ODExn}: 
the red saddle-node value $\lambda_\textrm{minmax}$ of \eqref{lamminmax} and the two blocking values $\lambda_\iota\,,\ \iota=0,1$ of \eqref{lam0}, \eqref{lam1}.
To recover the meaning for the full set of coefficients $\lambda, A, B, C$ in the general, unscaled center manifold setting \eqref{ODEgen}, we just have to revert the scaling \eqref{scale}.
The parameter values $\lambda$ in \eqref{ODEgen}, which correspond to each of the above reference values $\lambda_\iota,\ \iota\in\{\textrm{minmax},0,1\}$, for fixed $\alpha$, are then given by the asymptotic parabolas
\begin{equation}
\label{lABC}
\lambda = (\lambda_\iota(\alpha)/B) A^2 +\ldots \,.
\end{equation}
This shows how all bifurcation curves emanate from the 2-cluster singularity at $A=0$, $\lambda=0$, with horizontal tangent and curvatures given by the one remaining coefficient $c$ and the size ratios $\alpha$.
Higher order terms in $A$ transcend our third order truncation of the flow \eqref{ODEgen} in the center manifold, and also involve dependencies of the coefficients $A,B,C$ on $\lambda$.

See figures 1-4 in \cite{kf20}, for numerical illustrations of the 2-cluster singularity in Stuart-Landau oscillators \eqref{sln} with $\gamma=2$.
Specifically, size ratios $\alpha=N_1/(N-N_1)$ for the special case of $N=16$ oscillators and $N_1=1,\ldots,8$ are addressed there.
By \eqref{gammac}, the value $\gamma=2$ corresponds to the simplest case $c= \sqrt{3}-3/2$ of section \ref{resul7}.
The rebel flow is illustrated, for the equivalent cousin $c=1$ in the same interval $-1/2<c<+\infty$, in fig.~\ref{fig_6_8}. 
The complex value of the coupling constant $\beta$ at the 2-cluster singularity follows from \eqref{betac}.

\section{Conclusions}\label{conc}

Our results go beyond the discussion of 2-cluster equilibria and their local in-/stability.
In fact, we have studied rebel heteroclinic migrations between two large clusters, in the limit of large $N$.
We have encoded the dynamics of concatenated heteroclinic rebellions, in one-parameter families of vector fields, by the novel concept of rebel flows.
For each of the seven complementary intervals of the critical cubic coefficient $c$ in the center manifold dynamics \eqref{ODExn}, we have represented the resulting rebel flows of section \ref{rebdy}, between the two large clusters $(N_1\,, \xi_1)$ and $(N_3 \,, \xi_3)$, in figures \ref{fig_6_1}--\ref{fig_6_8} of sections \ref{resul1} -- \ref{resul7}. 
Since $N\nearrow +\infty$ is finite, in practice, we have to interpret these figures on the grid of rational values $N_1/N$, of course, for cluster sizes $N_1=1, \ldots, N-1$.
See figures \ref{fig_5_1}, \ref{fig_5_2} for the appropriate interpretation of heteroclinic rebel transitions.

The seven planar rebel flows encode the full heteroclinic rebel dynamics between large 2-clusters, for any bifurcation parameter $\lambda$, any nondegenerate cubic coefficient $c$ of \eqref{ODExn}, and arbitrary size ratios $\alpha$ between the large clusters.
All seven cases admit 2-cluster singularities.

In the setting \eqref{sln} of coupled Stuart-Landau oscillators, this establishes and explains the transient rebel dynamics of single oscillators between the two large clusters of synchronization, and the 2-cluster singularities of the Benjamin-Feir instability, as observed in simulations \cite{KeDiss, KeHaKr2019} and detailed further in our companion paper \cite{kf20}.

A quite analogous rebel approach can elucidate the $S_N$-equivariant heteroclinic dynamics of \eqref{ODExn} between 3-clusters, where the size of each cluster represents a nontrivial fraction of large $N\rightarrow\infty$.
This leads to two competing rebel flows which can be followed, alternatingly.
The situation then becomes reminiscent of linear, hyperbolic, second-order wave equations, with two vector fields of their associated characteristics.
Limited by brevity, regrettably, we cannot present further details here.

%%%%%%%%%%%%%%%%%%%%%%%%%%%%%%%%%%%%%%%%%%%%%%%%%%%%%%%
\bigskip
%\newpage

\end{document}